\documentclass[twocolumn]{aastex63}
\newcommand{\psj}{Planet. Sci. J.}

\submitjournal{The Planetary Science Journal}

\graphicspath{{./}{figures/}}

\usepackage{tikz}
\usetikzlibrary{arrows,shapes,positioning,shadows,trees}

\tikzset{
  basic/.style  = {draw, text width=4cm, drop shadow, font=\sffamily, rectangle},
  root/.style   = {basic, rounded corners=2pt, thin, align=center, fill=blue!10},
  level 2/.style = {basic, rounded corners=6pt, thin,align=center, fill=pink!40, text width=11em},
  level 3/.style = {basic, thin, align=center, fill=white!60, text width=8em}
}

\usepackage[utf8]{inputenc}
\usepackage{dirtytalk}
\usepackage{newtxmath}
\usepackage{newtxtext}

\usepackage{natbib}

\begin{document}

\title{An 800-Million-Year-Old Impact Shower on the Terrestrial Planets from the Breakup of the Eulalia Parent Body}


\author[0000-0002-1804-7814]{William F. Bottke}
\affiliation{Southwest Research Institute, 1301 Walnut St., Suite 400, Boulder, CO 80302, United States}
\author[0000-0002-6034-5452]{David Vokrouhlick\'y}
\affiliation{Astronomical Institute, Charles University, V Hole\v{s}ovi\v{c}k\'ach 2, CZ 18000, Prague 8, Czech Republic}
\author[0000-0001-6537-9339]{Melissa Dykhuis}\affiliation{Independent Researcher, Boulder, CO 80302, United States}
\author[0000-0002-2219-5043]{Nicolle Zellner} 
\affiliation{Department of Physics, Albion College, Albion, MI, USA}

\begin{abstract}

Multiple studies have proposed a substantial surge in large lunar impacts approximately 800 million years ago (Ma). Some are based on analyses of the ages of large lunar craters, such as the 93 km Copernicus crater. Others focus on the age distributions of impact glasses returned by lunar missions. A key challenge has been identifying and testing a plausible source for this putative impact spike. Here we use collisional and dynamical models to link this event to the formation of the Eulalia asteroid family, whose primitive carbonaceous chondrite-like parent body disrupted $\sim 800$~Ma near the 3:1 mean motion resonance with Jupiter (J3:1). Our simulations indicate that approximately three-quarters of the family’s fragments eventually entered the J3:1 over a $\sim 150$-million-year interval. While some fragments were injected into the resonance immediately after the disruption, others migrated more gradually via non-gravitational (Yarkovsky) thermal forces. Once in the J3:1, the fragments were dynamically transported into the planet-crossing region, leading to an elevated rate of bombardment on the Moon and terrestrial planets. Our results demonstrate that the Eulalia breakup can plausibly account for the observed lunar craters formed near $800$~Ma. Intriguingly, this event may also have had widespread repercussions across the inner Solar System. On Earth, its timing coincides with significant shifts in the biosphere, possibly linked to large impacts. On Mars, these impacts might have triggered a pulse of volcanic activity. Together, they showcase how certain catastrophic collisions in the main belt can have far-reaching consequences for the history of the terrestrial planets.

\end{abstract}


\section{Introduction}\label{intro}


The role that impacts have played in shaping the origin and evolution of life is poorly understood. While the heavily cratered surface of the Moon serves as a stark reminder that large bodies have struck in Earth’s past, so far only the Chicxulub impact event 66 million years ago (Ma) has been strongly linked to a mass extinction event \citep[e.g.,][]{alvarez1980,schu2010,nes2021}. A major challenge to any investigation lies in the fact that well-preserved craters (or associated impact signatures) in Earth’s geological record are relatively uncommon. Only a few tens of terrestrial craters larger than 20 km in diameter ($D_{\rm crat}$) have been identified, and almost all of them are younger than 650 Ma \citep[e.g.,][]{kel2019, maz2019}. This means that identifying concrete evidence that impacts played a role in astrobiological upheavals older than 650 Ma is a challenge. As Carl Sagan famously put it, ``extraordinary claims require extraordinary evidence."

\begin{figure*}[t!]
 \begin{center}
 \includegraphics[width=0.95\textwidth]{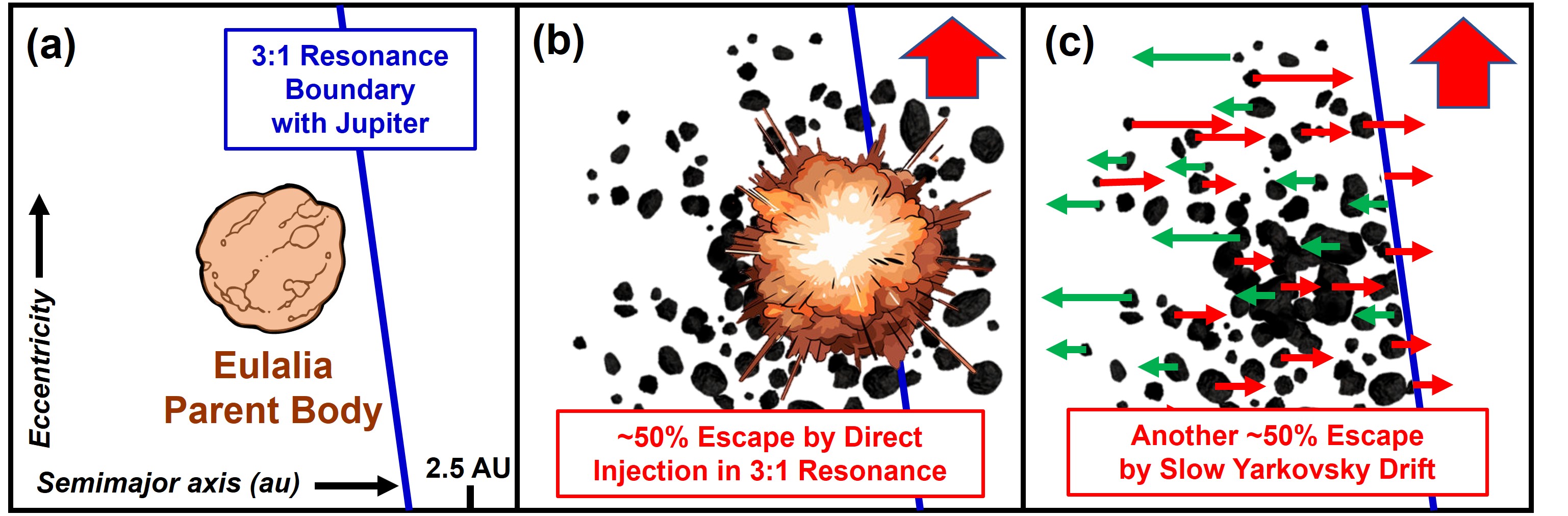} 
 \end{center} 
 \caption{Schematic illustrating the evolutionary stages of the Eulalia asteroid family. (a) The Eulalia parent body resides near the Jupiter 3:1 mean-motion resonance (J3:1). (b) Catastrophic disruption of the parent body occurs, with approximately 50\% of the fragments being directly injected into the J3:1, where they can be driven into the planet-crossing region. (c) Post-disruption dynamical evolution, during which an estimated ~50\% of the surviving fragments are expected to eventually migrate toward the J3:1 resonance via Yarkovsky thermal forces.} 
 \label{fig_eul}
\end{figure*}

One possible way to overcome this difficulty is to explore so-called asteroid showers. These rare events are triggered by large, well-positioned catastrophic collisions within the main belt. Fragments produced by these disruption events, if transported to planetary resonances either through direct injection or non-gravitational (Yarkovsky) forces, may find their way onto trajectories that allow them to hit the terrestrial planets. The timing and characteristics of asteroid showers can be reconstructed by examining the collisional and dynamical evolution of the fragments left behind in the main belt, and, where feasible, by determining the ages of large craters on the terrestrial planets.  

A significant advance on the latter was recently made by \citet{ter2020}. They estimated the ages of $D_{\rm crat} \geq 20$~km craters on the Moon by analyzing the smaller craters superimposed on top of them. Using a dataset of 59 such large craters that had formed over the last $3$ billion years (Gyr), they found $6–9$ with spatial crater densities similar to that of the $93$~km Copernicus crater, the largest crater formed during the Copernican era \citep[often described as less than $800$-1000~Ma; e.g.,][]{wilhelms1987}. The age of Copernicus is considered well-constrained, as one of its bright rays was sampled by the Apollo~12 astronauts. Materials attributed to the ray have yielded ages that are approximately $800$~Ma \citep[][]{sr2001,barra2006}. Additional support for an asteroid shower near this time comes from the abundance of lunar impact glass ages clustering near $800$~Ma \citep[see Sec.~\ref{back3} and][] {zellner2009,zell2015}. Formed by lunar cratering events, many of these lunar glasses have compositions that are not consistent with an origin near or in Copernicus crater. Taken together, \citet{ter2020} hypothesized that an asteroid shower occurred on the Moon sometime near $800$~Ma and lasted for approximately $100$ to $200$~Myr, depending on which craters were considered part of the event.

A general rule of thumb suggests that for every impact on the Moon, approximately $20$ similar-sized or larger impacts occur on Earth \citep[e.g.,][]{band1973,maz2019}. This estimate is based on the ratio of the gravitational cross-sections of the Earth and Moon and assumes that the projectiles behave similarly to the present-day population of Earth-crossing objects. Using this value, we predict that a significant impact shower should have struck the Earth around 800~Ma as well, with many of the projectiles being as large as, or larger than, the one responsible for forming Copernicus crater. The key question is whether any of these hypothesized impacts left either a detectable imprint on Earth or tangible evidence of their influence on Earth's biosphere.

The $800$~Ma impact shower occurred during the late Neoproterozoic Era. This period is notable for several ``Snowball Earth" glaciations, during which ice sheets likely covered vast areas of land and possibly the oceans \citep[e.g.,][]{hoff2009, rooney2015}. Continental glaciation extended to low paleolatitudes during the Sturtian ($660–717$~Ma), Marinoan ($635–641$~Ma), and Gaskiers ($\sim 580$~Ma) intervals, though the latter may not have been a global event \citep{pu2016}. The Gaskiers glaciation occurred shortly before the so-called ``Cambrian Explosion", a time period $\sim 541$~Ma when there was a sizable diversification of multicellular animals \citep[e.g.,][]{peters2012}. 

We find it intriguing and potentially significant that the putative $800$~Ma impact surge is coincident with several abrupt transitions in Earth's biosphere. Examples include major changes in $\delta\, ^{13}$C isotope records \citep{wor2019}, widespread ocean anoxia \citep{lu2017}, and a possible substantial diversification of marine eukaryotic life \citep{cohen2015,knoll2014,rs2017}. This period is considered a relatively warm period, with little evidence for large scale glaciation events. We will discuss these events further in Sec.~\ref {imp1}. While no direct link between the $800$~Ma events and impacts has yet been established, it is worth considering whether such a connection might exist.

To strengthen the argument that an $800$~Ma impact shower actually occurred, we need to identify a credible source capable of delivering the appropriate number and sizes of small bodies to both the Earth and Moon at the right time. We propose that this surge may have originated from the catastrophic disruption of the Eulalia parent body in the inner main belt. This breakup produced a cluster of low-albedo, carbonaceous chondrite-like asteroids that share similar semimajor axes, eccentricities, and inclinations. We define such clusters as asteroid families, and they are named after the largest remnant body in the group, which, for the Eulalia family, is (495) Eulalia.

The evolution of the Eulalia and New Polana families have been studied in detail \citep[e.g.,][]{bot2015} because both are considered a plausible source of the near-Earth asteroids Bennu and Ryugu, two C-complex asteroids from which samples were returned by the OSIRIS-REx and Hayabusa2 missions, respectively \citep{nakamura2023,yokoyama2023,lauretta2024}. The Eulalia and New Polana families are located within the Nysa-Polana family complex and have C- or B-type taxonomies \citep[e.g.,][]{cell2001,cell2002,cam2010,cam2013,gm2012,wal2013,bot2015,deleon2016,pin2016, deleon2018,arredondo2021,delbo2023,arredondo2025}. The spectral signatures of family members in both families are, at best, only modestly different from one another, and may in fact may be compositionally identical when factors such as porosity, space weathering, and surface grain size are taken into account. Given that Bennu and Ryugu have compositions that closely match CI chondrites, a primitive class of carbonaceous chondrites \citep{nakamura2023,yokoyama2023,lauretta2024}, we predict that members of the Eulalia and New Polana families have similar compositional characteristics. 

The Eulalia parent body had a likely diameter of $D \geq 100-200$~km, and disrupted $\sim 800$~Ma next to the inner boundary of the 3:1 mean-motion resonance with Jupiter (hereafter J3:1; Fig.~\ref{fig_eul}a) \citep{bot2015}. Such events are exceedingly rare; although large-scale disruption events have occurred across the main asteroid belt over the past several billion years, only a small fraction have occurred next to prominent resonances that can delivery that material efficiently to planet-crossing orbits. 

The singular location of the Eulalia breakup enabled a substantial portion of its fragments to be directly injected into the J3:1, where they were efficiently transported onto planet-crossing orbits (Fig.~\ref{fig_eul}b). Over time, additional fragments would have gradually entered the J3:1 through the influence of Yarkovsky-driven non-gravitational forces, further contributing to the planet-crossing population (Fig.~\ref{fig_eul}c). We postulate that a small fraction of these bodies eventually collided with the terrestrial planets, potentially explaining the proposed $800$~Ma impact shower first suggested by \cite{zellner2009} and observed by \citet{ter2020}.

To test this hypothesis, we reconstructed the evolutionary history of the Eulalia family and estimated both the duration and intensity of its impact shower on the terrestrial planets. Our procedure was as follows. In Sec.~\ref{methods}, we began by identifying the surviving remnants of the Eulalia family. We determined their size-frequency distribution (SFD), which serves as a critical parameter for understanding the family's collisional evolution and its contribution to impact events. Next, in Sec.~\ref{results}, we conducted a series of dynamical simulations to estimate the age of the Eulalia family. Additionally, we quantified the material lost through the J3:1. In Sec.~\ref{biases}, we addressed potential observational selection effects that could influence the inferred SFD. Then, in Sec.~\ref{impactflux}, we evaluated the collisional evolution of the Eulalia family's remnants and calculated the collision probabilities of these fragments with the terrestrial planets. By synthesizing these data, we were able to determine the flux of impactors striking the Earth, Moon, Mars, and Venus around $\sim 800$~Ma. Our lunar findings were then compared to similarly-aged craters identified by \citet{ter2020}, allowing us to assess their consistency with the predicted Eulalia impact shower. With these results in hand, we discussed the broader implications of the Eulalia impact shower in Sec.~\ref{implications}. This section discusses how the delivery of small and large debris from the Eulalia breakup event around $\sim 800$~Ma might have affected the geological and atmospheric evolution of Earth as well as the geological histories of Mars and Venus. Finally, in Sec.~\ref{conclusions}, we summarized the key findings and conclusions of this study. 

We caution the reader that Secs.~\ref{methods}, \ref{biases}, and much of Sec.~\ref{impactflux} are not easy reading for those without expertise in several topics related to small bodies (e.g., Eulalia family identification against a complicated background population, collisional and dynamical evolution of Eulalia family members, the treatment of observational biases for the Eulalia family, impact rates on the terrestrial planets). To make this paper more user-friendly for non-specialists, we offer the following suggestions. Readers primarily interested in the overarching narrative are encouraged to focus on the background material in Sec.~\ref{back}, the final subsection of Sec.~\ref{impactflux}, which presents our synthesis results, and the conclusions in Sec.~\ref{conclusions}. Additionally, we recommend exploring Sec.~\ref{implications}, which discusses the potential implications of the Eulalia asteroid shower for Earth, Mars, and Venus. It is worth noting, however, that Section~\ref{implications} contains more speculative content intended to inspire and guide future research.

\section{Background}\label{back}

In this section, we review key indicators supporting an $800$~Ma impact shower, namely the model ages of large lunar craters, the measured ages of lunar impact glasses returned by the Apollo missions, and shock degassing ages observed in meteorites. In Appendix A, we also discuss our current understanding of Martian crater ages near $800$~Ma and examine whether the seemingly constant lunar impact flux for sub-100-meter bodies presents a potential challenge to the impact shower hypothesis.

\begin{figure}[t!]
 \begin{center}
 \includegraphics[width=0.47\textwidth]{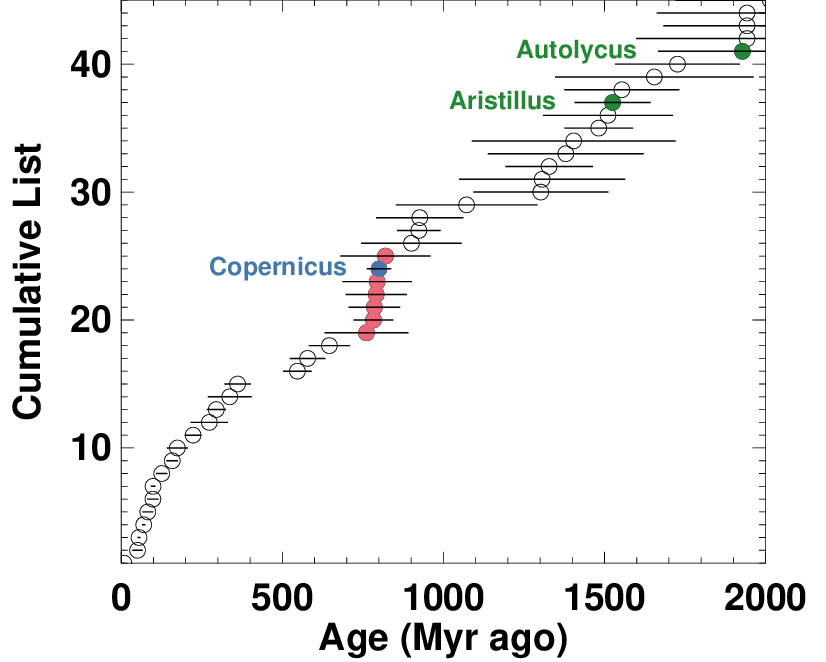} 
 \end{center} 
 \caption{A cumulative plot of the approximate ages of lunar craters larger than $D > 20$~km, according to \citet{ter2020}. The chronology used is provided in the text.  The sample-derived age of Copernicus (blue dot) is $800$~Ma, and it serves as our calibration point. The sample derived ages of Aristillus and Autolycus (green dots) $1400 \pm 60$~Myr and $1940 \pm 10$~Myr, respectively, match the model ages of $1510 \pm 140$~Myr and $1930 \pm 190$~Myr, respectively. The craters with red dots have nearly the same $N(1)$ values as Copernicus. They serve as our baseline for how the Eulalia impact shower affected the Moon.} 
 \label{fig_craters}
\end{figure}

\subsection{Estimates of Lunar Crater Ages}\label{back1}

Our primary evidence supporting the idea of an $800$~Ma impact surge is presented by \citet{ter2020}. Examining 59 relatively fresh lunar craters with $D_{\rm crat} \geq 20$~km, they measured the spatial densities of $D_{\rm crat} \approx 0.1-1$~km diameter craters superposed on their crater floors and ejecta blankets. From these data, they estimated the cumulative number of kilometer-sized or larger craters that formed on top of each large crater per square kilometer. These values were defined as $N(1)$.

By using the $N(1)$ values from \cite {ter2020}, we can establish a basic timeline for the formation of large lunar craters. As mentioned above, Copernicus crater (93 km in diameter) has a sample-derived age of $800$~Ma \citep[e.g.,][]{sr2001, stoff2006}. Assuming that the production rate of small craters has been constant for the last $3$~Gyr (see Appendix~\ref{appa2}), we can estimate the approximate absolute ages of other large lunar craters by scaling from their $N(1)$ values. Our results for $45$ of these craters are shown in Fig.~\ref{fig_craters}. 

To validate our model crater ages, we compared our results with the independently determined absolute ages of the lunar craters Aristillus (55~km) and Autolycus (39~km). Ejecta from these craters is thought to have thermally reset zircons collected at the Apollo 15 landing site. Based on the dating of these reset zircons, \citet{gra2013} determined the ages of Aristillus and Autolycus to be $1400 \pm 60$~Ma and $1940 \pm 10$~Ma, respectively. In comparison, our model yields ages of $1510 \pm 140$~Ma and $1930 \pm 190$~Ma, respectively. Given that our model age results are consistent with sample ages within uncertainties, we find no evidence of significant bias affecting the \citet{ter2020} findings.

Accepting the premise that our model crater ages are reasonable, we can now examine the cratering events that took place on the Moon approximately $800$~Ma. We find 7 craters whose mean ages are $760$ to $820$~Ma: Al-Khwarizmi (22.5~km), Stefan (26~km), Saha (29.4~km), Godin (35.1~km), Das (36.6~km), Lowell (65.5~km), and Copernicus (93.1~km).  An additional three craters have mean ages between $900$ and $930$~Myr old: 54S (19.7~km), Klute (30.2~km), and Stevinus (70.3~km). Their model ages with errors are shown in Fig.~\ref{fig_craters}. Collectively, they suggest numerous multikilometer impactors were striking the Moon in the $\sim 800$~Ma epoch. 

\subsection{Nature of the background impact flux 800 Ma}\label{back2}

Our next step is to analyze the nature of the background impact flux during the 800~Ma era. If it was particularly high, it would obscure the visibility of our putative asteroid shower. To quantify the nature of the background flux from 800 Ma, we need to interpret the meaning behind some of the features in Fig.~\ref{fig_craters}. For example, crater ages from Fig.~\ref{fig_craters} suggest that the production rate of $D_{\rm crat} \geq 20$~km lunar craters increased by a factor of $\sim 2$ to 3 approximately $300$~Ma. This change matches predictions made by \citet{maz2019} for large craters on both the Earth and Moon. We briefly discuss their results below, partly because their method for dating crater ages differs from that of \citet{ter2020} but also because their interpretation of the terrestrial crater record will be used in Sec.~\ref{implications}. 

When large lunar craters form, they excavate fresh rocks and boulders. These newly exposed rocks exhibit high thermal inertia, allowing them to retain heat and remain warmer during the lunar night compared to the surrounding lunar soil. By utilizing data from the Diviner thermal radiometer aboard the Lunar Reconnaissance Orbiter, crater-adjacent temperatures can be analyzed to estimate rock abundance. Since larger rocks degrade over time due to micrometeorite impacts and other surface processes, rock abundance can serve as a proxy for estimating the age of the craters. \citet{maz2019} used this method to estimate the ages of 111 lunar craters with $D_{\rm crat} \ge 10$~km. Their method was most effective for craters younger than several hundreds of Myr; limited data prevented them from detecting an asteroid shower dating back to 800 Ma. 

The trends identified by \citet{maz2019} in lunar crater ages were then compared with those found among 38 terrestrial craters with $D_{\rm crat} \ge 20$~km and radiometric ages $\lesssim 650$~Myr. Note that 85\% of these terrestrial craters are located on geologically stable cratons. These regions show minimal evidence of significant erosion until approximately $650$~Ma, a time that coincides with Earth’s last ``Snowball Earth" event \citep[e.g.,][]{black2012,maz2019,maz2019b,kel2019}. 

When put together, the combined lunar and terrestrial data indicated an increase in the recent impact flux of a factor 2.6, with a 95\% credible interval value of 1.7 to 4.7, and that the change occurred approximately $290$~Ma. Note that a lower impact flux beyond 290 Ma explains the relative scarcity of large terrestrial craters found between $290$ and $650$~Ma; their relative paucity would not be from erosion or observational bias \cite {maz2019,maz2019b}. 

The similarity in results between \citet{maz2019} and \citet{ter2020} give us reasonable confidence that we can predict the background impact flux using Fig.~\ref{fig_craters}. It shows that the impact flux between $650$ and $750$~Ma is as low as the levels predicted between $\sim 300$ and $650$~Ma. Assuming the background flux changes slowly, which we would argue is a likely scenario given our understanding of the collisional and dynamical evolution of main belt asteroid families, this implies the $800$~Ma event should stand out from the background flux in that era.  

\begin{figure}[t!]
 \begin{center}
 \includegraphics[width=0.47\textwidth]{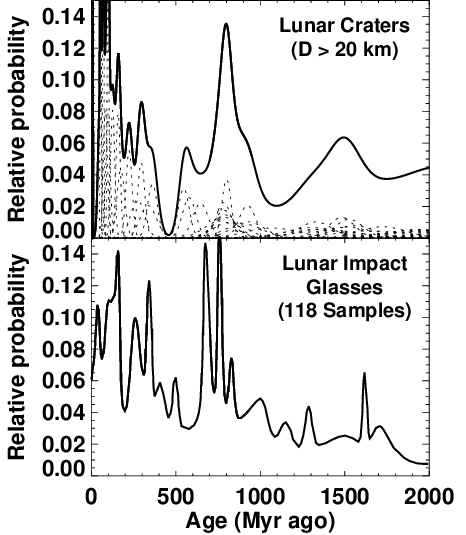} 
 \end{center} 
 \caption{A comparison between the model ages of lunar craters with diameter $D > 20$~km and those of $118$ lunar glasses taken from Apollo regolith samples. The craters ages are from Fig.~\ref{fig_craters}, while the ages for the glasses are provided by \cite {ghent2021}. We represent the mean age and 1~$\sigma$ age uncertainties as Gaussians distributions, with all of the distributions added to produce the observed distribution. Notable age spikes are observed around $800$~Ma in both distributions, along with an increase in events over the past $300$~Myr or so.} 
 \label{fig_glass}
\end{figure}

\subsection{Ages of lunar glasses near 800~Ma}\label{back3}

Lunar impact glasses are formed during cratering events. The heat generated from the impact is intense enough that it can melt and vaporize some of the ejecta thrown away from the site. As the molten material rapidly cools, it can solidify into tiny glass beads, typically less than a millimeter in size. These glasses exhibit a variety of shapes, including spheres, teardrops, dumbbells, and other irregular configurations. 

Many lunar impact glasses have been identified in regolith samples returned by the Apollo astronauts \citep[e.g.,][]{culler2000,zellner2009,zell2015} and by more recent missions like Chang’e-5 \citep{long2022}. Their formation times, determined using radiometric dating methods based on the $^{40}$Ar/$^{39}$Ar, U-Th-He, and U-Pb systems, reveal the ages of their host craters. In this manner, lunar impact glasses provide valuable insights into the history of lunar bombardment. 

For example, the impact ages of twelve individual glass samples from the Apollo~17 landing site indicate that they likely originated from nine distinct impact events, depending on the selected compositional groupings \citep {zellner2009b}. The challenge is to establish a connection between specific lunar glass ages and their corresponding host craters, as this linkage can provide context and help us better understand the significance of these ages.

Using compositional information from lunar glasses returned by Chang'e-5, which landed on a $\sim 2$~Ga basaltic region in north Oceanus Procellarum near the Mons R\"umker volcanic complex, \citet{long2022} argued that most lunar glasses travel less than $150$~km from their point of origin and are predominantly formed by small impact craters ranging from $1$ to $5$~km in diameter. If these findings are representative, the age distributions of lunar glasses derived from individual soil samples likely contain a sizable stochastic component, as they primarily reflect craters that formed near the sample collection site. Therefore, we would expect impact glasses returned from a variety of locations across the Moon to exhibit distinct age distributions unique to each site. This makes it surprising that certain trends have been identified in the $^{40}$Ar/$^{39}$Ar age profiles of lunar impact glasses collected from the Apollo~12, 14, 16, and 17 sites \citep[][]{culler2000,zellner2009,zell2015,zellner2019}.  

For example, using those glass ages known as of 2009, \citet{zellner2009} was the first to propose that a lunar impact shower took place around 800 Ma. They reported a clustering of $^{40}$Ar/$^{39}$Ar ages near $800$~Ma in nine glass samples from the Apollo~14, 16, and 17 landing sites. Similar ages were found in glasses retrieved from the Apollo~12 landing site \citep{levine2005,barra2006}. Evidence suggests that at least seven separate impact events contributed to the formation of these glasses \citep{zellner2009}. The diverse compositions of the glasses helped to mitigate concerns that all were derived from the Copernicus impact event \citep{huang2018}. 

The combined $^{40}$Ar/$^{39}$Ar age profiles of $118$ lunar impact glasses of varying sizes, presented as a probability distribution in \citet{ghent2021}, are shown in Fig.~\ref{fig_glass}b. This distribution was constructed by summing the Gaussian age profile of each sample, where the mean corresponds to the most probable age and the standard deviation ($1~\sigma$) reflects the uncertainty of the dating measurement.

For consistency in making comparisons, we applied the same method to convert the crater ages shown in Fig.~\ref{fig_craters} into the age probability distribution presented in Fig.~\ref{fig_glass}a. Both Figs.~\ref{fig_glass}a and \ref{fig_glass}b reveal ages near $800$~Ma, likely indicative of numerous cratering events occurring near that time.    

The probability distributions presented in Figs.~\ref{fig_glass}a and \ref{fig_glass}b also show an increase in the abundance of young glasses over the past few hundred million years. The trend in the age distribution of lunar glasses has been noted by several groups \citep[e.g.,][]{culler2000,levine2005,zellner2009,zell2015}. As discussed above, this pattern could reflect a genuine increase in the lunar impact flux taking place near $300$~Ma \citep{maz2019, ter2020}. Alternatively, it could partly or fully represent a survival bias, in which older glasses are progressively destroyed by impact-driven regolith gardening processes \citep{zell2015}. This potential selection effect may be exacerbated by the relatively shallow sampling conducted by the Apollo astronauts, who would preferentially collect younger samples located on or near the surface \citep{huang2018}. We note that after correcting for an argon diffusion bias in spherules from Apollo~14, 16, and 17 regolith samples, \citet{zell2015} observed a more uniform age distribution over the past billion years. 

Another complicating factor in comparing the age distributions in Figs.~\ref{fig_glass}a and \ref{fig_glass}b is that the lunar glasses may have been produced by impacts of different sizes. According to arguments made by \citet{long2022}, the ages in Fig.~\ref{fig_glass}b primarily reflect $D \ge 1-5$~km craters, while Figs.~\ref{fig_glass}a is derived from the model ages of $D \ge 20$~km craters reported by \citet{ter2020} \citep[see also][]{maz2019}. As discussed in Sec.~\ref{model2}, smaller projectiles have shorter collisional lifetimes, making it challenging to devise realistic scenarios in which their flux could be sustainably enhanced over hundreds of Myr. Nevertheless, the correlations observed in Figs.~\ref{fig_glass}a and \ref{fig_glass}b are evident, making it difficult to dismiss them solely as coincidence or bias \citep[e.g.,][]{zell2015,huang2018}. Ultimately, fully interpreting the age distribution of lunar spherules and shards requires both careful analysis and a good understanding of their contextual background.

\subsection{Meteorite shock degassing ages near 800~Ma}\label{back4}

Surprising evidence for impact events taking place near $800$~Ma can be found in the $^{40}$Ar/$^{39}$Ar shock degassing ages observed in several meteorite classes. These ages may arise from asteroid cratering events, during which argon is lost if the target rock is exposed to sufficiently high temperatures ($\gtrsim 1000$~K) for an extended period. Consequently, impacts that generate prolonged, intense heating are the most likely to leave a detectable record. The challenge lies in determining which specific impacts are capable of producing this type of thermal history.

Numerical hydrocode simulations indicate that impact velocities exceeding 10 km~s$^{-1}$ significantly enhance heating, with the volume of target material acquiring a $^{40}$Ar/$^{39}$Ar shock degassing age increasing by nearly three orders of magnitude compared to impacts at the more typical main belt collision velocities of $\sim 5$ km~s$^{-1}$ \citep[e.g.,][]{pier1997,marchi2013}. The reason is that only these high impact velocities can generate the necessary shock pressures and temperatures on a typical target surface \citep[unless the target material is highly porous, e.g.,][]{pm1998}. The problem is that impact velocities larger than $10$ km~s$^{-1}$ are rare between main belt asteroids \citep[e.g.,][]{bot1994,bot1996}. To achieve such impact speeds, projectiles typically need to follow highly eccentric and/or inclined orbits like those of objects on Earth-crossing trajectories \citep{bot1996}.

Another factor to consider when interpreting the $^{40}$Ar/$^{39}$Ar ages is the geological context in which the heating event occurred. For instance, when an impactor strikes a small asteroid, a significant portion of the ejecta, including material subjected to high temperatures, is typically lost to space. Conversely, when an impactor strikes a larger asteroid, with a high escape velocity, more of the highly heated materials stays behind within the breccia lens or ejecta blanket of the crater \citep[e.g.,][]{pier1997,marchi2013}. Similar trends are expected for impact melt. On smaller asteroids, melt pools cool rapidly during cratering events, leading to minimal melt production that generally has little to no effect on the surrounding rock. A comparable limitation arises when considering the catastrophic disruption of asteroids, as such events typically raise asteroid temperatures by only a few degrees \citep{keil1997}.

Consequently, we predict that significant impact heating events are more likely to occur in sizable crater events that occur on large asteroids, which are presumably better able to retain the hottest ejecta \citep {scott2011, scott2015}. These characteristics likely contribute to the observed rarity of impact melt features in ordinary chondrites, with less than $0.5$\% exhibiting such features \citep[e.g.,][]{keil1997,scott2011,scott2015}. It also potentially explains why so little impact melt appears to be associated with formation of the Rheasilvia and Veneneia basins on Vesta \citep {schenk2012}. These trends would also explain why few asteroid family members show signs that their bodies are inundated by impact melt, though the Baptistina family may be an exception \citep[e.g.,][]{reddy2011}.

We infer that to achieve the necessary conditions for meteorite shock degassing ages, it would be helpful to have high velocity impacts occur on larger bodies. These impacts would generate heat either within the buried breccia lens of substantial craters or in the surrounding ejecta blanket. These materials would presumably be ejected in a later collision event, with the debris making their way to Earth via a collisional cascade, Yarkovsky thermal drift, and main belt resonances \citep[e.g.,][]{bot2006}.

With this context in mind, we can try to interpret the $^{40}$Ar/$^{39}$Ar ages observed in various ordinary chondrite groups. Among the roughly two dozen shocked H chondrites with measured age data, the meteorites LAP 031308, Travis County, and Dimmett all exhibit $^{40}$Ar/$^{39}$Ar ages close to $800$~Myr \citep[e.g.,][see their Table~2]{swindle2009}. Of these, the first two meteorites provide the most reliable ages, while the age for Dimmett is considered more uncertain (J. Weirich and T. Swindle, personal communication). 

L chondrites shock ages are dominated by a prominent event occurring around $470$~Ma \citep[e.g.][]{haack1996,swindle2014}. Nevertheless, meteorites such as Cat Mountain, Yamato (Y) 74445, and Northwest Africa 091 suggest a significant impact event took place on the L chondrite parent body $\sim 700-800$~Ma \citep[e.g.,][]{swindle2011, swindle2014,ciocco2025}. Further discussion on the origin of the L chondrites can be found in \citep{nes2009, marsset2024, ciocco2025}, while an analysis of how the 470 Ma event may have affected the Earth can be found in Sec. 7.1.2.

In the case of LL chondrites, $^{40}$Ar/$^{39}$Ar ages younger than $4$~Ga are relatively scarce \citep {swindle2014}, but among the eight age measurements obtained from LL chondrites delivered by the Chelyabinsk bolide, one exhibits an age close to $800$~Myr \citep{righter2015}.

Taken together, we find that several $800$~Ma ages are found among the tested H, L, and LL meteorites. Based on our earlier discussion, this suggests that high-velocity projectiles were likely generating substantial craters on their parent bodies or perhaps on the largest remnants of their collisional families during that time. While it is possible that this could be a coincidental clustering, we favor the interpretation that the H, L, and LL parent bodies or their largest remnants were hit by material from the Eulalia family. In this scenario, Eulalia ejecta injected into the J3:1 would have been quickly transported to highly eccentric and inclined Earth-crossing orbits. While residing in these orbits, the fragments would have been well-positioned to bombard the main belt for a few Myr, potentially creating a limited number of craters on multiple large bodies. For these craters, hot material in the breccia lens or from ejecta landing near the crater would have produced the heating necessary to create the observed $^{40}$Ar/$^{39}$Ar shock degassing ages.  

Additional modeling work will be needed to test this hypothesis. 

\subsection{Summary}

Evidence from the ages of large lunar craters, lunar impact glasses, and shock degassing events in meteorites all suggests a potential spike in impact activity around $800$~Ma. This pattern is intriguing enough to warrant further investigation into possible sources of such an impact shower using collisional and dynamical modeling work. In the following sections, we detail our efforts to explore and evaluate this hypothesis in greater detail.

\section{Methods} \label{methods}

\subsection {A plausible source for an 800 Ma impact shower}\label{methods1}

Substantial changes in the flux of large impactors to the Earth-Moon system can be triggered by the formation of asteroid families within the main belt \citep[e.g.,][]{bot2007,vok2017}. Formally, a family is defined as the remnant of a substantial asteroid collision event that produced fragments with clustered proper semimajor axis $a$, eccentricity $e$, and inclination $I$ values \citep[e.g.,][]{netal2015,nov2022,nes2026}. Among the many tens of prominent asteroid families of varying sizes and ages \citep {nes2015, masiero2015}, we seek one capable of generating an impact surge starting around $\sim 800$~Ma. This means sifting the properties of candidate families against the following criteria.
\smallskip

\noindent{\it Asteroid family ages.} When a disruption event occurs, ejecta fragments large enough to be detected by ground-based observers are launched away from the impact site at velocities comparable to the parent body’s escape velocity \citep[e.g.,][]{betal2015}. From here, Yarkovsky thermal forces cause smaller asteroids to drift inward toward or outward away from the Sun faster than larger ones. The pattern created in semimajor axis vs absolute magnitude is diagnostic of the family’s formation age \citep[e.g.,][]{bot2006b,vetal2015}. Using methods discussed in \citet{wal2013}, we evaluated main belt families to find whether any formed near $800$~Ma.
\smallskip

\noindent{\it Location/Impact Probabilities.} Candidate families also had to (i) show orbital evidence they had lost numerous multikilometer-sized bodies (e.g., the family's orbital distribution has been truncated by a resonance capable of delivering many of those objects to planet-crossing orbits) and (ii) escaped members had to have reasonable probabilities of hitting the Moon \citep[e.g.,][]{bot2006b}.

One potential candidate that aligns with many of these characteristics is the Flora family, an S-type asteroid group located at the innermost region of the main belt adjacent to the $\nu_6$ secular resonance. The possibility that the Flora family produced an asteroid shower was investigated by \cite {vok2017}. Here we focus on their finding that members of the Flora family only replicate the observed distributions of semimajor axis, eccentricity, and inclination in the real family after undergoing $1.3 \pm 0.3$~Ga of dynamical evolution, assuming a bulk density for the family members of approximately $2.70 \pm 0.54$ g~cm$^{-3}$. This inferred dynamical age matches the estimated surface age of (951) Gaspra, a Flora family member observed by the Galileo spacecraft \cite {bot2020}, and the $^{40}$Ar/$^{39}$Ar ages of LL chondrite grains returned from (25143) Itokawa by the Hayabusa spacecraft \citep{nakamura2011, park2015, terada2018}. Since the age of the Flora family predates our target timeframe, we exclude it from consideration. The possibility that Flora family members produced some of the craters older than 1 Ga in Fig.~\ref{fig_craters} is left for future work. 

Our preferred candidate family is in the so-called Nysa-Polana complex of asteroids, a low inclination cluster of C-complex asteroids with C- and B-type taxonomies that overlaps the S-type Hertha family \citep[in what used to be called the Nysa family, though it is now recognized that Nysa is an interloper in this family e.g.,][]{cam2010,cam2013,dy2015}. Using Sloan Digital Sky Survey color data and WISE albedos, it can be shown that the Polana family is actually two overlapping families named Eulalia and New Polana by \cite{wal2013}. Extensive spectroscopic research has been conducted on the two families \citep[e.g.,][]{cell2001,cell2002,cam2010,cam2013,gm2012,wal2013,bot2015,deleon2016,pin2016, deleon2018,arredondo2021,delbo2023,arredondo2025}. When factors such as porosity, space weathering, and surface grain size are considered, it can be argued that the Eulalia and New Polana family members have reasonably similar spectroscopic signatures.

By modeling their origin and evolution, \citet{bot2015} found the following attributes for these two families: 
\begin{itemize}
\item Both likely came from $D > 100$~km parent bodies.
\item Both have possible connections to Bennu ($0.5$~km) and Ryugu ($1$~km), the carbonaceous chondrite-like targets of the OSIRIS-REx and Hayabusa2 sample return missions, respectively.  Other candidate families to produce these bodies were deemed to be too small, too young, or too poorly positioned (e.g., Clarissa, Erigone, Sulamitis).
\item New Polana was modestly favored over Eulalia as a source of Bennu and Ryugu by 70\% to 30\%.
\item Eulalia’s age was estimated to be $830^{+370}_{–100}$~Ma, while New Polana’s age was $1400^{+150}_{–150}$~Ma. 
\end{itemize}

Of the two families, the Eulalia family's age makes it the better candidate to produce the $\sim 800$~Ma impact surge, while remaining a plausible source for Bennu and Ryugu. This dual role for the Eulalia family may also offer an intriguing way to connect impactors from the family to the origin of Copernicus crater. Consider the following. 

As mentioned earlier, Apollo~12 samples appear to contain ejecta originating from Copernicus crater \citep[e.g.,][]{sr2001, stoff2006}. To that end, \citet{morgan1973} identified a minor component within Apollo~12 KREEP materials that exhibited a primitive meteoritic signature. Their conclusion was based on subtle but consistent enrichments of elements such as Ir, Re, Sb, Se, and Zn relative to Au. This observation prompted \citet{morgan1973} to suggest that the Copernicus impactor may have originated from a cometary nucleus. CI chondrites, the most primitive meteorites found in terrestrial collections, provide the closest match to this suggested projectile type. From a dynamical standpoint, however, the likelihood of a large Jupiter-family or nearly-isotropic comet forming Copernicus is extremely low \citep[e.g.,][]{bot2002,bot2015,granvik2018,neomod1,neomod2,neomod3}.

A possible solution to this conundrum may lie in the samples returned from the near-Earth asteroids Bennu and Ryugu by NASA's OSIRIS-REx and JAXA's Hayabusa2 missions, respectively. Both have compositions closely matching CI chondrites \citep{nakamura2023,yokoyama2023,lauretta2024}. Given that Bennu and/or Ryugu may have originated from the Eulalia or New Polana families \citep[e.g.,] [] {bot2015}, and both have comparable spectral signatures suggestive of CI chondrites \citep {broz2024}, this compositional connection would offer compelling albeit circumstantial evidence supporting the hypothesis that the Copernicus crater was formed by a member of the Eulalia family $800$~Ma. 

The largest remnant of the Eulalia family, (495) Eulalia, is located next to the sunward side of the J3:1 boundary (Fig.~\ref{fig_eul}a). If the ejecta from the parent body were dispersed isotropically, up to half of the family may have been directly injected into the J3:1 (Fig.~\ref{fig_eul}b). For the portion of ejecta launched away from the J3:1 boundary, assuming no initial preference for obliquity orientation, approximately half would be expected to drift outward from the Sun due to the Yarkovsky and YORP effects (Fig.~\ref{fig_eul}c). Over tens to hundreds of Myr, many of these fragments would eventually migrate to the J3:1, provided they avoided collisional disruption during their journey. Consequently, to a first approximation, approximately three-fourths of the Eulalia family meaningfully affected by the Yarkovsky effect may have entered the J3:1. In addition, numerous large family members formed near the family's center might have been sufficiently close to a chaotic zone adjacent to the J3:1, allowing them to enter the resonance as well. Our hypothesis is that this input was enough to trigger a significant impact shower on the terrestrial planets. 

In the next few sections, we revisit how the Eulalia family was shaped by collisional and dynamical processes, as well as exploring whether its ejecta contributed to an impact surge on the terrestrial planets. Achieving this goal requires reconstructing the original Eulalia family from its surviving members. Although our previous work in \citet{bot2015} provides a strong foundation for this task, further efforts were required to account for factors that were not fully addressed in the earlier modeling. Specifically:

\begin{itemize}
\item The Eulalia family overlaps the New Polana family and therefore must be carefully extracted to determine the current Eulalia size frequency distribution \citep{wal2013, bot2015}.
\item (495) Eulalia likely migrated $\sim 0.007$~au toward the J3:1 over $\sim 800$~Ma \citep[see][]{vok2016}. To find the quantity of ejecta injected into the J3:1 vs. how much migrated in afterwards, we must re-calculate the parent body’s original location using the latest Eulalia family data.
\item Collisional and dynamical losses over $\sim 800$~Ma within the Eulalia family must be accounted for in a self-consistent manner. 
\item The impact signature produced by the Eulalia family on the terrestrial planets has yet to be modeled. For the Eulalia fragments to have formed Copernicus crater (Fig.~\ref{fig_craters}), several thousand bodies with $D \approx 4-5$~km had to reach the J3:1 within timescales consistent with producing the impact surge.
\end{itemize}

In the sections that follow, we will explore the details of the Eulalia family’s evolution.

\subsection{Family identification and basics of Yarkovsky chronology} \label{ident}

\subsubsection{Pre-selection of the Eulalia family members}

We began our investigation of the Eulalia family region by updating the available databases of asteroid properties. They are defined as proper orbital elements of semimajor axis $a$, eccentricity $e$, sine of inclination $\sin I$, reflectance properties $a^\star$ and $i-z$ colors from the Sloan Digital Sky Survey, and albedo values from the WISE database. We obtained proper elements for 777,353 objects in the catalog provided by \citet{propel2024}, as well as albedo measurements for 134,314 objects from the Minor Planet Physical Properties Catalog (MP3C) of the Observatoire de la Côte d'Azur. For the color information, we made use of a new study by \citet{sc2021} that optimized the Sloan Digital Sky Survey original data in search of moving objects. Combined and cross-checked with the fourth release of the SDSS Moving Objects Catalog \citep{ivezic2002}, we increased the database of objects with SDSS colors to 306,856 objects. The results of this database compilation effort nearly double the number of inner main belt objects with reliable proper elements and increase the number with reliable albedos and colors by factors of 1.4 and 1.3, respectively.

\subsubsection{Yarkovsky chronology elements and final identification of the Eulalia family members} 

As described above, the Yarkovsky effect describes the secular change of an asteroid's heliocentric semimajor axis $a$ due to the absorption of sunlight and how that energy is re-radiated away in a non-isotropic fashion as heat \citep[e.g.,][]{bot2006,vetal2015}. The corresponding drift rate produced by this effect is $da/dt\propto D^{-1}$, where $D$ is diameter and the coefficient of proportionality may be positive or negative depending on whether the object's obliquity is positive or negative, respectively. In general, Yarkovsky thermal drift is most noticeable in $D \le 30$~km objects.  

The Yarkovsky effect is capable of modifying the orbital distribution of asteroid families. Family members with $D \le 30$~km will spread in proper element space in the $a$-direction, with the speed of each object determined by its size. Over a given family's age, smaller asteroids disperse more than the larger ones. This characteristic distribution pattern forms the basis of Yarkovsky chronology, where the breakup times of asteroid families can be deduced by modeling the dynamical evolution of their members \citep[e.g.,][]{vok2006,netal2015}. 

Smaller asteroids are also affected by the Yarkovsky–O'Keefe–Radzievskii–Paddack effect, or YORP effect. YORP modifies the spin vectors of asteroids, reorienting them over time to values near $0^\circ$ or $180^\circ$ that promote maximum Yarkovsky $da/dt$ drift \citep[e.g.,][]{bot2006,vetal2015}. This behavior has been verified by observations of thousands of asteroids by the Gaia spacecraft, most which show extreme obliquity values \citep[e.g.,][]{dh2023}. 

In an asteroid family, the combination of YORP torques working to optimize Yarkovsky drift inward and outward will eventually cause smaller family members to evacuate the family center. These objects become concentrated near the furthest semimajor axis values of the family. Assuming we know the initial ejection velocity spread of the family members, the family's distribution in $a$ can often be used to constrain its age \citep[e.g.,][]{nes2003,vok2006}. 

If the drifting objects can avoid being captured in mean motion or secular resonances with the planets, which can lead to their diffusion in eccentricity or inclination, family age modeling reduces down to an analysis of how the objects are distributed in $(a,D)$ space. In family chronology studies, it is common to exchange diameter $D$ for absolute magnitude $H$, changing the analysis to $(a,H)$ space, because $H$ is more directly available in the asteroid catalogs. This requires the user to make a conversion from $H$ to $D$ at a later stage in the analysis. This method works well if family members have similar albedos to one another, which is usually the case \citep[e.g.,][]{masiero2013,masiero2015}.    

Given that Yarkovsky-induced semimajor axis drift depends on diameter $D$, or magnitude $H$, asteroid family chronology can be further reduced from two dimensions (i.e., $(a,D)$ or $(a,H)$) to a one-dimensional parameterization. Such a reduction was introduced by \citet{vok2006}, who proposed that the family structure could be represented using the distribution of $C(a,H)$ values defined implicitly by
\begin{equation}
 H(C,a;a_{\rm c})=5\,{\rm log}\left(\frac{a-a_{\rm c}}{C}\right) \; . \label{cline}
\end{equation}
The map assigns each asteroid in the $(a,H)$ plane with a unique $C$ value, provided the family center $a_{\rm c}$ has been selected (note that $C$ is negative for $a<a_{\rm c}$ and vice versa). For families that are far from major mean motion resonances, $a_{\rm c}$ is typically very close to the proper semimajor axis value of the largest member. 

In the Eulalia case, however, the choice of $a_{\rm c}$ is uncertain. The largest member of the Eulalia family, (495) Eulalia, is very close to the J3:1. Numerical integration tests indicate that asteroids on such orbits can be affected by chaotic diffusion over hundreds of Myr \citep[e.g.,][]{wal2013,vok2016}. This means its initial orbit immediately after the family forming event could have had a range of semimajor axis values.

A secondary, potentially related group of objects was identified at lower $e$, $\sin I$ values than the primary family. This cluster, centered on (495) Eulalia, was observed in both the $e$-$\sin I$ and $a$-$H$ parameter spaces. The distribution in $a$-$H$ space suggests a similar spread, consistent with a collisional structure of similar age \cite[][see their Fig. 15b]{dy2015}. If this cluster represents a diffuse second half of the Eulalia family, dispersed to lower $e$, $\sin I$ and higher semimajor axis values as a result of anisotropic ejection velocities, it is possible that a larger percentage of the original collision fragments were lost to immediate dispersion via the J3:1. This scenario would lead to a modest increase in the impactor flux across all size ranges, potentially by a factor of two. While intriguing, this possibility lies beyond the scope of the current work, though it could contribute to an overall enhancement of the impactor flux. Additionally, this cluster influences the potential range of $a_{\rm c}$ values considered in our modeling, as discussed further below.

It is also worth noting that the Eulalia family may have experienced the loss of even larger members through dynamical diffusion into the J3:1, though we cannot verify this possibility using existing knowledge. Given these uncertainties, our focus will remain on the observed members of the Eulalia family, working with the data currently available to us to draw meaningful insights.

Identification of asteroid family members in the main belt often requires a multiple step process. For the Eulalia family, the first step is to select a population of low-albedo and low-inclination asteroids in the inner main belt region, defined as those bodies with $a < 2.5$~au. 

Using the SDSS color information as a guide, we isolated several clusters in the Eulalia dynamical region that have low $a^\star$ colors. The value $a^\star$ is defined as:
\begin{equation}
a^\star = 0.89 (g-r) + 0.45 (r -i) -0.57. \label{astar}
\end{equation}
with $g, r, i$ representing three of the five SDSS observed colors based on their  optical filters \cite {ivezic2002}. We analyzed their Yarkovsky signatures using a parameter search in $a$ and $C$.  This allowed us to identify several clusters that display comparable levels of spreading, with their $C$ values being statistically indistinguishable within the margin of uncertainty. These observations align with the interpretation that the clusters likely originated from the same collision event. 

The magnitude of the offsets in proper orbital elements among the clusters is consistent with ejection velocities of approximately $120$ m~s$^{-1}$. From this, we infer that a major, anisotropic collision occurred on the Eulalia parent body in the vicinity of the J3:1. Furthermore, a substantial portion of the collision fragments appears to have been dispersed and evacuated from the region.

Here we defined the parameter ``E0" to correspond to the extensive $(e,\sin I)$ box region that spans the outer boundaries of the Eulalia family, with $0.125 \le e \le 0.167$ and $0.037 \le \sin I \le 0.063$. It contains the dark-albedo objects ($p_V \le$ 0.125) previously designated as ``Eulalia" in \citet{wal2013}. Using a parameter search primarily constrained by objects within the magnitude range of $15 < H < 17$, we calculated best-fit values for E0 that have $a_{\rm c} = 2.4645 \pm 0.0118$~au and $C = -0.0741^ {+0.0067}_{-0.0017}$. 

These values, however, are not sufficient to define the Eulalia family because some E0 asteroids may be interlopers, defined as bodies in the $(a, e, I)$ cluster unrelated to the Eulalia family. Based on previous work, we infer that some interlopers belong to the overlapping (new) Polana family \citep[e.g.,][]{wal2013,bot2015}. 

This brings us to the second step in our process of identifying Eulalia, namely distinguishing between the Eulalia and New Polana families. Here we will focus on analyzing the families in $(a,H)$ space. Recall that asteroid family members are driven by the Yarkovsky effect over time into a ``polarized" dynamical state, defined as many of the smaller objects trending towards extreme values in $a$. As a consequence, the family takes on an identifiable structure in $(a,H)$ space often called ``the V-shape" \citep[see, e.g., Sec.~4 of][]{netal2015}. The $C$-parameter in (\ref{cline}) was designed to help identify the expected boundary of the family. Objects with $C$ values larger than some critical value $C_\star$, which would be a characteristic value defining a given family, would be classified as outside the family and therefore considered interlopers.

\begin{figure}[t!]
 \begin{center}
 \includegraphics[width=0.47\textwidth]{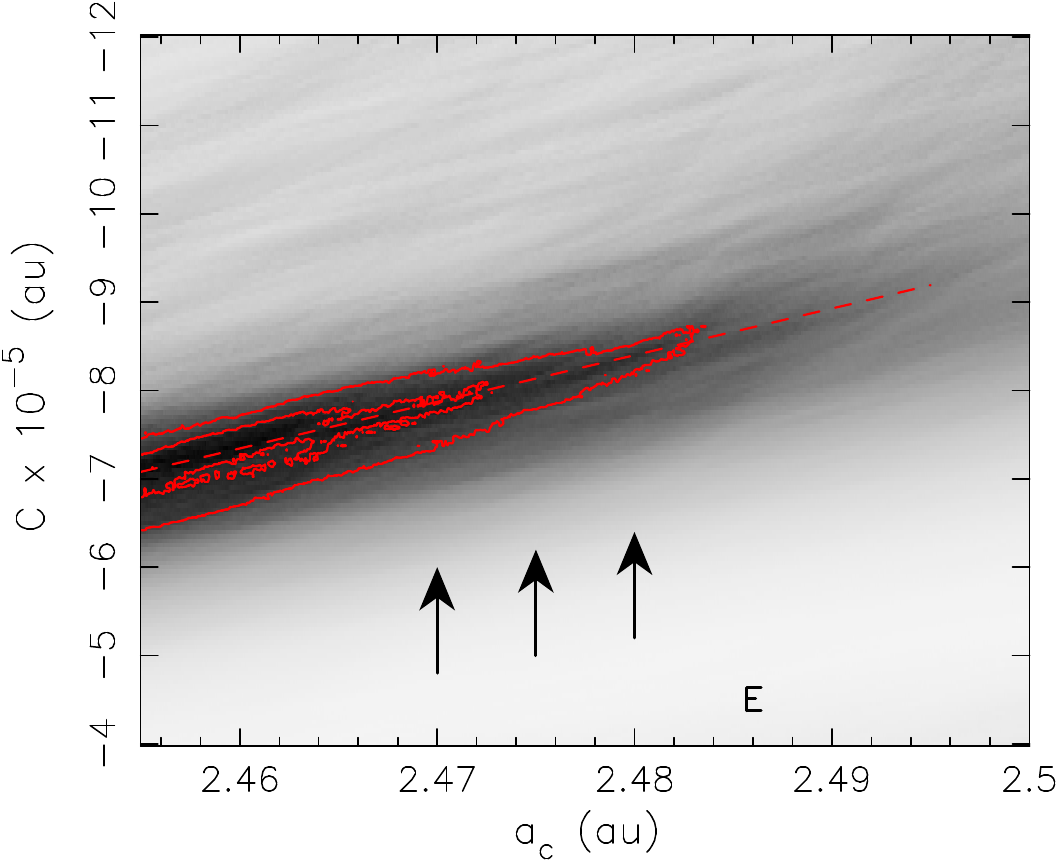} 
 \end{center} 
 \caption{The value of the Eulalia cluster contrast function $r(C,a_{\rm c};\Delta C)$ in the $(a_{\rm c},C)$ space for $\Delta C=1.5\times 10^{-5}$~au. Darker pixels correspond to larger $r$-values with a maximum of $\simeq 7.2$. Red isolines show values $r=(7.1,6.1,5.1)$, and the dashed line is the correlation axis of $a_{\rm c}$ and $C$ with close $r$ values. The current proper semimajor axis of (495) Eulalia is indicated by label~E, and the arrows show our tested Eulalia family realizations, including the canonical choice of $a_{\rm c}=2.475$~au (i.e., both \citet{wal2013} and \citet{vok2016} argue that (495) Eulalia might have been plausibly transported to its current semimajor axis from values between $\simeq 2.475$ and $\simeq 2.48$~au via Yarkovsky drift and chaotic diffusion near the J3:1).}
 \label{fig1}
\end{figure}
\begin{figure*}[t!]
 \begin{center}
 \includegraphics[width=0.95\textwidth]{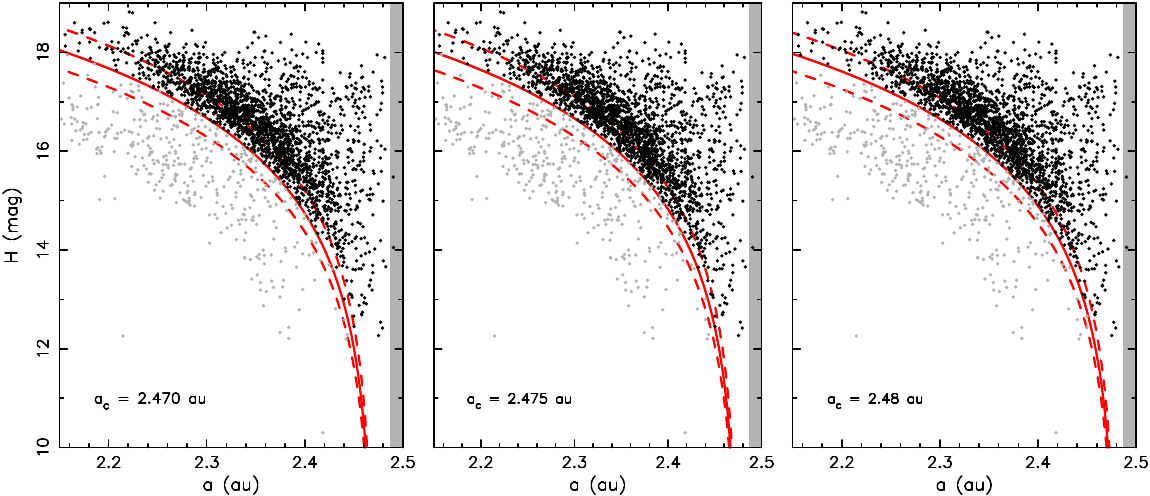} 
 \end{center} 
 \caption{Eulalia family members (black symbols) for three center choices $a_{\rm c}$: (i) $2.47$~au (left panel), (ii) $2.475$~au (our canonical choice; middle panel), and
  (iii) $2.48$~au (right panel). The solid red $C_\star$-isoline corresponds to the family
  border from Fig.~\ref{fig1} and Eq.~(\ref{ccrit}).  The two neighbor and dashed $C$-isolines are $C_\star\pm \Delta C$ with $\Delta C=1.5\times 10^{-5}$~au used in the contrast $r$-function calculation. The gray symbols are populations assigned to the New Polana family. The darker-gray zone at $\simeq 2.5$~au indicates the border of the J3:1 mean motion resonance at the characteristic eccentricity of Eulalia-family asteroids.}
 \label{fig2}
\end{figure*}
\begin{figure*}[t!]
 \begin{center}
 \includegraphics[width=0.95\textwidth]{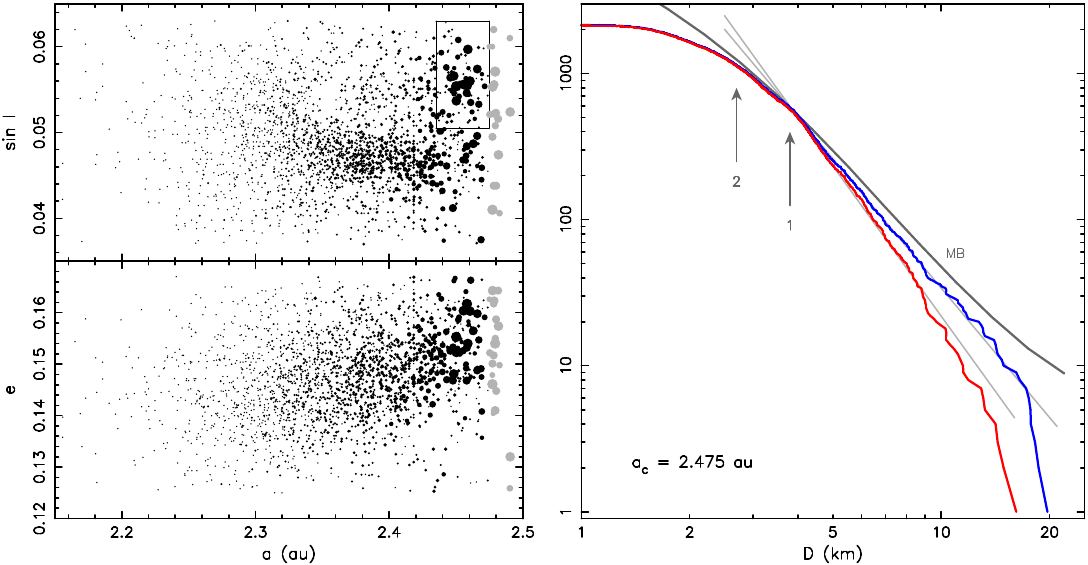} 
 \end{center} 
 \caption{Low albedo asteroids in our canonical classification of the Eulalia family with $a_{\rm c}=2.475$~au and limiting $C_\star=-8.14\times 10^{-5}$~au. The left panels show the projection of family members to the planes defined by proper semimajor axis and sine of inclination (top), and proper semimajor axis and eccentricity (bottom). The symbol sizes scale with asteroid diameter. The black and gray symbols are cluster members with $a<a_{\rm c}$ and $a>a_{\rm c}$, respectively. The right panel shows the cumulative SFD of the family population with $a<a_{\rm c}$, assuming the objects have a mean albedo value of $p_V=0.055$. The blue curve corresponds to the entire cluster.  The red curve corresponds to the population where $H<14$ objects in the rectangle shown in the top and left panel were excluded (these objects may belong to the older and larger Polana family, centered on (142) Polana with proper eccentricity of $0.1574$ and sine of proper inclination of $0.0577$). The light gray lines are power law fits to the populations in the $(4,11)$~km size range with slope values $-2.95$ (blue population) and $-3.42$ (red population). The dark gray curve (labeled MB) is the main belt SFD from \citet{bot2020}. We shifted the main belt SFD downward by a factor of $\simeq 200$, such that it approximately matches the Eulalia SFD over the $\simeq (2.8,4)$~km interval. This similarity indicates the slope change at $\simeq 4$~km stems from collision evolution (arrow labeled~1), while the second slope change at $\simeq 2.8$~km is from observational incompleteness (arrow labeled~2).}
 \label{fig3}
\end{figure*}

A possible approach to determine $C_\star$ was suggested by \citet{wal2013} in relation to the Eulalia family specifically and later applied by \citet{bol2017,bol2018} for a number of other main belt families \citep[see also][]{delbo2017,delbo2019,ferrone2023}. The idea was to identify the critical $C_\star$ isoline by the limit to which small members in the family piled up over time. Here we denote $N(C_1,C_2;a_{\rm c})$ as the number of asteroids between the $C_1$ and $C_2$ isolines (\ref{cline}) for a given choice of the family center $a_{\rm c}$. From there, \citet{wal2013} and \citet{bol2017} defined a contrast function $r(C,a_{\rm c};\Delta C)$ by
\begin{equation}
 r(C,a_{\rm c};\Delta C)=\frac{N(C-\Delta C,C;a_{\rm c})}{N(C,C+\Delta C;a_{\rm c})}\; , \label{rcrit}
\end{equation}
where $\Delta C$ is a predefined step.  

For our analysis of the Eulalia family, we chose $\Delta C=1.5\times 10^{-5}$~au. The highest contrast value $r_{\rm max}$ defines the pile-up limit and thus the optimum family border at the associated value of $C=C_\star$. This solution is unique if $a_{\rm c}$ is exactly known. In cases where $a_{\rm c}$ is uncertain, such as with the Eulalia family, however, one needs to explore a range of values, each which will produce a different $C_\star$ value and contrast function (\ref{rcrit}).

An additional complication with the Eulalia family relates to its orbital location adjacent to the J3:1 (see Fig.~\ref{fig2}). This right portion of the ``V-shape" has been lost to this resonance. This means that our study of the family structure is restricted to the portion with $a\leq a_{\rm c}$. In turn, this implies that the $C$-parameter can only take on negative values (Eq.~\ref{cline}).

Figure~\ref{fig1} shows the $r(C,a_{\rm c};\Delta C)$ function represented in gray-scale; the larger value, the darker the pixel. Searching in the $(2.455,2.5)$~au range for $a_{\rm c}$ and $-(4,12)\times 10^{-5}$~au range for $C$, we find a maximum $r_{\rm max}\simeq 7.2$ for $a_{\rm c}\simeq 2.456$~au and $C=C_\star\simeq -7.1\times 10^{-5}$~au. 

These values also indicate an acceptable solution could potentially exist for $a_{\rm c} < 2.455$~au using the contrast method. This highlights the need for caution when assigning higher $a_{\rm c}$ values in an effort to match the current proper semimajor axis of (495) Eulalia ($\sim 2.486$~au). However, when considering the possibility that the Eulalia family could consist of two distinct components separated in $a_{\rm c}$, we find it reasonable to explore a broader range of $a_{\rm c}$ parameter space to identify potential solutions. 

We find that acceptably high contrast values are obtained along a correlation line defined by 
\begin{equation}
 C_\star(a_{\rm c}) = -[7.08+53\,(a_{\rm c}-2.455)]\times 10^{-5}\;\;{\rm au} \label{ccrit}
\end{equation}
It is shown by dashed red line in Fig.~\ref{fig1}. Choosing $a_{\rm c}$ over a reasonable range of values, Eq.~(\ref{ccrit}) provides the family critical $C$-line. 

We note that the highest contrast solutions do not extend to the center $a_{\rm c}$, which would correspond to the current proper semimajor axis of (495)~Eulalia. After extensive numerical simulations, however, both  \citet{wal2013} and \citet{vok2016} argued that (495)~Eulalia might plausibly be transported to its current orbital location from proper-$a$ values ranging between $\simeq 2.475$~au and $\simeq 2.48$~au. 

As a consequence, we will test the histories of Eulalia family realizations for three $a_{\rm c}$ choices: $2.47$~au, $2.475$~au and $2.48$~au (arrows in Fig.~\ref{fig1}). The family with $a_{\rm c}=2.475$~au will be called the ``canonical choice".

Figure~\ref{fig2} shows our three choices for the Eulalia family plotted in the $(a,H)$ plane. In each panel, we plot the critical $C_\star$-isoline from (\ref{ccrit}) obtained by maximizing the contrast $r$-function for the specific choice of $a_{\rm c}$. Each choice is found to exclude most of the $H>14$ asteroids, as discussed above. The extension of low albedo inner belt asteroids beyond this boundary is arguably a distinct population belonging to the older New Polana family \citep[e.g.,][]{wal2013,bot2015,netal2015}. 

The Eulalia population hugs the J3:1 (gray rectangle), with relatively few members in close proximity to its periphery (Fig.~\ref{fig2}). This implies that a substantial number of former Eulalia members escaped through this resonance in the past. The remaining family members were predominantly left with retrograde spin axes, causing them to gradually drift inward toward the Sun over time due to the Yarkovsky effect.

Figure~\ref{fig3} shows the structure of the canonical Eulalia family in proper element space (two left panels) and its SFD, assuming the bodies have the family's mean geometric albedo of $p_V=0.055$. At this time, we only use the highly biased population of low albedo E0-cluster objects whose family membership was tested via the $(a, H)$ analysis discussed above. Our attempt to compensate for various biases is postponed to Sec.~\ref{biases}. 

As also seen in Fig.~\ref{fig2}, the $C$-criterion selection of family members implies that the largest objects are near the family center at $a_{\rm c}$. They are shown as having the larger symbols in the left two panels of Fig.~\ref{fig3}. Both \citet{wal2013} and \citet{vok2016} studied the chaotic diffusion of asteroids near the J3:1 under the assumption that some were initially located at eccentricity and inclination values consistent with family members further from the resonance. They found that as soon as the bodies made contact with a chaotic buffer associated with the J3:1, they underwent substantial diffusion in eccentricity and inclination (see Fig.~5 in \citet{wal2013} or Fig.~9 in \citet{vok2016}). As a consequence, it is possible that some of the large objects with higher inclinations belong to the Eulalia family (see the box in the top and left panel of Fig.~\ref{fig3}). On the other hand, it is also possible that some, or most, of them are members of the overlapping New Polana family. 

This potential spread led us to consider two options for the SFD shown in the right panel of Fig.~\ref{fig3}. For the blue and red curves, we include and exclude, respectively, the large objects discussed above in our formulation of the Eulalia family SFD. While the difference mainly affects the large end of the SFD, excluding the larger bodies means that our estimate of the Eulalia fragment SFD becomes steeper between $4 < D < 11$~km (i.e., see the light gray lines). 

The shape of the main belt SFD from \cite {bot2020} is included as a dark gray curve. It has been scaled so it hugs the Eulalia SFD over the $(2.8,4)$~km size intervals. The observed match in this range implies that the first slope change (label~1) is not an expression of the observational incompleteness, but rather is of collisional origin. If so, we can potentially derive the age of the Eulalia family independently using a collisional evolution model (Sec.~\ref{model2}).

Finally, Eq.~(1) in \citet{netal2015} provides a zeroth order estimate of the Eulalia family age. Adopting the canonical realization, with $C_\star\simeq -8.14\times 10^{-5}$~au (Eq.~\ref{ccrit}), and assuming the fragments have a bulk density of $\simeq 1.3$ g~cm$^{-3}$ and geometric albedo $\simeq 0.055$, the family's age is expected to be $0.8-1$~Gyr old. This ballpark estimate is in agreement with the results of \citet{wal2013} and \citet{bot2015}. We will further substantiate it with modeling work in the next section. We will also determine the flux of Eulalia family members reaching the J3:1 since the family-forming event.

\subsection{Application to the Eulalia family} \label{model1}

We will now apply our Yarkovsky chronology approach to each of the Eulalia family variants identified above to see how they compare. The methodology we will use has been previously described in \citet{vok2006} and \citet{bot2015}.  For that reason, here we only provide a list of its principal features, relegating the reader to those references for more information. 

Asteroid family chronology models need to account for three components when considering the semimajor axis spread of family members:
\begin {enumerate} 
\item The initial ejection velocity of each member, which is not known a prion,
\item Yarkovsky drift of that member toward the estimated $C_\star$-isoline from the family center, and 
\item How Yarkovsky drift for each member is affected over time by YORP thermal torques, with their semimajor axis drift rate $da/dt$ maximized for obliquity values near $0^\circ$ or $180^\circ$.
\end {enumerate}

As described above, this process pushes family members to pile up near the extremes in $(a, H)$ or $(a, D)$, creating the ``V-shape". The YORP contribution also helps to break the initial degeneracy that exists in semimajor axis between (i) the initial ejection orbits of the fragments at the family-forming event, and (ii) the semimajor axis spread produced by the Yarkovsky effect over the family's age. Asteroid family chronology models that neglect the first and third components may potentially overestimate the family's age \citep[e.g.,][]{spo2015,mil2017}. In addition, these kinds of models cannot provide information that is critical to testing our hypothesis, namely the fraction of family fragments directly injected into the J3:1 by the family-forming event and the timing of those transported to the resonance at later times by Yarkovsky drift (which is affected by the YORP effect).  

Next we list major components of the model.  In the first component, we assume the observed family, projected into the $(a,H)$ plane, has been mapped onto the one-dimensional $C$-parameter space as described above. The information about the family structure is comprised in the differential distribution $dN(C)$ (i.e., the number of family members in the interval $(C,C+dC)$). In the Eulalia family case we consider $C$ in the range $(0,C_\star)$, where the infinitesimal $dC$ is replaced with a finite $\Delta C=5\times 10^{-6}$~au value. The distribution function is then represented by a finite number of values in bins $(C,C+\Delta C)$. Given the $C_\star$ values determined for the different Eulalia family realizations, each with different $a_{\rm c}$ choices, the number of these bins is typically $19$. 

Many Eulalia members are found in the bins near $C_\star$, as we will show below. This implies the distribution function has a maximum away from its center near $C=0$. 

In constructing the distribution $dN(C)$, we consider only Eulalia members with $H\leq 17.5$. The population beyond this limit is compromised by observational biases. Fortunately, this constraint does not affect our goal of describing the fate of multikilometer Eulalia fragments.

For the purpose of model optimization, we represent the formal uncertainty of the bin occupancy with $\sigma_{dN}=\sqrt{dN}$. In the first six bins, which have the smallest $C$ values and are only sparsely populated, we increase this formal uncertainty by a factor $1-2$. This may also reflect a potentially larger fractional interloper population from the overlapping (new) Polana family in the bin.
 
In the second model component, we try to match the observed-family $dN(C)$ distribution with a prediction function $dM(C;{\bf p})$. This is accomplished by minimizing a target function defined as
 \begin{equation}
   \chi^2 = \sum_i \left(\frac{dN_i-dM_i({\bf p)}}{\sigma_i}\right)^2 \; , \label{tg}
 \end{equation}
 where the summation goes over the $C$-bins. 
 
The minimization is achieved by determining the optimum (best-fit) internal model parameters ${\bf p}$. Following the method in \citet{vok2006}, we choose ${\bf p}=(v_5,T,c_{\rm YORP})$, where (i) $T$ is the family age, (ii) $v_5$ is a characteristic initial ejection velocity of $D = 5$~km size fragments, and (iii) $c_{\rm YORP}$ is an empirical strength parameter for YORP torques, which can be adjusted using a simple template from \citet{cv2004}.
 
For the third component, the model operates in the $(a,H)$ plane. At the family forming event, a synthetic population of Eulalia fragments is created from an isotropic velocity field. This leaves the bodies centered around $a_{\rm c}$ and distributed in $a$. The initial number of synthetic members is chosen to double the population currently found in the $a<a_{\rm c}$ zone. 

Fragments with estimated size $D$ (using albedo $p_V=0.055$) are assumed to be ejected with a characteristic velocity $v_{\rm ej}(D)=v_5\,(5\,{\rm km}/D)$. The fragments are also given initial rotation-state parameters, namely (i) rotation period $P$ from a Maxwellian distribution peaked at $6$~hr, and (ii) obliquity $\gamma$ corresponding to an isotropic distribution. To cover the family's dynamical evolution, the integration time $t$ advances in steps of $dt=0.2$~Myr. 

During this process, the function $(a;P,\gamma)$ for each fragment is evolved as follows: (i) semimajor axis $a$ is modified by the Yarkovsky effect, and (ii) $(P,\gamma)$ is changed by the YORP effect, with details of this modeling effort described in \citet{vok2006} and \citet{bot2015}. At regular timesteps $dt'=2$~Myr, the synthetic family population is mapped to $C$-space. The corresponding distribution function $dM(C;{\bf p})$ is determined from the new values, while the target function (\ref{tg}) is computed to determine the quality of the fit between model and observational data. When making these calculations, the number of modeled Eulalia asteroids is always re-calibrated so it is equal to the number of observed family members. This change is made by scaling the $dM_i({\bf p})$ values.

For the fourth component, we register the flux of synthetic family members that reach the J3:1. This kind of calculation was not attempted in our previous works \citep[e.g.,][]{vok2006,vok2006a,vok2006b}. We assume the J3:1 boundary is met whenever a fragment's $a$ becomes larger than $2.487$~au \citep[an approximate resonance border for the Eulalia-family mean eccentricity; e.g., Fig.~9 in][]{vok2016}. This condition may be satisfied at the family-formation event stage, with some objects directly injected into the resonance, or during the family evolution stage.

The model, as set above, depends on a number of free parameters whose values are only known to reasonable approximation. This uncertainty led us to choose a range of values for the parameters ${\bf p}$ and then run our model across the grid of possible parameter choices. Our method allows us to highlight the role of these parameters in the solution, but it prevents us from evaluating their correlation with the three internal parameters in ${\bf p}$. As we learn more about these parameters in Eulalia family members, this approximation can be revisited in future work.

In choosing parameters for the Eulalia fragments, we take advantage of what has been learned from other asteroids analogs from spacecraft missions, etc. In this respect, we lean on asteroid (101955)~Bennu, which was explored in detail by the OSIRIS-Rex mission and whose origin was plausibly linked to the Eulalia family \citep[see][]{bot2015}. 

Bennu's mean bulk density was found to be $1.2$ g~cm$^{-3}$ \citep[e.g.,][]{lau2019,sche2020}, and the mean thermal inertia of its surface was $300\pm 30$ J~m$^{-2}$~K$^{-1}$~s$^{-1/2}$ \cite[denoted SI units below for short;][]{roz2020}. As specific regions of Bennu, such as its center and equatorial ridge zone, were found to be less dense than Bennu's mean bulk density, we find it plausible that slightly higher bulk density values may also be possible for generic Eulalia family members. For this reason, we explore bulk density values up to $1.5$ g~cm$^{-3}$. 

Similarly, certain fine-grained regions or unusual rock populations on Bennu's surface exhibit thermal inertia values as low as $\simeq 150-200$ SI units \citep[e.g.,][]{roz2020,rozitis2022}. Furthermore, the larger heliocentric distance of Eulalia members implies lower mean surface temperatures. This motivates us to consider thermal inertia values within the range of $100$ to $300$ SI units. For simplicity, we do not account for potential size-dependent variations in the thermal inertia of the fragments in this study \citep[see][for further discussion and results]{bol2018}.

Regarding the YORP effect, \citet{bot2015} introduced a ``variable (or stochastic) YORP" approach, which was found to better model the Eulalia family compared to the traditional ``static YORP" approach. In the static YORP model, the characteristic strength of YORP torques remains constant over an asteroid's lifetime. In contrast, the variable YORP model allows the strength of YORP torques to change over a characteristic timescale. This distinction arises because the magnitude of YORP torques on a small asteroid is highly sensitive to its small-scale surface topography \citep[e.g.,][for a review]{vetal2015}. Consequently, processes such as the formation of new craters, surface shaking, landslides, and boulder mobility can significantly alter how the YORP effect shapes the rotation-state evolution of a small asteroid.

\citet{bot2015} thus considered a characteristic YORP-change timescale $\tau_{\rm YORP}$ over which they changed the YORP-strength coefficient pre-computed by \citet{cv2004}. Given that the essence of the effect may be associated with small (sub-catastrophic) impacts, we additionally assume $\tau_{\rm YORP}=\tau_0\,\sqrt{D}$, with $\tau_0$ a free parameter and $D$ the size in kilometers \citep[the power-exponent of the size dependence was motivated by analysis in][]{fvh1998}. We considered $\tau_0$ in the $0.5$ to $10$~My range.  The reader should note that $\tau_0\rightarrow \infty$ is the formal limit to the static approach of YORP modeling. 

A key implication of the variable YORP model is the decoupling of the evolution timescales for obliquity and rotation rate \citep[see][]{bot2015,s2015}. While the obliquity evolves on timescales similar to those predicted by the static YORP model, the evolution of the rotation rate is slowed due to the effects of a stochastic random walk. This mechanism helps to explain why small asteroids tend to accumulate near the extreme semimajor axis values within the family, thus contributing to the characteristic ``V-shape" distribution.

Finally, we also ran simulations for the three possible choices of the Eulalia family center $a_{\rm c}$ listed above, paying most attention to the canonical value $a_{\rm c}=2.475$~au.

As for the estimated Eulalia age $T$, the general trends of the solutions may be summarized as follows. Overall, $T$ closely follows the scaling relationship $T\propto \rho$, where $\rho$ is the adopted bulk density, assuming all other parameters are held constant. This behavior is anticipated, as both the Yarkovsky accelerations and YORP torques scale inversely with bulk density, following $\propto \rho^{-1}$.  

Similarly, we find that $T$ approximately scales as $\propto \Gamma$, where $\Gamma$ is the surface thermal inertia. This scaling arises because the Yarkovsky drift rate satisfies $da/dt\propto \Gamma^{-1}$ in the regime where the thermal parameter exceeds unity \citep[e.g.,][]{bot2006,vetal2015}. As a result, the model exhibits a strong correlation between its dependence on $\rho$ and $\Gamma$. Higher bulk density can be offset by lower surface thermal inertia, and vice versa. 

Finally, our simulations suggest that $T$ scales as $T\propto \tau_0^\alpha$, where the exponent $\alpha\simeq 1/3$ in the tested range of $\tau_0$ values. This empirical relationship provides an additional opportunity to adjust $\tau_0$ in correlation with changes in either $\rho$ or $\Gamma$ while still preserving the estimated age of the Eulalia family. It is important to note, however, that pushing these parameters toward the upper limits of their respective value ranges tends to degrade the quality of the $C$-distribution fit, as it causes the target function (\ref{tg}) to increase to unacceptably high values.

\begin{figure*}[t!]
 \begin{center}
 \begin{tabular}{cc}
  \includegraphics[width=0.47\textwidth]{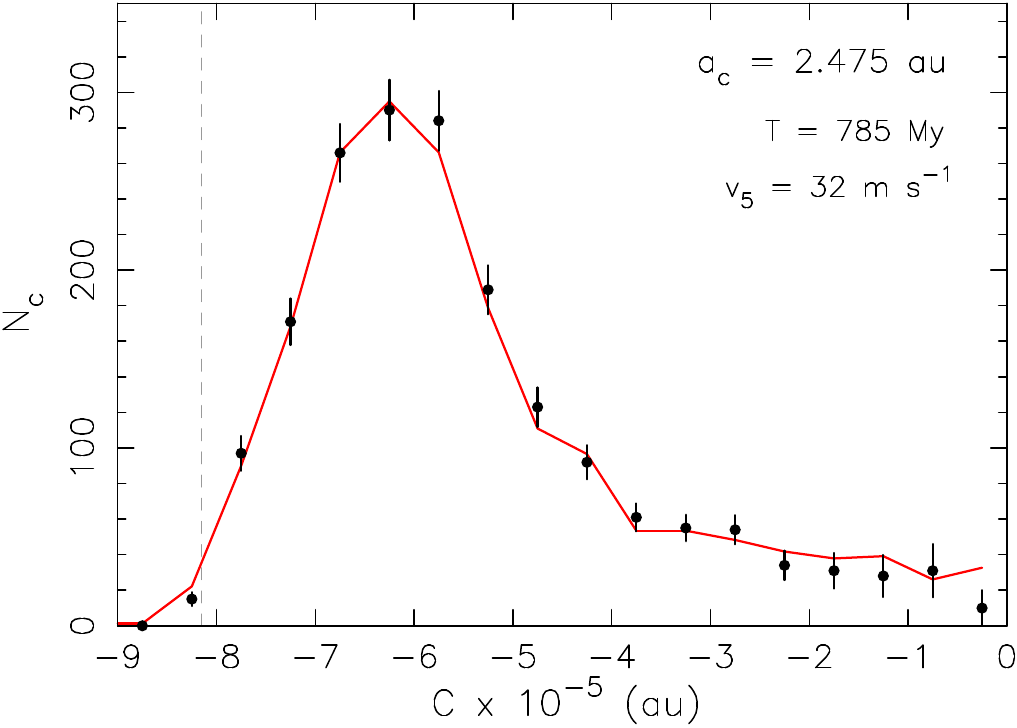} &
  \includegraphics[width=0.47\textwidth]{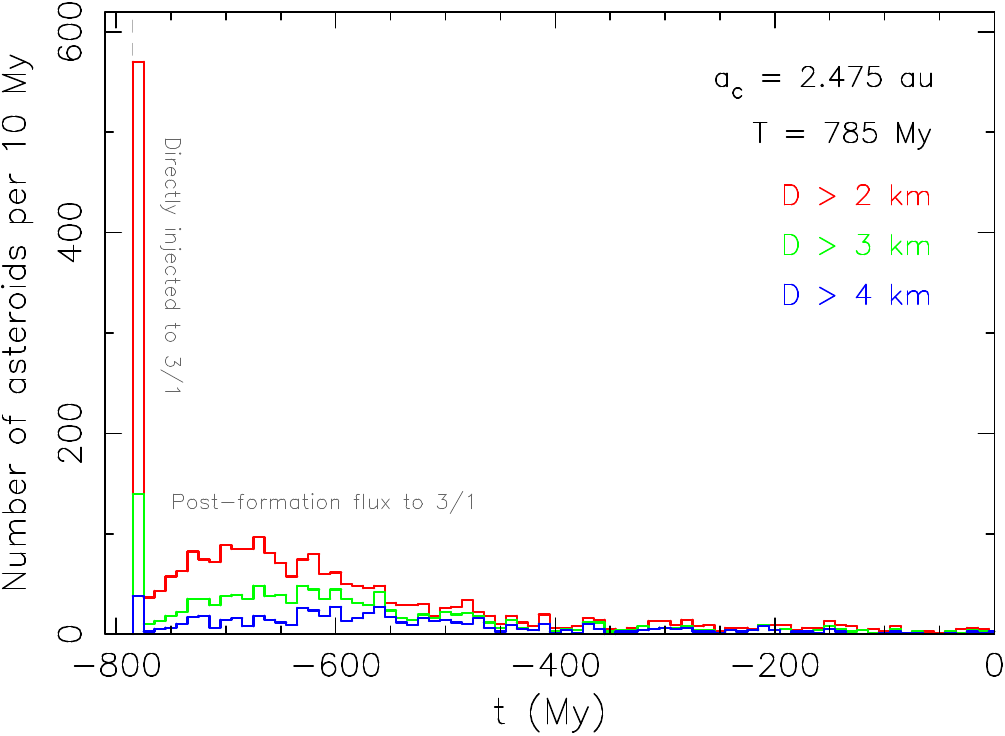} \\ 
 \end{tabular}
 \end{center}  
 \caption{Modeling the dynamical evolution of the canonical Eulalia family with the center $a_{\rm c}=2.475$~au. This solution used the bulk density $\rho=1.3$ g~cm$^{-3}$, the surface thermal inertia $\Gamma=200$ SI units and the variable YORP timescale $\tau_0=1$~My. The best match to the data (left panel) is obtained for $T=785$~My and $v_5=32$ m~s$^{-1}$ (recall that the characteristic initial ejection velocity is $v_{\rm ej}(D)= v_5\,(5\,{\rm km}/D)$ in meters per second). The vertical dashed line on the left panel shows the $C_\star$ value, though our data (black symbols with formal uncertainties) continue formally farther due to finite binsize. This solution provides a significant amount of $D\leq 3$~km fragments directly injected to the J3:1 (right panel). Still they represent $\simeq 22$\% of the uncorrected population of fragments reaching the J3:1 in total. Both ordinates are calibrated to the adopted family identification without correcting for incompleteness.}
 \label{fig4}
\end{figure*}
\begin{figure*}[t!]
 \begin{center}
 \begin{tabular}{cc}
  \includegraphics[width=0.47\textwidth]{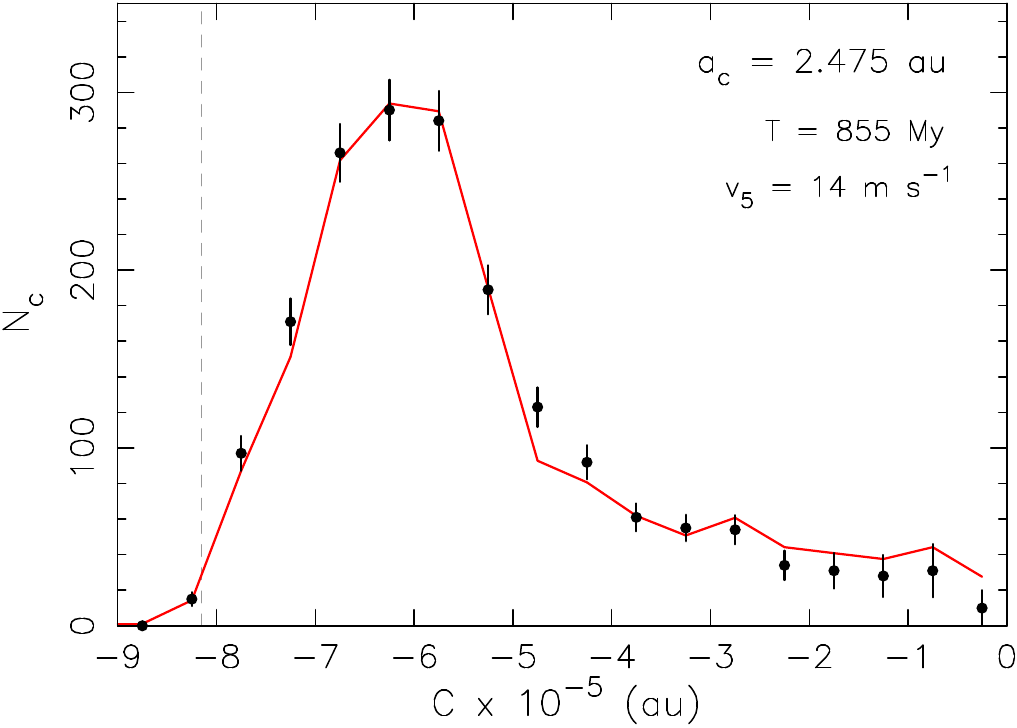} &
  \includegraphics[width=0.47\textwidth]{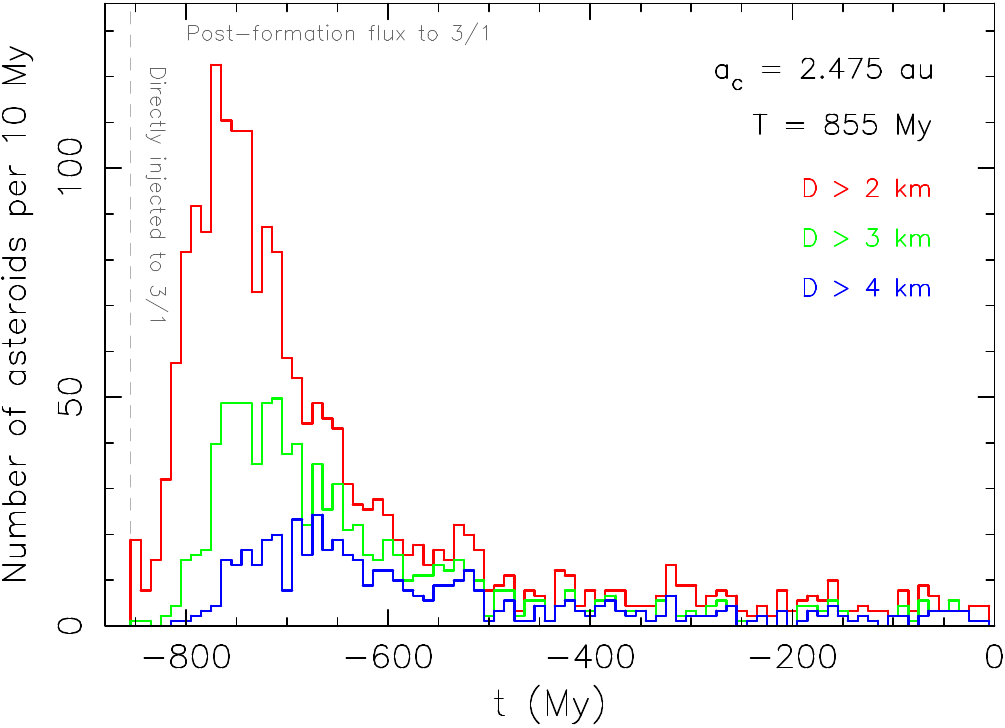} \\ 
 \end{tabular}
 \end{center}  
 \caption{Modeling the dynamical evolution of the canonical Eulalia family with the center $a_{\rm c}=2.475$~au. This solution used the bulk density $\rho=1.3$ g~cm$^{-3}$, the surface thermal inertia $\Gamma=230$ SI units and the variable YORP
  timescale $\tau_0=1$~My. The best match to the data (left panel) is obtained for $T=855$~My and $v_5=14$ m~s$^{-1}$ (recall that the characteristic initial ejection velocity is $v_{\rm ej}(D)= v_5\,(5\,{\rm km}/D)$ in meters per second). This solution provides fewer fragments directly injected into the J3:1. Most of the flux into the resonance is due to post-formation transport of fragments by the Yarkovsky effect (right panel).  The total number of transported $D\simeq 2-3$~km fragments to J3:1 is about 
  $75$\% of that provided by the solution shown in Fig.~\ref{fig4}. Both ordinates are calibrated to the adopted family identification without correcting for its
  incompleteness.}
 \label{fig5}
\end{figure*}
\begin{figure*}[t!]
 \begin{center}
 \begin{tabular}{cc}
  \includegraphics[width=0.47\textwidth]{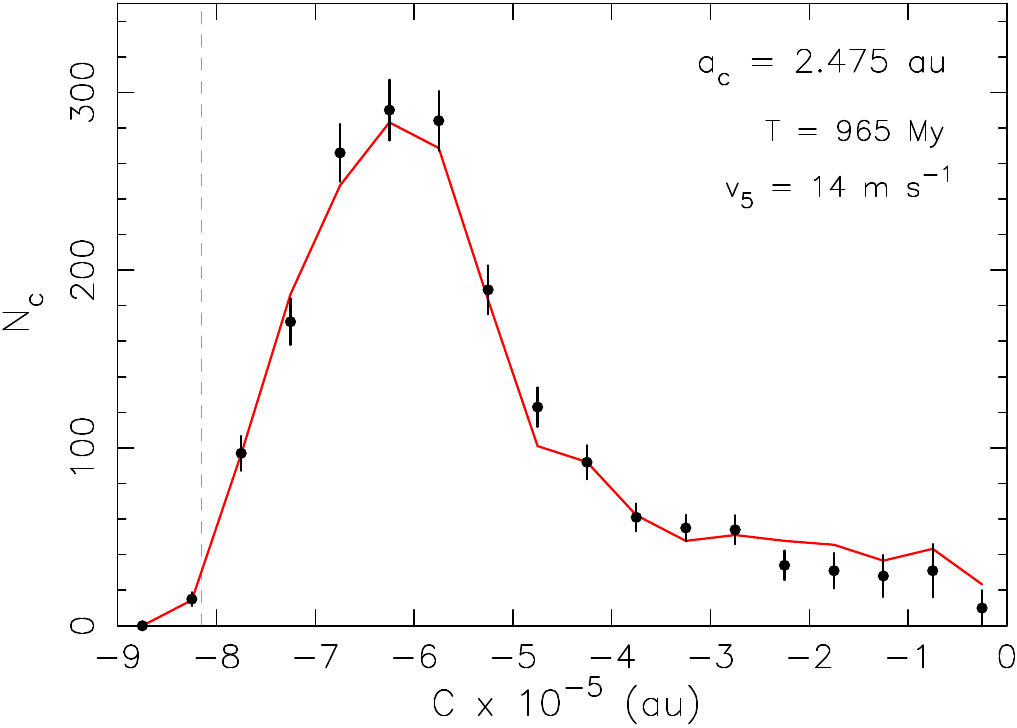} &
  \includegraphics[width=0.47\textwidth]{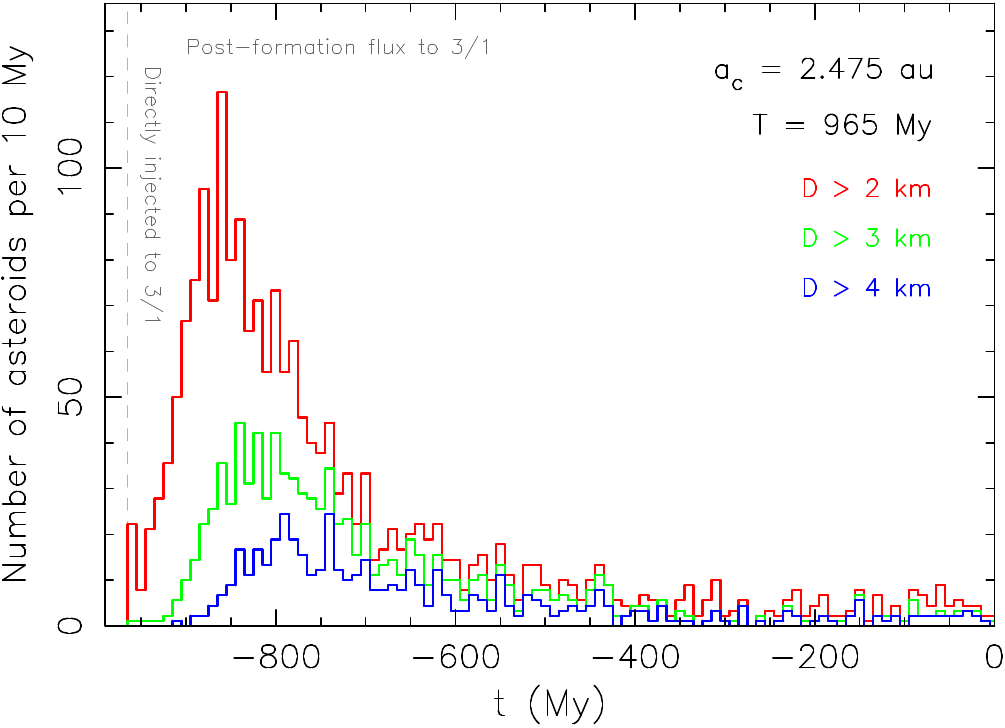} \\ 
 \end{tabular}
 \end{center}  
 \caption{Modeling the dynamical evolution of the canonical Eulalia family with the center $a_{\rm c}=2.475$~au. This solution used the bulk density $\rho=1.5$ g~cm$^{-3}$, the surface thermal inertia $\Gamma=200$ SI units and the variable YORP timescale $\tau_0=1$~My. The best match to the data (left panel) is obtained for $T=965$~My and $v_5=14$ m~s$^{-1}$ (recall that the characteristic initial ejection velocity is $v_{\rm ej}(D)= v_5\,(5\,{\rm km}/D)$ in meters per second). The solution is similar to that shown in Fig.~\ref{fig5}.  It shows degeneracy of the possible solution in the choice of the bulk density and the surface thermal inertia. Both ordinates are calibrated to the adopted family identification without correcting for its incompleteness.}
 \label{fig6}
\end{figure*}
\begin{figure*}[t!]
 \begin{center}
 \begin{tabular}{cc}
  \includegraphics[width=0.47\textwidth]{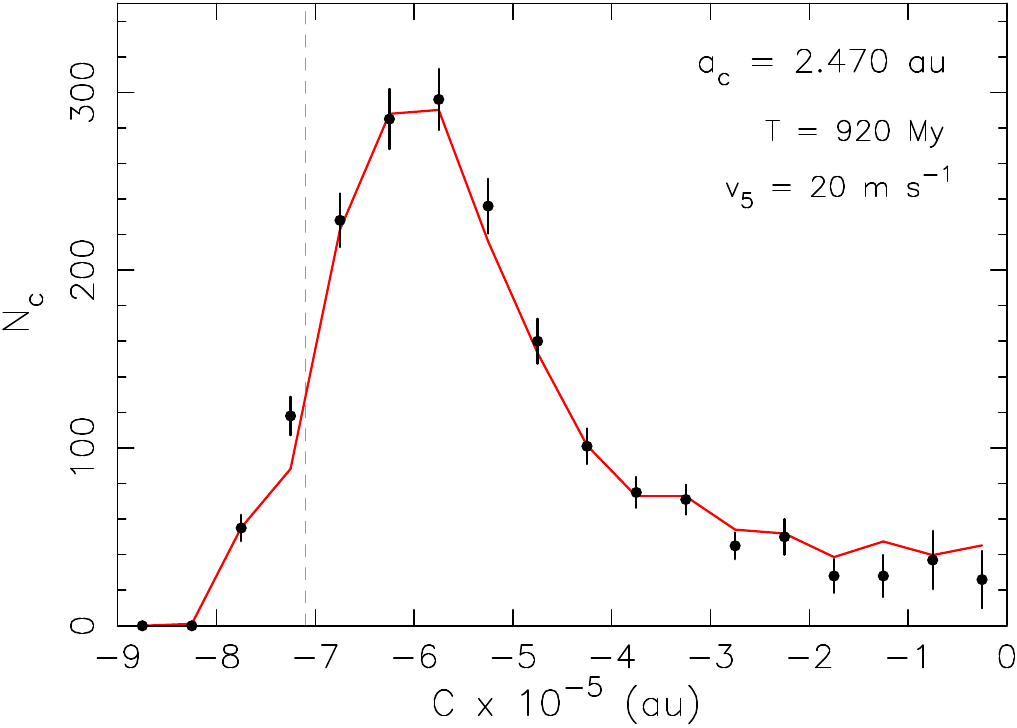} &
  \includegraphics[width=0.47\textwidth]{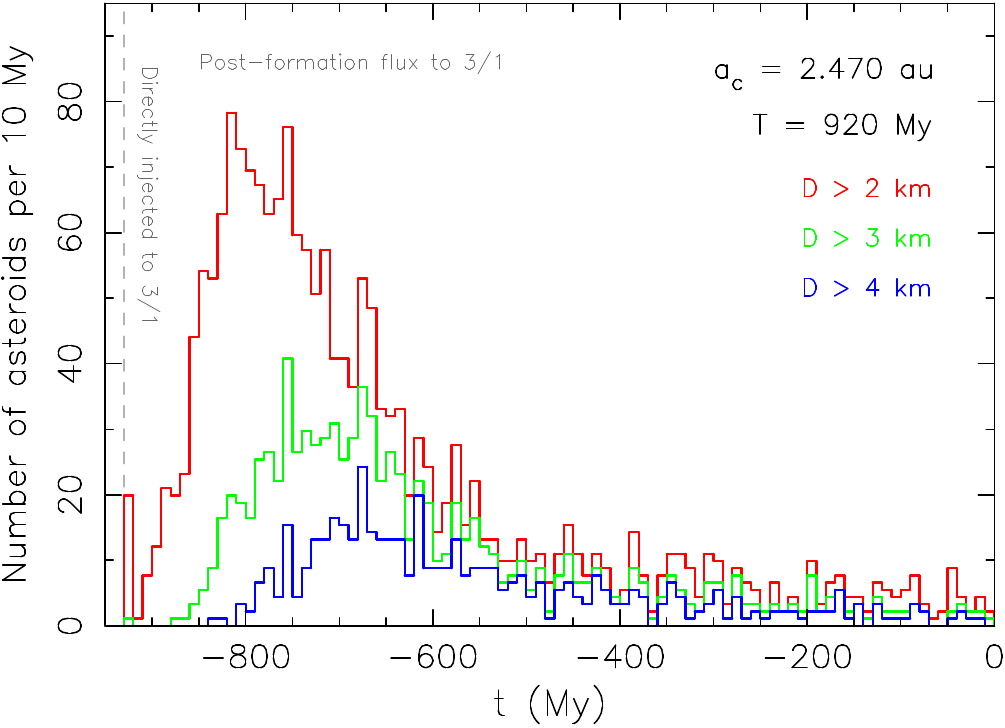} \\ 
 \end{tabular}
 \end{center}  
 \caption{Modeling the dynamical evolution of the Eulalia family with the center $a_{\rm c}=2.47$~au, more distant from the J3:1 than our canonical choice. The vertical dashed line on the left panel shows the $C_\star$ value, though our data (black symbols with formal uncertainties) continue formally farther due to finite binsize and a limited mobility population of objects in the bin containing the $C_\star$ value.  This solution used the bulk density $\rho=1.5$ g~cm$^{-3}$, the surface thermal inertia $\Gamma=200$ SI units and the variable YORP timescale $\tau_0=1$~My. The best match to the data (left panel) is obtained for $T=920$~My
 and $v_5=20$ m~s$^{-1}$ (recall that the characteristic initial ejection velocity is $v_{\rm ej}(D)=v_5\,(5\,{\rm km}/D)$ in meters per second). Here a larger initial distance from the J3:1 is partially compensated by a higher initial ejection velocity if compared to the solution shown in Fig.~\ref{fig6}. The total number of transported fragments to the J3:1 is only modestly smaller than in solutions shown in Figs.~\ref{fig5} and \ref{fig6}. Both ordinates are calibrated to the adopted family identification without correcting for its incompleteness.}
 \label{fig7}
\end{figure*}
\begin{figure*}[t!]
 \begin{center}
 \begin{tabular}{cc}
  \includegraphics[width=0.47\textwidth]{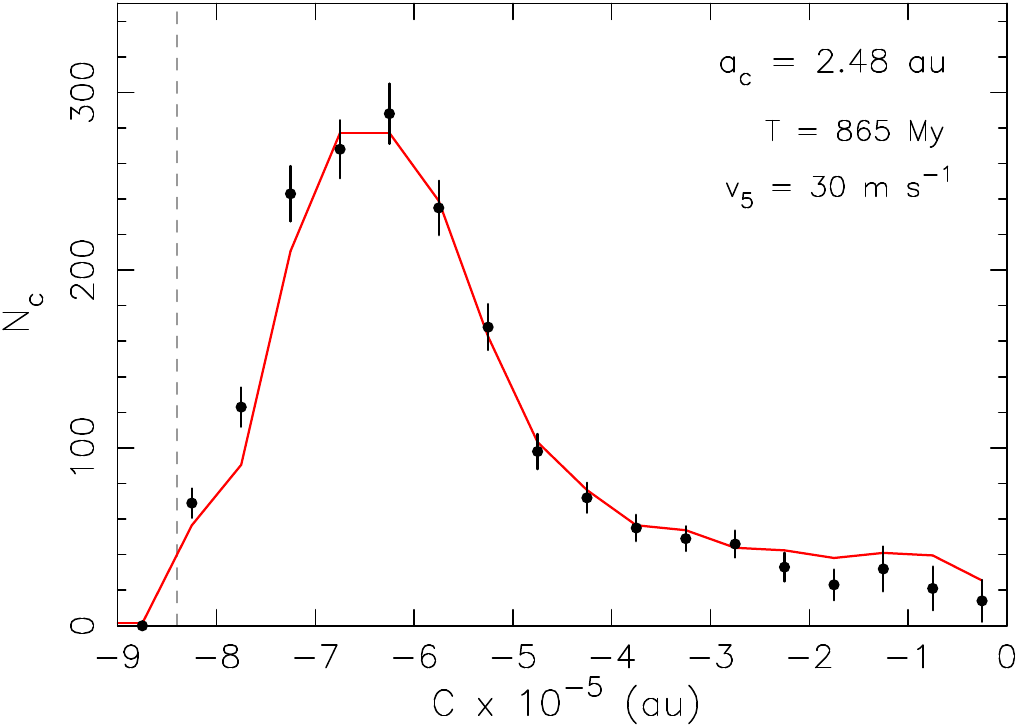} &
  \includegraphics[width=0.47\textwidth]{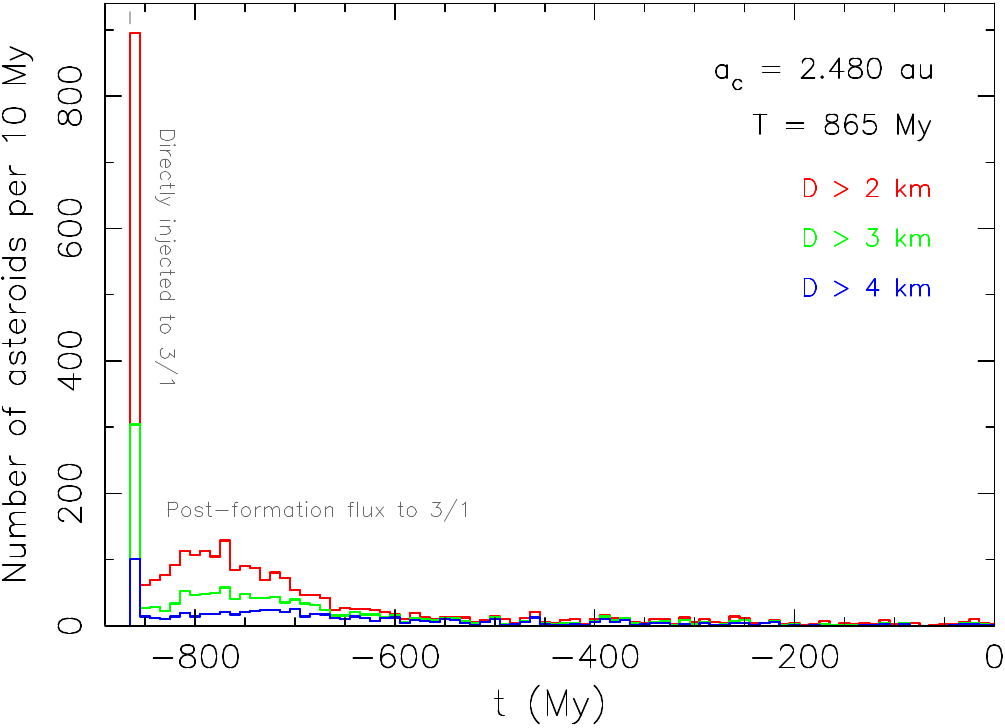} \\ 
 \end{tabular}
 \end{center}  
 \caption{Modeling the dynamical evolution of the Eulalia family with the center $a_{\rm c}=2.48$~au, closer to the J3:1 than our canonical choice. The vertical dashed line on the left panel shows the $C_\star$ value, though our data (black symbols with formal uncertainties) continue formally farther due to finite binsize and a limited mobility population of objects in the bin containing the $C_\star$ value. This solution used the bulk density $\rho=1.3$ g~cm$^{-3}$, the surface thermal inertia $\Gamma=230$ SI units and the variable YORP timescale $\tau_0=1$~My. The best match to the data (left panel) is obtained for $T=865$~My and $v_5=30$ m~s$^{-1}$ (recall that the characteristic initial ejection velocity is $v_{\rm ej}(D)=v_5\,(5\,{\rm km}/D)$ in meters per second). Here the small initial distance to the J3:1, combined with a high initial ejection velocity, produces a significant spike in fragment delivery at the family formation event (right panel). With that said, more than twice as many fragments drift to the J3:1 over the next $\simeq 200$~My. The overall yield to the resonance is the largest from all solutions shown. Both ordinates are calibrated to the adopted family identification without correcting for its incompleteness.}
 \label{fig8}
\end{figure*}

\section{Model Results} \label{results}

In this section, we present several successful solutions for the age of the Eulalia family, estimated to be within the range of $800–900$~Ma. Our work shows that varying parameter choices can lead to diverse outcomes regarding the fragment flux delivered to the J3:1 during this time period.
\smallskip

\subsection {Solution 1}

Figure~\ref{fig4} shows results for the canonical Eulalia family and the following set of physical parameters: $\rho=1.3$ g~cm$^{-3}$, $\Gamma=200$ SI units and $\tau_0=1$~My. Given that $v_5=32$ m~s$^{-1}$ for this case, many multikilometer-sized fragments were ejected with velocities high enough to reach the J3:1 immediately following the family-forming event. Additionally, approximately half of the remaining fragments migrated to the resonance at later times, primarily within the first $\simeq 50-150$~Myr after the family-forming event.

After that early epoch, the remaining Eulalia fragments migrated towards the presently observed population, becoming concentrated along the critical $C_\star$ isoline (i.e., left left ear of the V-shape in $(a, H)$). The number of $D = 2$, 3 and 4~km fragments reaching the J3:1 shown in the right panel of Fig.~\ref{fig4} is calibrated to ensure that the final simulated population of fragments precisely matches the observed (albeit biased) family at the conclusion of the simulation. These values will be recalibrated in Sec.~\ref{model2} to account for (i) various observational biases that limit our ability to detect the complete Eulalia population down to kilometer-sized fragments and (ii) the impact of collisional disruption events among these bodies over the estimated age of the family.

As expected, the flux of the $D = 2$~km size fragments is about five times larger than the $4$~km ones. Due to their higher initial ejection velocities and faster Yarkovsky-driven mobility, their flux reaches its peak within the first $100$~Myr following the formation of the Eulalia family. The trend in Figure~\ref{fig4} is consistent across all solution examples presented below.

\subsection {Solution~2}

Figure~\ref{fig5} shows results from a simulation similar to that described above, but with a slightly higher assumed thermal inertia $\Gamma=230$ SI units. As a result, the estimated age of the family is now slightly older.

A more substantial difference comes from the lower required initial ejection velocities of the fragments ($v_5=14$ m~s$^{-1}$). It implies that fewer fragments were directly injected into the J3:1 immediately after the family-forming event. Instead, the majority of the flux into the J3:1 occurred some time later, moderately after the family's formation. During this period, the region near the family center gradually became cleared of members. From Figure~\ref{fig5}, we observe that the flux into the J3:1 reaches its peak shortly after $800$~Ma.

\subsection {Solution~3}

This simulation used a higher bulk density ($\rho=1.5$ g~cm$^{-3}$) but smaller thermal inertia ($\Gamma=200$ SI units) compared to the results shown in Solution~2. Based on the scaling rules discussed previously, the family age $T$ should be comparable to the two cases. The age obtained, $T=965$~Myr, however, suggests that the scaling rules are only approximate for our parameter set and/or that the data fit exhibits a rather flat minimum in the target ($\chi^2$) function. 

Despite these differences, the right panel of Fig.~\ref{fig6} shows that the fragment flux to the J3:1 in both Solution~2 and Solution~3 remains broadly similar.

\subsection {Solution~4}

In this case, the Eulalia family was given a more distant family center at $a_{\rm c}=2.47$~au.  This choice has two consequences (Fig.~\ref{fig7}). First, in order for the family's age to be between $T = 800$ and $900$~Myr, the semimajor axis drift rate of the fragments needs to be smaller. This can be accomplished by giving these bodies a higher density or a higher thermal inertia. In this case, we assumed they had  $\rho=1.5$ g~cm$^{-3}$. 

Second, the larger distance to the J3:1 implies few Eulalia fragments are directly injected into the resonance.  The majority of them are instead transported to the resonance over several hundreds of Myr following the family formation event.

\subsection {Solution~5}

The last solution shown here corresponds to moving the family center close to the J3:1, or $a_{\rm c}=2.48$~au (Fig.~\ref{fig8}). Here we find that the influx trend is opposite to that shown in Solution~4. 

The family's close proximity to the J3:1, and consequently the highest $C_\star$ value necessitates a non-negligible initial velocity dispersion among the ejected fragments ($v_5=30$ m~s$^{-1}$). This increases the fraction of bodies that are directly injected into the J3:1.  It also allows many Eulalia asteroids to be transported to the J3:1 at later epochs, but fewer remain available for delayed transport, as a substantial number were already lost through direct injection. As a result, the flux into the resonance reaches its highest peak among all our solutions (right panel). Unsurprisingly, the total influx into the J3:1 is also the largest across all the presented solutions.
\vfill\null

\section{Incompleteness factors affecting the Eulalia family} \label{biases}

Up to now, we have based our analysis on the Eulalia family population as identified among the observed inner main belt asteroids. Wherever possible, we have taken advantage of the available albedo and color data to distinguish family members from background objects. The sample of objects we are using, however, has been influenced by observational selection effects.

In this section, we attempt to compensate for these biases and estimate a more complete Eulalia population. This will allow us to update the influx that potentially reached the J3:1.
\begin{figure*}[t!]
 \begin{center}
 \includegraphics[width=0.95\textwidth]{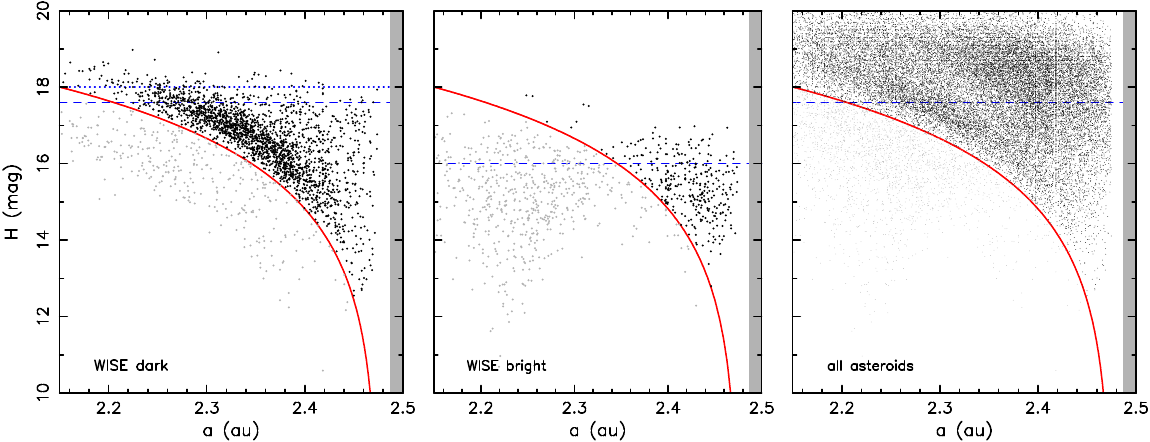} 
 \end{center} 
 \caption{Asteroids in our canonical E0-box in $(e, \sin i)$ projected onto the proper semimajor axis $a$ (abscissa) and absolute magnitude $H$ (ordinate) plane. The population has been restricted to $a\leq 2.475$~au. The left and middle panels show $3,322$ asteroids for which the WISE mission has estimated albedo values. In the left panel, we show dark objects with $p_V<0.125$, while in the middle panel, we show bright objects with $p_V>0.125$. The median albedo values in the two samples are $0.057$ and $0.26$, respectively. The right panel shows all asteroids for which \citet{propel2024} provided proper orbital elements. The red curve is the $C_\star=8.14\times 10^{-5}$~au critical isoline of the Eulalia family with the center at $2.475$~au. The blue dashed lines shown on the left and middle panels, at $H=17.6$ and $H=16$ magnitudes, respectively, indicate absolute magnitude values corresponding to the same size, $1.7$~km, for their median albedo values. The blue dotted line in the left panel, which in the middle panel would coincide with the dashed line, shows the absolute magnitude/size value in the albedo category that would have a 10.2 magnitude value in the WISE W3 passband at opposition and $2.5$~au heliocentric distance.}
 \label{fig9}
\end{figure*}
\begin{figure}[t!]
 \begin{center}
 \includegraphics[width=0.47\textwidth]{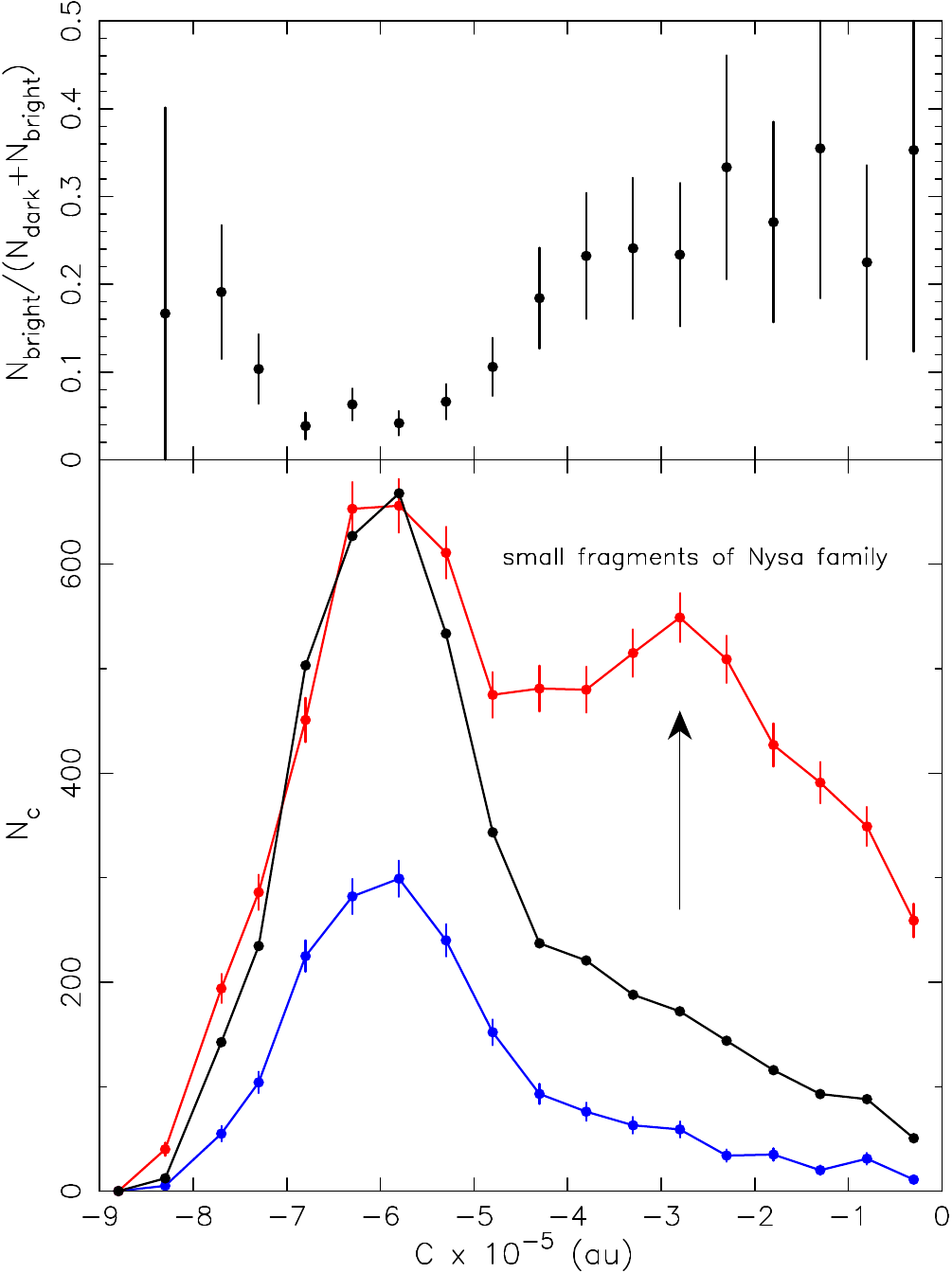} 
 \end{center} 
 \caption{Top panel: Fraction $f$ of bright asteroids ($p_V>0.125$) in $C$ bins, binsize $5\times 10^{-6}$~au and center $a_{\rm c}=2.475$~au, in the zone of the canonical Eulalia family introduced in Sec.~\ref{ident}. We use a sample of $244$ bright and $1,784$ dark objects observed by WISE with $H\leq 16.0$ and $H\leq 17.6$, respectively (see the left and middle panels in Fig.~\ref{fig9}). The magnitude limits make the two samples correspond to the same size limit of $\simeq 1.7$~km for the respective median albedo value in the bright and dark categories. Bottom panel: Blue and red data show the $C$-distribution of $H\leq 17.6$ magnitude objects in the $C_\star$ limit of the canonical Eulalia family: (i) blue for WISE observed dark objects (constituting the Eulalia family itself, left panel in Fig.~\ref{fig9}), and (ii) red for all known asteroids (right panel in Fig.~\ref{fig9}). The black distribution has been computed using Eq.~(\ref{unb}) and $F_2=3.02$, $F_1=2.3$ and $f$ values inferred from data in the top panel individual to each of the bins. We postulate that the excess of the total asteroid population in the $|C|< 4\times 10^{-5}$~au bins is due to an interloping population of small Nysa asteroids with $H>16.0$ magnitudes (see the middle and right panels of Fig.~\ref{fig9}).}
 \label{fig10}
\end{figure}

To assess the incompleteness of the currently identified Eulalia population, we first need to estimate the fraction of bright interloping asteroids within the orbital zone of the Eulalia family.  Using the catalog of WISE observations \citep{wise2011}, we found $3,322$ objects that matched the orbital zone restriction for the Eulalia family. The albedo values of these objects display a distinct bimodal distribution, which can be separated at $p_V=0.125$. Objects with $p_V<0.125$ represent a population of dark objects, while those with $p_V>0.125$ represent a population of bright objects. The distribution of these two groups, projected onto the $(a,H)$ plane, is shown in the left and middle panels of Fig.~\ref{fig9}. 

The dark class is primarily composed of Eulalia family members, while the bright class includes a mixture of Flora-region and Nysa/Hertha complex asteroids \citep{dg2015}. To refine our analysis, we focus on sub-samples of both dark and bright objects that fall within the $C_\star$ limit, which defines our canonical realization of the Eulalia family, and that have $a\leq a_{\rm c}=2.475$~au (Fig.~\ref{fig9}). Applying this additional constraint resulted in a total of $2,386$ objects with albedo values provided by WISE, of which $2,064$ belong to the dark class and 322 to the bright class. The median albedo value of the dark objects was $0.057$, whereas that of the bright objects was considerably higher at $0.26$. 

Next, we analyzed the photometric detection limit of WISE observations in the W3 passband, which is $\simeq 10.25$ magnitude \citep{wise2011b}. Using the Near-Earth Asteroid Thermal Model (NEATM) by \citet{h1998} with a beaming parameter set to $\eta=1.08$, and applying a color correction from \citet{w2010}, we mapped this detection efficiency limit in the infrared band to the absolute magnitude value $H$. This value is different for dark and bright objects. For simplicity, we used the median albedo values in each class and assumed opposition conditions at a heliocentric distance  of $2.48$~au.  From these parameters, we calculated the absolute magnitude limits of $H=18.0$ for the dark class and $H=16.0$ for the bright class. Under these assumptions, we propose that asteroids with absolute magnitudes lower than these limits in each class should have detection probabilities of one (or very close to it).

To align the samples from both albedo classes to the same limiting size, we determined that dark objects with magnitudes $H\leq 17.6$ correspond to approximately $D\geq 1.7$~km, matching the size of the bright objects with $H\leq 16$. These limits are indicated by the blue dashed lines in Fig.~\ref{fig9}. The application of these constraints yields a sample of $1,784$ dark objects and 244 bright objects. We assume that both samples are photometrically complete in the W3 band and are subject to the same incompleteness factors discussed below. 

Based on their respective numbers, we deduce that bright interlopers within the Eulalia family zone account for approximately 12\% of the total population. This fraction seems reasonable, given that Eulalia family members overwhelmingly dominate this region of proper element space.

We further refined this estimate by analyzing the fraction of bright interlopers within individual bins of the $C$-parameter distribution across the family zone, as shown in the top panel of Fig.~\ref{fig10}. Unsurprisingly, in the outer bins with $|C|>5\times 10^{-5}$~au, the fraction of bright interlopers drops to approximately $5$\%. In the bins closer to the center of the Eulalia family ($|C|<4\times 10^{-5}$~au), however, the interloper fraction increases to about $25$\%. This increase reflects the dynamics of Eulalia family members, many of which have either drifted into the J3:1 or have become concentrated in higher-$C$ bins (near the left ear of the V-shape in $(a, H)$).  Notably, the $C$-distribution for the bright objects remains relatively uniform.

In the following analysis, we use an approximate method to estimate the incompleteness factor $F_1$ of WISE observations within the orbital zone of the Eulalia family for asteroids with $D\geq 1.7$~km. This size threshold corresponds to $H\leq 17.6$ for dark objects and $H\leq 16$ for bright objects. At these sizes, asteroids from both categories are expected to be photometrically complete in the infrared band (i.e., their W3 passband magnitudes are smaller than WISE's limiting value $\simeq 10.2$), and they should have similar orbital distributions. As a consequence, the same incompleteness factor $F_1$ should apply to both dark and bright asteroids. This incompleteness arises from the fact that some asteroids within the Eulalia orbital zone, regardless of whether they are dark or bright, were missed by the field of view of WISE observations.

To estimate $F_1$, we consider a sample of all asteroids observed in the Eulalia orbital zone with $H\leq 17.6$ (right panel of Fig.~\ref{fig9}). These objects were obtained from the updated catalog of proper elements presented in \citet{propel2024}, which we assume to be complete \citep[e.g.,][]{hm2020}. This sample includes, as a subset, the WISE-identified dark asteroids with the same magnitude limit. For both groups, all asteroids and WISE-dark objects, we constructed distributions of the $C$-parameter (Eq.~\ref{cline}) up to the critical $C_\star$ values that defines our canonical Eulalia family (i.e., corresponding to the center $a_{\rm c}=2.475$~au). The resulting distributions are shown in Fig.\ref{fig10}, with the red curve representing all asteroids and the blue curve representing the WISE-identified dark objects.

The discrepancy between the red and blue curves provides insights into the incompleteness factor $F_1$. Additional care is needed, however, to estimate it with reasonable accuracy. The red curve, corresponding to all asteroids, was constructed uniformly for the $H\leq 17.6$ limit, which corresponds to dark objects with $D\geq 1.7$~km. In contrast, bright objects of the same size correspond to a sample with a magnitude limit of $H=16$. For this group (WISE-bright asteroids), their $C$-parameter distribution needs to be adjusted not only by the incompleteness factor $F_1$ factor but also by an additional factor that we define as $F_2$. The $F_2$ factor accounts for the expected increase in the bright asteroid population as the magnitude limit is extended from magnitude $H = 16$ to $17.6$.

To estimate $F_2$, we consider the characteristic slope of the cumulative magnitude distribution for the main belt population in the range $H\simeq 15$ to $H \simeq 17$, whose value is $\gamma_{\rm MB}\simeq 0.3$ \citep[e.g.,][]{jed2002}. The increase in the bright asteroid population between $H=16.0$ and $17.6$~magnitudes, corresponding to a magnitude difference of $\Delta H =1.6$, is represented by the relation $F_2=10^{\gamma_{\rm MB} \Delta H}$. Substituting the values, we find $F_2 \simeq 3.02$. It represents the average factor by which the bright population increases over this magnitude range.

Finally, for each bin in the $C$-parameter distribution, we determined the fraction $f$ of bright objects within the Eulalia region (Fig.~\ref{fig10}, top panel). Using these values, we estimated the $C$ distribution of the complete asteroid sample $N$ expected in each bin. This distribution is constructed as a linear combination of $N_{\rm dark}$, the blue distribution in Fig.~\ref{fig10} corresponding to WISE-identified dark asteroids, and $N_{\rm bright}$, which represents the distribution of WISE-observed bright asteroids in the Eulalia zone with $D\geq 1.7$~km:
\begin{equation}
 N = F_1\,\left[\left(1-f\right) N_{\rm dark} + F_2 f N_{\rm bright}\right] \; . \label{unb}
\end{equation}
We treat $F_1$ as a knob, adjusting it until $N$ matched the observed population (red distribution in Fig.~\ref{fig10}). After experimenting with the data, we found that $F_1\simeq 2.3$ provided the best agreement. The resulting predicted distribution is shown as the black curve in Fig.~\ref{fig10}.

The adopted $F_1$ value produces a strong match between the modeled and observed populations in the region defined by $|C|\geq 5\times 10^{-5}$~au. This is to be expected, given that this region is the most densely populated area within the Eulalia family's defined zone. The modeled population at $|C|\leq 4\times 10^{-5}$~au, however, underestimates the observed asteroid population shown by the red distribution. After examining the right panel of Fig.~\ref{fig10}, we believe this discrepancy arises from the contribution of small fragments originating from the Nysa/Hertha family, which likely migrated into the Eulalia orbital zone \citep[e.g.,][]{dg2015,netal2015}.

\begin{deluxetable}{lcccc}[t]
\tablecaption{\label{bins}
 Total yield of Eulalia fragments of different sizes entering the J3:1 in our model simulations. Note that the values have been re-calibrated using the $F_1$ factor from the text.}
\tablehead{
  & \colhead{$D\geq 2$~km} & \colhead{$D\geq 3$~km} & \colhead{$D\geq 4$~km}  & \colhead{$D\geq 5$~km}  }
\startdata
 solution~1 & $6090$ & $2932$ & $1464$ & $750$ \\
 solution~2 & $4510$ & $2158$ & $1100$ & $529$ \\
 solution~3 & $4554$ & $2198$ & $1132$ & $533$ \\
 solution~4 & $4400$ & $2120$ & $1074$ & $531$ \\
 solution~5 & $6995$ & $3326$ & $1645$ & $833$ \\ 
\enddata
\end{deluxetable}

\smallskip 

\section {The Impact Flux of Eulalia Fragments on the Terrestrial Planets}\label{impactflux}

\subsection {The number of Eulalia fragments reaching the J3:1}\label{model1bis}

Assuming that observational selection effects in the Eulalia family have been properly addressed through the $F_1\simeq 2.3$ parameter, we now have a reasonable estimate of the present-day Eulalia family SFD. Using our dynamical results from Solutions~1 through 5, we can use the family SFD to estimate the total number of Eulalia fragments that migrated into the J3:1 over their lifetimes. The effects of collisional evolution will be addressed in the following section.

Our results are shown in Table~\ref{bins}. The largest yield is produced by Solution~5 (Fig.~\ref{fig8}), whose family center $a_{\rm c}=2.48$~au is closest to the J3:1 among our test solutions. Here many objects are directly injected into the J3:1, leaving fewer available for delayed transport. The age for Solution~5 is $T = 865$~Ma, which is reasonably consistent with the ages of lunar craters, lunar impact glasses, and meteorite shock-degassing events thought to be associated with the $800$~Ma impact spike (see Sec.~\ref{back} as well as Figs.~\ref{fig_craters} and \ref{fig_glass}). 

The next highest yield of objects comes from Solution 1, where $a_{\rm c}=2.475$~au is slightly further from the J3:1. This solution has an estimated age of $T = 785$~Ma, which is close to the inferred age of Copernicus crater. This proximity in age, however, might not necessarily favor this scenario. A substantial number of objects would leave via the J3:1 over the subsequent $100–200$~Myr, potentially shifting the end of the impact shower closer to $\sim 600$~Ma. This timeline may not align well with the observed data.

As a cautionary note, it is important to remember that the ages for Solutions 1–5 could be adjusted by modifying parameters such as the thermal conductivity or bulk densities of the individual family members. We consider this fine-tuning that would have minimal impact on the overall yield of fragments, which is the primary focus of this section.

\begin{figure}[t!]
 \begin{center}
 \includegraphics[width=0.47\textwidth]{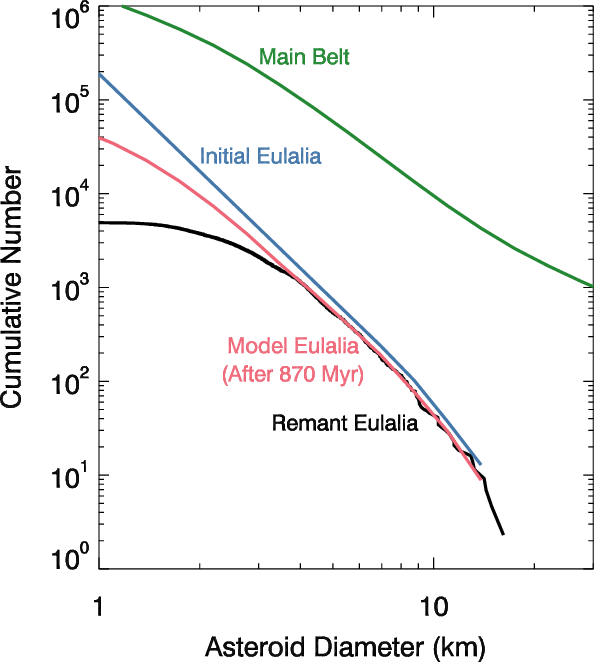} 
 \end{center} 
 \caption{Collisional evolution of the Eulalia family size frequency distribution (SFD). We define it as the remnant SFD, with considerable material lost to the J3:1. The green curve represents the main belt SFD as modeled by \citet{bot2020}. The blue curve illustrates our estimated initial SFD for the Eulalia family. The red curve shows the Eulalia family's SFD after $870$~Myr of collisional evolution. The black curve corresponds to the observed Eulalia family SFD as derived using the methods described in the text. For family members between $3 < D < 8$~km, the difference between the blue and red SFD is an approximate factor of 1.5.} 
 \label{fig_collisions}
\end{figure}

\subsection{Collisional evolution for Eulalia family fragments}\label{model2}

Our next objective is to understand how the Eulalia family SFD was shaped by collisional evolution over the family's lifetime (as defined by Solutions~1 to 5). Here we concentrate on the portion of the Eulalia family that did not escape via the J3:1, but instead remained in the main belt. We define it as the remnant Eulalia family. Shortly after the remnant family's formation, it began experiencing collisions with asteroids from the significantly larger main belt SFD, which exceeded it in size by several orders of magnitude. Collisions between remnant family members can be considered negligible for this exercise. Given enough time, main belt impacts should reshape the remnant family's SFD, causing it to more closely take on the wavy shape of the main belt SFD \citep[e.g.,][]{coddem1,coddem2,betal2015}. As shown in Fig.~\ref{fig_collisions}, the remnant Eulalia family SFD for $D \lesssim 2$~km bodies may have already reached this state. 

To investigate how collisions affected the remnant Eulalia family, we employ the Collisional and Dynamical Depletion Evolution Model (CoDDEM) described by \citet{coddem1,coddem2}. CoDDEM was designed to accurately track how collisional evolution affected the main belt SFD for the last several Gyr. Using this 1D code, \citet{coddem1,coddem2} followed how main belt impacts modified the number of objects within specific diameter bins between 0.0001 to $1,000$~km. The bins were defined by logarithmic intervals of $d\log D = 0.1$. Details on the initial assumptions and computational methodology can be found in the referenced papers. 

CoDDEM was used by \citet{bot2020} to reproduce the crater SFDs observed on spacecraft-visited asteroids with diameters $D > 10$~km, such as Ceres, Vesta, Lutetia, Mathilde, Ida, Gaspra, and Eros. As part of this work, \citet{bot2020} updated and refined the formulation of the main belt SFD over what has been done in \citet{coddem1,coddem2}. This provides us with reasonable confidence in CoDDEM's capability to effectively simulate the collisional evolution of the remnant Eulalia family's SFD. 

Our approach using CoDDEM involved the following steps. First, we input an estimated initial SFD for the remnant Eulalia family into CoDDEM. Second, we modeled how this SFD evolved over the family's estimated lifetime, as determined by Solutions~1 through 5. Third, we compared the simulated results to the observed SFD of the remnant Eulalia family to assess whether the two align reasonably well. By repeatedly iterating through this procedure, we eventually found a satisfactory match. From there, we used the chosen initial SFD to estimate how collisions have altered the remnant population, particularly for asteroid sizes relevant to the Eulalia impact shower.  

While we analyzed Solutions~1 through 5, our discussion will focus exclusively on Solution 5. It offers the most favorable conditions for producing a lunar impact shower $\sim 800$~Ma, as shown in Fig.~\ref{fig_glass}. Specifically, Solution 5 satisfies two key criteria: (i) a significant portion of the Eulalia family's mass is lost relatively close to $800$~Ma, and (ii) it delivers more large objects to the J3:1 compared to Solutions~1–4. We argue that if Solution~5 cannot produce an adequate impact shower, the results from Solutions~1–4 are unlikely to be relevant. As shown in Fig.~\ref{fig8}, the Eulalia family's estimated age for Solution~5 is $865$~Ma.  

Our collisional evolution results for Solution~5 are presented in Fig.~\ref{fig_collisions}. The black curve represents the observed Eulalia family SFD, scaled by a factor of $F_1\simeq 2.3$. The blue curve depicts our estimate of the initial remnant SFD; recall that it does not include family members that were lost to the J3:1. To infer the characteristics of the initial remnant family at small sizes, we calculated the power-law slope for family members with $3 < D < 10$~km and extrapolated it to smaller sizes. Although this approach is approximate, it represents the best we can do with current information. Finally, the red curve represents the modeled remnant SFD after $865$~Myr of collisional evolution. It aligns well with the observed family SFD in the size range of $3 < D < 15$~km. A growing discrepancy becomes apparent at smaller sizes, where the observed family SFD is increasingly influenced by the effects of observational incompleteness.

Our results indicate that collisions had a substantial effect on the number of $D \lesssim 10$~km asteroids within the remnant family. We predict that $3 < D < 10$~km objects lose a factor of $\approx 1.5$~in population. For objects with $D \lesssim 3$~km, the loss factors are modestly larger because smaller objects in the family become increasingly easy to disrupt \citep[e.g.,][]{coddem1,coddem2,bot2020}. These values will be used in our efforts to estimate the total number of Eulalia family members that successfully migrated into the J3:1.

Our estimated loss factors lead to an increase in the initial remnant SFD of the Eulalia family, as shown in Fig.~\ref{fig_collisions}. To incorporate these adjustments into our impact shower calculations, the values in Table~\ref{bins} need to be scaled by these loss factors going forward.

\begin{figure}[t!]
 \begin{center}
 \includegraphics[width=0.47\textwidth]{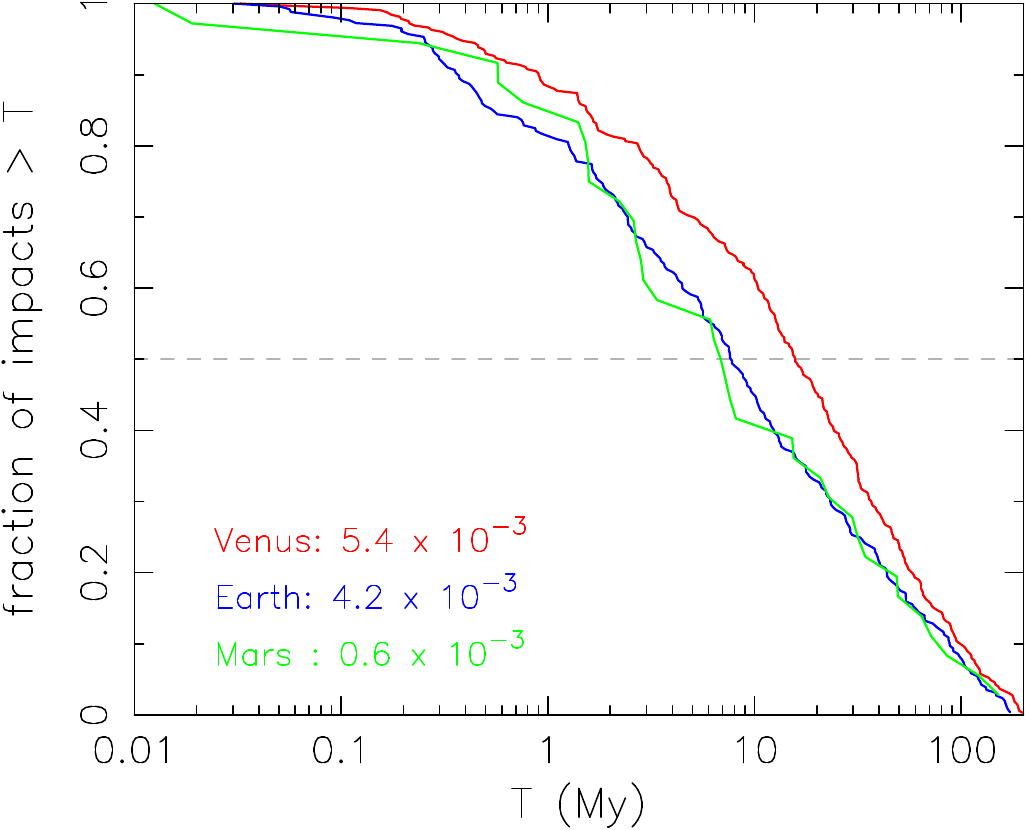} 
 \end{center} 
 \caption{Fraction of impacts on terrestrial bodies over time. We define $T$ as the interval from injection of the particle into the J3:1 to impact on a planet. The total impact probability is $5.4\times 10^{-3}$ for Venus, $4.2\times 10^{-3}$ for Earth, and $0.6\times 10^{-3}$ for Mars. Only particles entering the resonance with osculating inclination $\leq 7.5^\circ$ were considered. Data constructed using simulations performed by \citet{neomod1}.} 
 \label{figimp}
\end{figure}

\subsection{Collision probabilities with the Moon and terrestrial planets}\label{model3}

We are now prepared to estimate how many Eulalia family fragments impacted the Moon and the terrestrial planets. This involves calculating the collision probabilities of those objects with those worlds.

Here we took advantage of the extensive numerical simulations conducted by \citet{neomod1}. In their study, 100,000 synthetic objects were injected into various resonant locations responsible for feeding the near-Earth asteroid population. For our analysis, we focused specifically on results for the J3:1. From the 100,000 particles integrated by \citet{neomod1}, we selected 61,065 that entered this resonance with osculating inclinations below $7.5^\circ$, consistent with the low-inclination nature of the Eulalia family. Given that we used such a large number of particles, we can directly calculate impacts on planetary surfaces within the simulation, eliminating the need for approximate statistical methods \citep[e.g.,][]{bot1994}. We found that 257 impacts were recorded on Earth, corresponding to a total impact probability of $4.2 \times 10^{-3}$. Similarly, the impact probabilities for Venus and Mars were found to be $5.4\times 10^{-3}$ and $0.6\times 10^{-3}$. 

An additional advantage of directly resolving planetary impacts is the ability to accurately characterize their distribution in time $T$ after fragments enter the J3:1. This is illustrated in Fig.~\ref{figimp}, where the x-axis represents time $T$ and the y-axis shows the cumulative fraction of impacts occurring at times greater than $T$. The cumulative distributions are similar across all terrestrial planets, with a slight tendency for impacts on Venus to occur at later times. The median values of $T$ are $6.4$~Myr for Mars, $7.5$~Myr for Earth, and $15.7$~Myr for Venus. The simulations did not explicitly include the Moon, but we adopt a characteristic impact probability that is approximately 20 times smaller than Earth's, based on the gravitational cross sections of both worlds \cite[e.g.,][]{maz2019}.

Combining these results with those from Solution~5, we find that the majority of the Eulalia impact shower, as shown in Fig.~\ref{figimp}, would take place within the $\sim 700-850$~Ma time window. This aligns reasonably well with the lunar crater and impact glass data presented in Figs.~\ref{fig_glass} and \ref{fig_glass}), both which suggests a peak near $800$~Ma. It is important to note that the start time of the Eulalia impact shower can be shifted by adjusting the initial parameters of Solution~5. As such, further efforts to refine the exact timing are not necessary at this stage. 

\begin{figure}[t!]
 \begin{center}
 \includegraphics[width=0.47\textwidth]{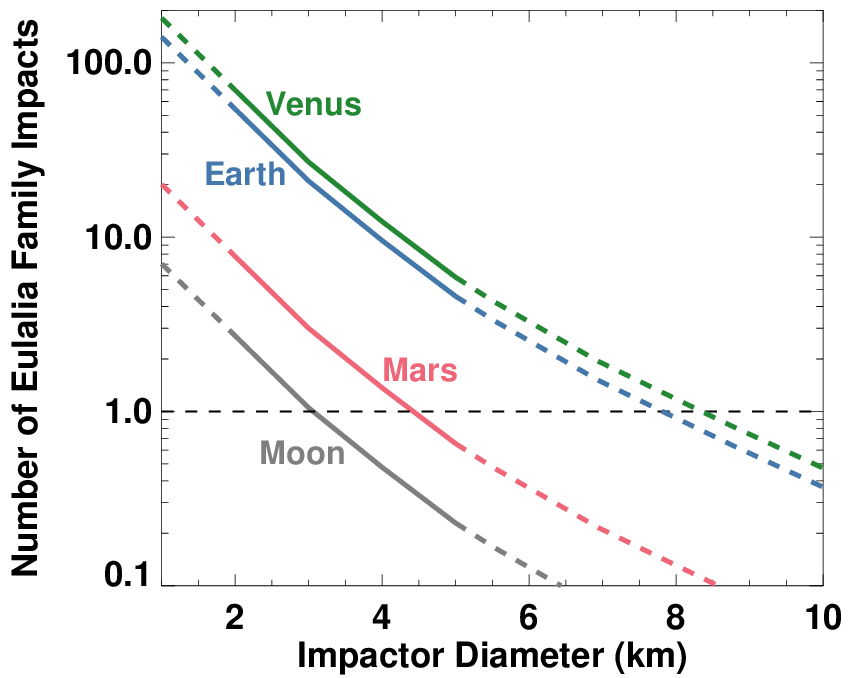} 
 \end{center} 
 \caption{The number and sizes of Eulalia family members impacting the terrestrial planets and the Moon. The impact flux for Venus (green) is the highest, followed by the Earth (blue), Mars (red), and the Moon (gray). The impact flux was calculated using the data from Table~\ref{bins}, the collisional evolution results shown in Fig.~\ref{fig_collisions}, and the collision probabilities outlined in Fig.~\ref{figimp}. We assume that the lunar impact flux is $20$ times lower than that of Earth. The solid lines use the values from Table~\ref{bins}, while the dashed lines are extrapolated from the shape of the observed Eulalia SFD on the large end (black curve in Fig.~\ref{fig_collisions}) and from the shape of the estimated initial Eulalia SFD on the small end (blue curve in Fig.~\ref{fig_collisions}).} 
 \label{fig_impacts}
\end{figure}

\subsection{Eulalia family impact rates with the Moon and terrestrial planets}\label{model4}

We are now ready to bring all the components together to calculate the scale of the Eulalia impact shower across different worlds. This involves multiplying the dynamical results from Solution~5, represented in Table~\ref{bins}, by the collisional evolution enhancement factors from Sec.~\ref{model2} (i.e., the differences between the blue and red curves) and the collision probabilities outlined in Sec.~\ref{model3}. Our results are presented in Fig.~\ref{fig_impacts}.

In this figure, the Eulalia family model results for family members between $2 \le D \le 5$~km are shown as solid curves, while the dashed curves represent extrapolations based on our understanding of the family SFD shape shown in Fig.~\ref{fig_collisions}. It is important to note that for larger asteroid sizes, the dashed curves are modestly conceptual, as the J3:1 can exhibit highly diffusive behavior for objects located near its boundaries \citep[e.g.,][]{vok2016}. We may never know the exact number of large objects produced by the family forming event in that region, as many of these objects likely migrated into the J3:1.

We find that, on average, the largest impactor to strike Earth and Venus is approximately $D \approx 8$~km. This size is comparable to the estimated $D \sim 10$~km projectile that created Chicxulub crater $66$~Ma on Earth \citep[e.g.,][]{schu2010,nes2021}. The Eulalia impact shower would also produce the order of 5 $D > 5$~km impactors on those worlds. Each one would produce craters as large or larger than Popigai crater on Earth, which is about $100$~km in diameter \citep[e.g.,][]{grieve2001, schm2020}. For Mars, the largest impactor to strike on average is $4.5$~km. Such an impact would likely create a crater comparable to the size to Copernicus crater on the Moon.    

The largest projectile to hit the Moon on average is $D > 3$~km. To convert that projectile size into a crater size, we use the Shoemaker crater scaling formula discussed in \citet{shoemaker1990, stuart2004,sheward2025}. Our justification for this choice is provided in Sec.~\ref {appa3}. We find that the size of this projectile would be sufficient to form Lowell (65.5~km) within the $800$~Ma timeframe shown in Fig.~\ref{fig_impacts}. The formation of Copernicus crater, however, would require a larger projectile. Note that Copernicus represents a relatively unique impact on the lunar nearside. Over the past $\sim 1$~Gyr of lunar history, only Tycho crater (86 km) is close in size to Copernicus \citep {wilhelms1987}. 

\begin{figure}[t!]
 \begin{center}
 \includegraphics[width=0.47\textwidth]{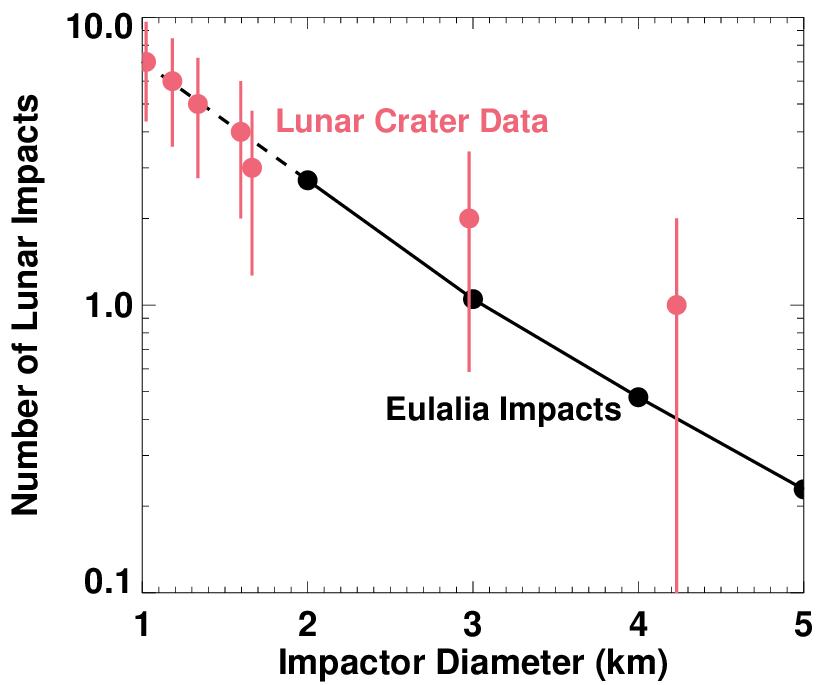} 
 \end{center} 
 \caption{The number and sizes of Eulalia family impacts on the Moon are compared to the estimated projectile sizes responsible for creating the large lunar craters with ages comparable to that of the Copernicus crater. The lunar craters (represented as red dots) are from Fig.~\ref{fig_craters}, while the black curve represents the lunar impact flux calculated in Fig.~\ref{fig_impacts}. The error bars for the number of craters reflect $\sqrt{2}$ uncertainties. Our analysis shows that the modeled Eulalia impact shower is consistent with the observed data within the stated uncertainties.} 
 \label{fig_compare}
\end{figure}

In Fig.~\ref{fig_compare}, we present a comparison between the projectile sizes that produced the lunar craters from Fig.~\ref{fig_craters} that formed near 800 Ma and the impactor flux associated with the Eulalia family. The craters considered are Al-Khwarizmi (22.5~km in diameter), Stefan (26~km), Saha (29.4~km), Godin (35.1~km), Das (36.6~km), Lowell (65.5~km), and Copernicus (93.1~km). Using the Shoemaker crater scaling law (see Sec.~\ref{appa3}), we converted crater diameters into their corresponding projectile sizes and plotted these as red points on the cumulative distribution. Standard $\sqrt{2}$ error bars were included to represent the uncertainties in these estimates. The Eulalia family impactor flux is overlaid in black. Notably, the calculated impact flux is consistent with the uncertainties of the crater data.  

Based on these results, it appears unlikely that Stevinus crater (70.3~km), whose age uncertainty margins approach the age of the Eulalia family for Solution 5 (i.e., $865$~Ma) was made by the Eulalia impact shower unless we are underestimating the net flux. The two smaller craters with comparable ages to Stevinus, 54S (19.7~km) and Klute (30.2~km), however, could potentially be included with no change to our model impact flux. At this time, we opt to remain conservative and exclude all three from consideration. 

\begin{figure}[t!]
 \begin{center}
 \includegraphics[width=0.47\textwidth]{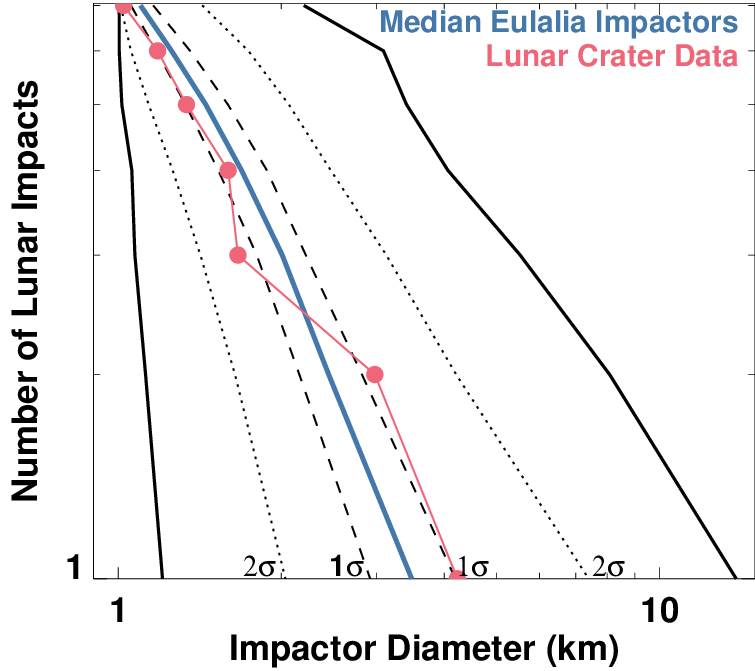} 
 \end{center} 
 \caption{The range of stochastic impacts on the Moon from the Eulalia impact shower. Using a Monte Carlo code, we performed 10,000 simulations to track seven projectiles striking the Moon. The solid black curves represent the maximum range of impactor diameters determined from these simulations. The long-dashed and short-dashed black lines indicate the 1~$\sigma$ and 2~$\sigma$ ranges from the ensemble, respectively, while the blue curve represents the median impactor diameter. The red dots correspond to large lunar craters from Fig.~\ref{fig_craters} that have ages similar to that of Copernicus crater. A comparison between model and observations show the model impact shower generally falls within the 1~$\sigma$ range.}
 \label{fig_monte}
\end{figure}

While Fig.~\ref{fig_compare} is useful, we would like to get a better feeling for whether an impact event capable of making Copernicus crater is statistically probable. To do so, we put together a Monte Carlo simulation. In each trial, we used random deviates to generate 7 projectiles with $D > 1$~km from the impactor SFD shown in Fig.~\ref{fig_monte}. After running 10,000 trials, we sorted the resulting projectile sizes and constructed an envelope representing the range of possible impacts. The black solid lines define the full range of impact outcomes across all 10,000 trials, while the black dashed lines mark the $1\sigma$ and $2\sigma$ envelopes indicating where the impactor sizes fall below or exceed a given value. The blue line represents the median impactor SFD from Fig.~\ref{fig_collisions}, while the red line represents the projectile sizes needed to make the lunar craters from Fig.~\ref{fig_craters}.  

We find that the results of our Eulalia impact shower simulations generally fall within the $1\sigma$ envelope for producing the observed lunar craters near $800$~Ma. Considering the uncertainties inherent in our analysis, we regard this as a reasonable and satisfactory outcome. It is also the best we can do with the available data at the present time. More definitive proof may have to wait for some future era when sample return from the Moon is relatively commonplace.


\section{Implications}\label{implications}

In the analysis presented above, we have argued that a pronounced impact shower linked to the breakup of the Eulalia parent body occurred around 800 Ma. This conclusion is directly supported by evidence from collisional and dynamical evolution models as well as age data from lunar craters, lunar impact glasses, and meteorite shock degassing records. 

The next question to consider is whether this putative impact shower had any meaningful influence on the evolution of the terrestrial planets. Our challenge lies in the fact that evidence connecting various events to the impact shower is often ambiguous and/or difficult to obtain. This presents a classic ``chicken-and-egg" problem: while we seek more evidence of impact-related events on various worlds from around 800 Ma, future researchers interested in pursuing this work need an engaging motivating problem to take on such studies.

Accordingly, to encourage new research activities, we use the following sections to explore plausible, though necessarily speculative, connections to events occurring around $\sim 800$~Ma on Earth, Mars, and Venus. Our goal is to present scenarios that inspire the development of new data or modeling studies capable of supporting or refuting these ideas.

\subsection{Examining potential connections between events on Earth and the Eulalia impact shower}\label{imp1}

\subsubsection {Impact structures} \label{imp1a}

Perhaps the most intriguing question to ask is whether there is any evidence for the Eulalia impact shower on Earth. The most direct way to investigate this would be through the record of Precambrian impact craters. Unfortunately, the majority of terrestrial craters preserved today are younger than $650$~Myr old, whereas most impacts from the Eulalia shower are $\gtrsim 100$~Myr older. This apparent deficit in older terrestrial craters coincides with the onset of major episodes of globally extensive ``Snowball Earth'' glaciations occurring between approximately $650$ and $720$~Ma \citep[e.g.,][]{hoff2009, rooney2015}. They are believed to have caused substantial subglacial erosion, removing many kilometers of material from the continents. This erosional process likely obliterated the vast majority of impact craters that existed prior to this period \citep {kel2019}. This underscores the geological challenges in linking the Eulalia impact shower to Earth's record, as nearly all craters from $800$~Ma would almost certainly have been erased.

As an aside, we note that there some exceptions to the apparent paucity of ancient craters, with the most notable being the Sudbury and Vredefort impact structures \citep[$1850$ and $2023$~Myr old, respectively, e.g.,][]{schm2020}. These craters, with diameter $D_{\rm crat} > 130$~km, were sufficiently deep to survive the intense geological processes that followed their formation, though both exhibit signs of extensive erosion.  

This leaves us to examine indirect evidence for the Eulalia shower. 

\subsubsection {The Bitter Springs Anomaly}\label{imp1b}

Perhaps the most notable occurrence coincident with the timing of impacts from Eulalia fragments is the so-called Bitter Springs Anomaly (BSA). The BSA refers to a significant carbon isotope excursion in the geological record, characterized by an abrupt and prolonged negative shift in carbonate $\delta ^{13}$C values of $\sim 8$\textperthousand. It occurred around $800-810$~Ma and persisted for $5–10$~Myr before $\delta\, ^{13}$C values returned to pre-anomaly levels \citep{wor2019}. The BSA was initially discovered in the Bitter Springs Formation of the Amadeus Basin in central Australia but has since been recognized at several locations across the globe (e.g., Svalbard's Akademikerbreen Group, Canada's Mackenzie Mountains). 

The BSA is thought to have been driven by abrupt changes in ocean and atmospheric chemistry, and is potentially linked to a temporary collapse in primary productivity. For example, \citet{lu2017} find geochemical evidence of fluctuating ocean redox conditions, including initial expanded anoxia or euxinia, followed by transient surface ocean oxygenation, as indicated by elevated iodine-to-calcium+magnesium (I/[Ca+Mg]) ratios reaching up to $8$~$\mu$mol/mol during the recovery phase. These changes could have disrupted the balance of organic carbon burial and oxidation, thereby influencing the carbon isotope record. Low values of $\delta\, ^{13}$C are also commonly associated with episodes of cooling \citep[e.g.,][]{och2012}. 

The end of the BSA is coeval with multiple indicators of a rise of oxygen levels in the ocean and atmosphere \citep[e.g.,][]{plan2014,cole2016,lu2017,crock2019,lyons2021}. It may even be associated with accelerated eukaryotic diversification \citep[e.g.,][]{cohen2015,knoll2014,rs2017}, though see \citet{porter2025} for a contrasting view. If the latter statement is true, the events producing the BSA may have led to the emergence of new metabolic pathways for life or changes in ecological structures.

The origin of the BSA and its connection to large-scale environmental and biological changes remain poorly understood. The leading hypothesis is that the BSA is somehow linked to the breakup of Rodinia, a supercontinent that existed during the late Mesoproterozoic and early Neoproterozoic eras. Rodinia began to fragment between $850$ and $750$~Ma, likely as a byproduct of plume-related extension \citep{li2008}. The young end of this time window coincides with several Snowball Earth glaciations events ($\sim 635-720$~Ma), which are in turn associated with dramatic shifts in global geography, ocean circulation, atmospheric composition, and negative shifts in carbonate $\delta \, ^{13}$C values \citep[e.g.,][] {hoff2009, rooney2015}. Enhanced continental weathering during these times may have increased the influx of nutrients like phosphorous to the oceans, stimulating biological activity and altering carbon cycling. Alternatively, weathering processes could have introduced isotopically lighter carbon into the system, contributing to the negative $\delta\, ^{13}$C shifts.

The connection between Rodinia's breakup and the earlier Bitter Springs Formation (BSA), however, remains speculative. At $\sim 810$~Ma, Rodinia’s breakup was likely only in its infancy, while the Bitter Springs Formation predominantly consists of shallow marine sedimentary rocks, such as carbonates, shales, siltstones, and dolostones \citep{wor2019}. These lithologies are indicative of a warm, stable depositional environment and show no evidence of regional or global glaciation, further reducing the likelihood of a direct link between the two events.

In this context, we sought to investigate whether the Eulalia breakup event could explain the origin of the BSA. This influence might have occurred predominately through a single very large impact, multiple large impacts, and/or the delivery of substantial amounts of dust and small debris produced by the breakup. While direct modeling of how these phenomena affected Earth's atmosphere and climate is beyond the scope of this paper, we present two historical analogies from Earth's past that suggest such events may have played a strong contributory role.

\paragraph {Analogy \#1: Chicxulub impact event}

For the first analogy, consider that at the Cretaceous-Paleogene (K-Pg) boundary (KPB), a 
$\sim 2$\textperthousand~negative excursion in $\delta\,^{13}$C has been reported in both terrestrial organic carbon \citep[e.g.,][]{sdn1984,ari1999,aj2000,maruoka2007,therrien2007,gradpre2013,bourque2021} and marine inorganic carbon \citep[e.g.,][]{kl1989,schu2010,sepul2019}. This $\delta\, ^{13}$C anomaly was likely triggered by the large K-Pg impactor that struck the Yucatán Peninsula $66$~Ma \citep[e.g.,][]{archi2010,schu2010,hull2020}. That event released vast amounts of $\text{CO}_2$, sulfate aerosols, and other gases into the atmosphere, which in turn led to sudden global cooling, acidic rain, and disruptions to the carbon cycle \citep {morgan2022}. The negative $\delta\, ^{13}$C anomaly observed in marine sediments would have been a byproduct of the collapse of primary productivity in ocean ecosystems, as photosynthetic plankton preferentially incorporate the lighter carbon isotope $^{12}\text{C}$ during carbon fixation. Additionally, the rapid influx of carbon from decaying organic matter, and the potential destabilization of seafloor methane hydrates, could have further amplified the release of $^{12}\text{C}$-enriched carbon into the ocean-atmosphere system, reinforcing the negative $\delta\, ^{13}$C anomaly. Together, these processes reflect the dramatic impact of the Chicxulub event on Earth's carbon cycle and global climate.

Our model results suggest that the largest projectile to strike Earth from the Eulalia asteroid shower could have been comparable in size to the Chicxulub impactor. Its effects on the early Neoproterozoic era remain unclear, however, given the vastly different environmental conditions that existed on Earth at that time compared to $66$~Ma. Modeling work on the Chicxulub impact indicates that large terrestrial impact events eject enormous amounts of dust and debris into and beyond Earth's atmosphere, with the subsequent reentry of this material radiating enough thermal energy to potentially subject Earth's surface to extreme heating for several hours \citep[e.g.,][]{morgan2022}. Afterwards, the lingering dust and sulfate aerosols in the atmosphere would have formed dense stratospheric clouds, reflecting sunlight and triggering significant global cooling. This so-called ``impact winter" appears to have triggered widespread environmental upheaval. Although it remains uncertain whether the largest Eulalia impact alone was big enough to cause similarly dramatic effects, it is plausible that such an event, perhaps in combination with numerous smaller impacts, may have prolonged the disruption and hindered the recovery of oceanic ecosystems over an extended period. Collectively, these cascading effects may have plausibly produced or contributed to the BSA.

\paragraph {Analogy \#2: L chondrite impact shower from 466 Ma}

For the second analogy, we examine an asteroid breakup event that shares several similarities with the Eulalia breakup, namely the disruption of the L chondrite parent body approximately $466$~Ma \citep[e.g.,][]{haack1996,swindle2014,schmitz2001,liao2020,ts2021,schmitz2022,zhang2024,zhang2025}. The timing of L chondrite breakup event (LCBE) is well documented by multiple lines of evidence, including isotopic dating of mid-Ordovician L chondritic meteorites \citep[e.g.,][] {haack1996,swindle2014,walton2023}, fossilized meteorites preserved in sedimentary rocks \citep{schmitz1997,schmitz2001,heck2004}, and layers of micrometeorites recovered from sediments deposited shortly after the breakup \citep{schmitz2001,schmitz2003,schmitz2008,schmitz2022,zhang2024,zhang2025}. Zircons embedded into the same Ordovician limestone as the fossil meteorite indicate the LCBE occurred $465.8 \pm 0.3$~Ma \citep {liao2020}. The specific asteroid family associated with the LCBE remains debated; some researchers link it to the Gefion family, disrupted near the J5:2 mean motion resonance with Jupiter \citep{nes2009}, while others favor the Massalia family, located near the J3:1 \citep{marsset2024} (\citep[see also][]{ciocco2025}). We defer a detailed discussion of this debate to a future paper. For the purposes of this study, we will focus on how debris from the LCBE may have affected Earth’s environment.

As with the Eulalia breakup event, there is intriguing evidence that the LCBE triggered an asteroid shower during the mid-Ordovician period. Sub-kilometer impactors produced by the LCBE are compelling candidates to explain the multitude of $D_{\rm crat} \lesssim 10$~km craters that formed on Earth between $440–470$~Ma \citep{schm2020, schmitz2022,osinski2022}. They may also be responsible for an increase in the production rate of $D_{\rm crat} \sim 10$~km lunar craters near this time \citep{maz2019}. For this analogy, however, we emphasize the prolonged and enhanced flux of meteoroids, dust, and micrometer-scale particles that the LCBE delivered to Earth over millions of years. Empirical evidence suggests that the flux of small debris during this time was as much as four orders of magnitude higher than the average background flux reaching Earth over the entire Phanerozoic eon \citep{schmitz2019}. Such a massive influx of debris also helps explain the presence of L chondrite fossil meteorites preserved in the Thorsberg limestone quarry in Sweden \citep{schmitz2001,schmitz2008}. To put fossil meteorites into perspective, consider the immense effort required to locate meteorites on the ground today, while here researchers are literally digging them out of the ground, with the fossil meteorites embedded within large blocks of excavated limestone.

The prolonged influx of debris from the LCBE appears to have had substantial atmospheric effects on Earth. For example, atmospheric cooling may have been driven by high-altitude dust, which in turn affected the planet's climate and ocean systems. \citep{schmitz2019} propose that the ice age conditions of the mid-Ordovician were either initiated or intensified by this event. The shower may have fertilized the oceans with phosphorus and other nutrients, either by direct delivery or through materials scavenged from the continents via accelerated erosion. This influx of nutrients could have stimulated biological activity, which, in turn, may have contributed to a reduction in atmospheric $\text{CO}_2$ \citep{fil2008,reiners2018} \citep[see also][]{pasek2005,pasek2008}.  

We find it plausible that these changes, together with the numerous sub-kilometer and occasional multi-kilometer-sized impacts associated with the LCBE, played a significant role in driving broader ecological and evolutionary shifts during the mid-Ordovician period ($487–443$~Ma). One such event is commonly referred to as the Great Ordovician Biodiversification Event (GOBE) \citep[e.g.,][]{servais2004}. The GOBE represents a 20-30 Myr period of unequaled marine biodiversification that reshaped marine ecosystems, fostering the development of complex and specialized food webs that formed the foundation of modern marine life. It does not appear to have been a single event, but rather an extended sequence of biotic events that were driven by factors such as rising oxygen levels, nutrient-enriched oceans, tectonic activity, and global cooling \citep{servais2021}. This vibrant period ended with the Late Ordovician Mass Extinction (LOME) ($443 \pm 0.4$ Ma), one of Earth’s largest extinction events, which coincided with a period of dramatic climatic cooling \citep {benton2003, trotter2008, harper2014, harper2023, zhang2025}. 

While the GOBE and LOME were each pivotal to the history of life, their origins remain poorly understood. Intriguingly, analyses of Scandinavian-Baltic and Chinese strata suggest that the faunal and climatic shifts associated with GOBE and LOME may coincide with the LCBE and its aftermath. For example, the $465.8 \pm 0.3$ Ma age of the LCBE occurs near the start of GOBE’s main pulse of marine life radiation \citep[e.g.,][] {deng2023}. The LCBE’s asteroid shower also persisted for at least tens of Myr \citep{schm2020, schmitz2022,osinski2022}, suggesting a plausible causal link to the GOBE and LOME.

The remarkable increase in small debris delivered to Earth from the LCBE appears to have required favorable conditions. While numerous large asteroid breakup events have occurred within the main belt over the past $500$~Myr \citep[e.g.,][]{netal2015}, and extensive searches have been conducted to identify relict extraterrestrial chromite grains in sediments from various intervals of the Phanerozoic \citep{ts2021}, no other event has shown anything approaching the mid-Ordovician micrometeorite flux increase. A possible explanation is that the LCBE injected substantial amounts of material directly into a powerful nearby resonance, enabling rapid transport to Earth-crossing orbits. This mechanism may account for the short cosmic ray exposure ages of chromite grains originating from small L chondrite meteorites in this era, with some ages being as brief as 50,000 years. \citep {heck2004,nes2009}. A collisional cascade among the larger fragments that reached Earth-crossing orbits through the same mechanism would help to replenish this population of debris over time as well.

In most main belt breakup events, small ejected particles reach Earth more gradually, primarily via slow sunward drift driven by Poynting-Robertson drag. This process is sufficiently slow that many small particles undergo extensive collisional evolution en route, with many destroyed prior to reaching Earth \citep[e.g.,][]{nes2006,farley2006}. The replenishment of small debris via a collisional cascade would rely on impacts among millimeter-sized and larger fragments that remain near the disruption site, as such fragments are generally unaffected by Poynting-Robertson drag. Accordingly, most breakup events cannot beat the background dust flux for very long (see Appendix~\ref{appa2}). In fact, for much of Earth's history, the delivery of small particles has been primarily driven by the disruption of Jupiter-family comets, with only minor contributions from main belt collisions \citep{nes2008}.

This sets up an intriguing prediction for the Eulalia impact shower. Our analysis suggests that a significant fraction of the Eulalia parent body was directly injected into the J3:1 orbital resonance, potentially including a substantial amount of small particles. If the size distribution of debris from the Eulalia breakup contained sufficient small particles, it might have delivered a dust flux to Earth comparable to the LCBE shortly after disruption. While some differences would be expected (e.g., produced by variations in the size distributions of the two breakups, differences in the specific resonances that delivered the ejecta to Earth), comparable atmospheric and climatic effects may still have occurred. These effects could include atmospheric cooling due to high-altitude dust, the acceleration of ice age conditions, the deposition of dust enriched with phosphorus and other nutrients into the oceans, and increased continental erosion that similarly delivered phosphorus and nutrients to marine environments. Collectively, these processes had the potential to stimulate biological productivity and significantly influence Earth's ecosystems. Taken together, we propose that the former mechanisms may offer an alternative explanation for the onset of the BSA, while the latter may have contributed to the putative diversification of eukaryotes during this period \citep[e.g.,][]{cohen2015,knoll2014,rs2017}.

\paragraph {Concluding remarks}

While we have suggested many provocative correlations in this section, considerable work will be needed to prove these connections via genetic, paleontological, geochemical, geophysical, and dynamical lines of evidence. Our goal here is to highlight the possible interplay between certain extraterrestrial breakup events and evolutionary milestones on Earth. The bottom line is that major, well-timed disruption events, such as the Eulalia breakup event and the LCBE, may have brought about substantial changes in our biosphere and potentially influenced the complexity and diversity of eukaryotic life.


\subsubsection {Snowball Earth events of the Neoproterozoic era}\label{imp1c}

As briefly mentioned in previous sections, the Snowball Earth events of the Neoproterozoic era, known as the Sturtian ($660–717$~Ma), Marinoan ($635–641$~Ma), and Gaskiers ($\sim 580$~Ma) glaciations, are among the most extreme ice ages in Earth's history \citep[e.g.,][]{hoff2009, rooney2015}. These events are thought to have been initiated by a combination of geological and climatic factors. Once a critical threshold of ice accumulates, the ice-albedo feedback mechanism (i.e., ice reflects sunlight, causing further cooling and promoting the formation of more ice) can propel the planet toward a Snowball Earth state. The eventual termination of each Snowball Earth episode is believed to have been driven by sustained volcanic activity that released significant amounts of greenhouse gases (e.g., carbon dioxide) over tens of Myr.

Numerous mechanisms have been proposed to explain the onset of Snowball Earth events, but no single factor has yet been identified as the primary cause. A few notable examples are:

\begin {itemize}

\item Equatorial continents. The positioning of continents near the equator may have altered global weathering processes and atmospheric carbon dioxide levels, leading to cooling. \citep[e.g.,][]{hoffman2002snowball, kirschvink1992snowball}. 

\item Rising oxygen levels. Increased oxygen levels likely reduced atmospheric methane, a potent greenhouse gas, contributing to cooling \citep[e.g.,][]{kopp2005paleoproterozoic,holland2006oxygenation}.

\item Volcanic activity. Widespread volcanic eruptions, including large igneous provinces, could have released aerosols like sulfur dioxide into the atmosphere, reflecting sunlight and promoting planetary cooling \citep[e.g.,][]{schrag2002initiation,godderis2003longterm,hoff2009}.

\end {itemize} 

A less studied mechanism to date is whether large impacts produced sufficient global cooling to initiate a Snowball Earth event. It has been argued that the impact of a $5-10$~km diameter asteroid could produce a plume capable of injecting large quantities of rock vapor, fine particulate matter, and other volatile compounds (e.g., sulfur dioxide, water vapor, carbon dioxide) into the stratosphere and beyond \citep[e.g.,][]{koeberl2019}. Modeling studies further indicate that impacts from the larger end of this diameter range could produce sufficient cooling to push an already cold climate into a Snowball Earth state through the amplification effects of the ice-albedo feedback mechanism \citep{fu2024}.

The question for this paper is whether the Eulalia impact shower played any role in Snowball Earth events of the Neoproterozoic era. In support of this possibility, Fig.~\ref{fig8} shows that the Sturtian and Marinoan glaciations coincide with a period of heightened flux in the post-family formation era, such that it is plausible that numerous asteroids from the Eulalia family struck Earth during these periods. Our model cannot provide the precise timing of the largest impacts, but some could have indeed occurred during either glaciation event. For the Gaskiers glaciation, which took place during the waning phase of the impact flux, our model suggests a lower likelihood of it being directly connected to the Eulalia impact shower. 

On the opposing side, there is currently no direct geological evidence to suggest that the Sturtian or Marinoan glaciations were triggered by an impact event. The best available evidence to date, which is arguably ambiguous, comes from \cite{bod2005}, who investigated iridium anomalies found at the base of the cap carbonates associated with these glaciations. Iridium is considered a strong tracer of extraterrestrial materials because it is rare in the Earth's crust but relatively abundant in asteroids, comets, and cosmic dust. \cite{bod2005} proposed that during periods when the Earth was covered in ice, cosmic dust accumulated on or within the ice layers. Upon the rapid melting of the ice sheets at the end of the glaciation event, this cosmic dust, enriched with iridium, would have been deposited, leaving a distinct iridium signature in the geological record. \cite{bod2005} argued that this record could potentially serve as a ``clock," offering insights into the duration of the glaciation period. Alternatively, we find it possible that most of the observed iridium anomaly was instead delivered by one or more large asteroid impacts.  At this time, however, there is no way to tell the difference between the two possibilities.     

A challenge to both scenarios is that the findings of \cite{bod2005} may be region-specific, as similar anomalies were not confirmed at other locations. While it is possible that modestly large impacts produce  regional rather than global iridium deposits, it is also conceivable that the conclusions drawn are based on non-representative or incomplete samples. Ultimately, further data and analysis from the Sturtian and Marinoan eras will be required to confirm or rule out the possibility that impacts played some role in these Snowball Earth events.

\begin{figure}[t!]
 \begin{center}
 \includegraphics[width=0.47\textwidth]{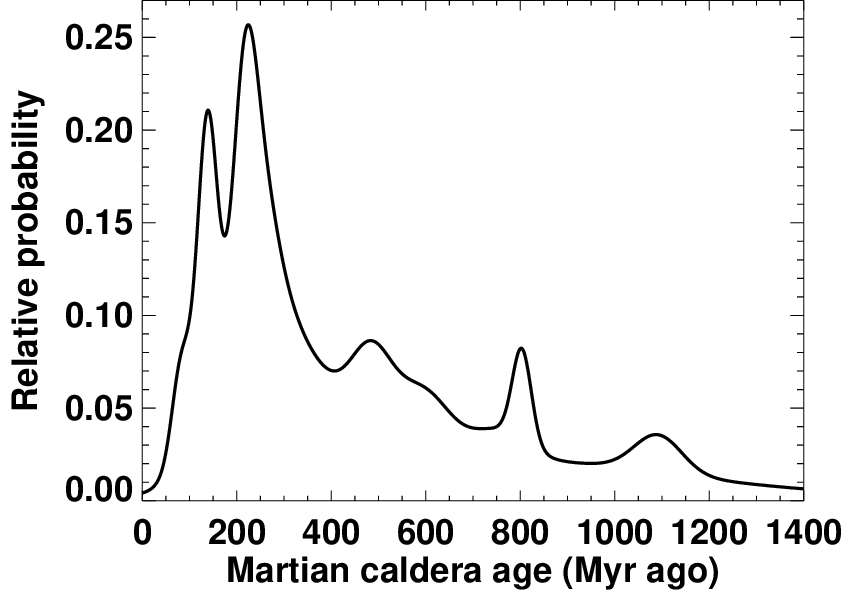} 
 \end{center} 
 \caption{The estimates ages of Mars calderas as derived using data from \citet{rob2011}. Here we assume the ratio of Martian to lunar impacts per square kilometer is $R_b = 2.8$. The mean age and $1~\sigma$ age uncertainties are represented as Gaussian distributions, which are combined to produce the observed age distribution. A substantial increase in caldera ages is observed near $800$~Ma, the time of the Eulalia impact shower. Additional increases are observed for ages younger than $300$~Ma, coinciding with the period when the lunar impact flux, as shown in Fig.~\ref{fig_glass}, also exhibits a notable rise.} 
 \label{fig_caldera}
\end{figure}

\subsection{Evidence for increased Martian volcanism $800$~Ma}\label{imp2}

Surprising evidence that may point to a potential increase in the Martian impact flux around $800$~Ma comes from \citet{rob2011}. In their work, they examined the calderas of Mars's 20 largest volcanoes using high-resolution images from the Mars Reconnaissance Orbiter’s ConTeXt Camera and measured the crater SFDs superposed on each one. Using a crater production function, they then estimated the ages of the most recent volcanic activity associated with each of the major volcanoes' calderas. Efforts were also made to count craters as small as $60$ meters in diameter to account for and examine the influence of secondary cratering on caldera ages.  

To determine Martian surface ages from crater counts, it is necessary to adapt the Moon's crater chronology to Mars. A critical parameter for this calculation is the impact rate ratio between Mars and the Moon, a value called ``R bolide" ($R_b$). It is formally defined as the number of impactors striking each planet per unit surface area per unit time \citep[e.g.,][]{ivan2002}. It assumes that the projectile SFDs impacting both worlds have the same shape. We consider this assumption reasonable because most sub-km impactors striking the Moon and Mars come from the inner and central main belt \citep[e.g.,][]{bot2002,bot2015,granvik2018,neomod1,neomod2,neomod3}. For our work, we assume $R_b = 2.8$ \citep{ivan2002}, a value commonly used to determine Martian surface ages (e.g., results assuming a range of uncertainties from \cite{hartmann2005} yield $R_b = 2.6 \pm 0.7$; \citet {williams2014}).
  
The ages of 36 Martian volcanic calderas, derived from data reported by \citet{rob2011}, are shown in Fig.~\ref{fig_caldera}. Individual age distributions were calculated by creating a Gaussian profiles for each caldera, with the profile centers and widths corresponding to their most probable ages and associated $1\sigma$ uncertainties. All distributions were then summed, with the black curve representing relative probability.  

Several peaks are found in caldera age timeline. Relatively sharp peaks are found near $140$, $220$, and $800$~Ma. More subdued local maxima are found near $480$ and $1090$~Ma.  Naturally, we find the $800$~Myr peak to be intriguing, given the expected timing of the Eulalia asteroid shower. This correspondence could be considered a fluke, but several of the other peaks also appear to be associated with times when the inner solar system impact flux may have been higher.  

For example, the subdued peak near $480$~Ma closely aligns with the predicted breakup age of the L chondrite parent body, estimated to occur at $466$~Ma \citep[e.g.,][]{haack1996,schmitz2001,swindle2014,ts2021,schmitz2022,zhang2024,zhang2025} \citep[see also][for discussions of possible source families]{nes2009,marsset2024}. As discussed in Sec.~\ref{imp1}, this event triggered a substantial spike in the number of small bodies falling on Earth \citep[e.g.,][]{schmitz1997, schmitz2001,schmitz2003,schmitz2008,schmitz2011}. Evidence further suggests a substantial increase in the flux of modest-sized impactors following this event. This is supported by the rise in the number of terrestrial craters with $D_{\rm crat} \lesssim 10$~km and lunar craters with $D_{\rm crat} \sim 10$~km that formed near this time \citep{maz2019,schmitz2022}. As shown in Fig.~\ref{fig_glass}b, it may also be seen in the ages of lunar impact glasses.

Although the peaks near $140$ and $220$~Ma have not yet been definitively linked to an increase in the Martian impact flux or to a particular asteroid disruption event, they align with slope changes observed in Fig.~\ref{fig_craters} for $D > 20$~km lunar craters, as reported by \citet{maz2019} and \citet{ter2020}. While we acknowledge that this correlation could be coincidental due to small number statistics, it is nonetheless worth highlighting for further consideration. 

Even the relatively subdued peak around $1090$~Ma could reflect an increase in the impact flux, as this age coincides with the estimated disruption of the Flora parent body. As discussed above, multiple lines of evidence suggest that the Flora parent body fragmented approximately $1300 \pm 300$~Ma \citep{vok2017,bot2020}. Dynamical simulations indicate that the peak of inner solar system impacts from the Flora asteroid shower would likely have occurred $100$ to $300$~Myr after the family-forming event \citep{vok2017}. Within the margin of error, this timeline aligns with a slope change observed in Fig.~\ref{fig_craters} in the ages of lunar craters with $D_{\rm crat} > 20$~km. Supporting evidence may also be found in the ages of lunar impact glasses (Fig.~\ref{fig_glass}b).

The question of why asteroid impacts might be associated with increased volcanic activity on Mars revolves around their potential to induce seismic shaking. The idea of a connection between earthquakes and volcanic eruptions dates back to Charles Darwin's observations during his expedition to Chile in 1835. After witnessing the Concepci\'{o}n earthquake, Darwin noted increased activity in some nearby volcanoes, sparking one of the earliest discussions on this relationship \citep {darwin1840}. Today, this phenomenon is well-documented on Earth, with numerous studies confirming that significant seismic events can disrupt magmatic systems and trigger or enhance existing volcanic eruptions \citep[e.g.,][]{gonz2021}. The precise mechanisms behind this interaction, which include dynamic stress changes, lithostatic pressure reduction, permeability changes, bubble nucleation and growthin magmas, advective overpressure associated with bubble rise and falling crystal roofs continue to be active areas of research \citep {seropian2021}.

The mechanical energy released by impacts are thought to produce dynamic stress changes that propagate through the crust and mantle, affecting magmatic and volcanic systems \citep[e.g.,][]{mb2006,ph2018}. Such stress changes can temporarily reduce lithostatic pressure or disrupt the stability of magma reservoirs. In turn, they help promote the ascent of magma that was previously held in equilibrium \citep[e.g.,][]{brodsky1998,manga2012}. 

Seismic waves can also fracture surrounding rock, enhancing permeability and allowing the migration of volatiles such as water and carbon dioxide into magma chambers \citep{manga2012}. The influx of volatiles reduces magma viscosity, promoting faster magma rise toward the surface. Even moderate seismic disturbances can reset a system's pressure equilibrium by altering magma chamber walls or gas pocket dynamics, ultimately creating pathways for volcanic eruptions \citep{ew2009}.

A compelling example of an impact triggering a new round of volcanic activity on Mars was proposed by \citet{horvath2021}. They point out that the Cerberus Fossae mantling unit, believed to be Mars's youngest volcanic deposit, has an estimated age of $46-222$~ka. Notably, the eruption site is located approximately $10$~km from Zunil crater, a $D_{\rm crat} = 10$~km crater with a closely matching estimated lower bound age of $53\pm 7$~ka. \citet{horvath2021} argue that the proximity of these features and their near-synchronous formation are so statistically improbable that they suggest a causal relationship. Here seismic shaking generated by the formation of Zunil crater, which predates the Cerberus Fossae mantling unit, may have destabilized subsurface magma reservoirs, thereby triggering the eruption of molten rock onto the surface through pre-existing fissures. 

Given the apparent connection between the Eulalia impact shower and Martian volcanism, we propose that impact-induced seismic shaking acts as a key catalyst for mobilizing magma and influencing volcanic activity on Mars, both on short- and long-term timescales. Investigating this phenomenon further could yield valuable insights into several interconnected fields, including the evolution of asteroids, the flux of asteroid impacts over time, and the geological and volcanic evolution of Mars during the Amazonian epoch. Such studies offer a promising avenue for advancing our understanding of the complex interplay between external impacts and internal planetary processes.


\subsection{Possible coincidences regarding the estimated surface age of Venus}\label{imp3}

We have not yet addressed whether the Eulalia impact shower had any effect on Venus. This omission is partly due to the uncertainty surrounding Venus's surface age, which some estimates suggest is only $300$ to $600$~Ma \citep[e.g.,][]{strom1994,nmk1998}. If this age range is accurate, impacts from the Eulalia family would predate the observable surface history of Venus by many hundreds of Myr, placing them beyond its event horizon. On the other hand, some studies propose that Venus may have a crater retention age as old as $1$~Ga \citep{mckinnon1997}, which would imply that its current surface could have been shaped, at least in part, by impacts from the Eulalia family. To investigate this possibility further, we must estimate the approximate age of Venus's surface. 

Here we use a relatively straightforward calculation that builds on previous work. By leveraging Earth's crater record and solar system dynamics, it is possible to estimate the age of Venus’s surface while folding potential variations in the impact flux over the last several hundreds of Myr (e.g., Fig.~\ref{fig_craters}) (see Sec.~\ref {back}). On Earth, there are 38 craters with diameters of $D_{\rm crat} \ge 20$~km that formed within the last $650$~Myr (PASSC-EID). \citet{maz2019} argued that most of these craters are located on stable cratonic surfaces that experienced minimal erosion during this time. As supporting evidence, they demonstrated that these same terrains contain an abundance of intact kimberlite pipes—volcanic features that serve as indicators of limited erosion. Building on this observation, \citet{maz2019} calculated that cratonic cratered surfaces cover approximately $10.7 \pm 3.1$\% of Earth's total area. Using this fraction, 38 $D_{\rm crat} \ge 20$~km craters correspond to $355^{+145}_{-80}$ craters over the entire surface of the Earth that formed in the last $\sim 650$~Ma.  

From here, we can use dynamical models to extend our analysis to Venus. Models suggest that Earth and Venus are impacted at approximately the same rate by planet-crossing objects \citep[e.g.,][]{granvik2018,neomod1,neomod2,neomod3}. Assuming that craters on Earth and Venus follow similar scaling laws, which seems probable given their similar sizes and gravitational accelerations, we estimate that a Venusian surface of similar surface age ($\sim 650$ Ma) should also exhibit approximately $355^{+145}_{-80}$ craters with $D_{\rm crat} \ge 20$~km.

Using this model value, we can make comparisons to the Venus crater record. Venus has a population of more than 900 craters distributed randomly across the surface of the planet \citep[e.g.,][]{strom1994}. The rims of most of these craters have not been substantially modified by volcanism and tectonics. It is debated whether this population was produced in the aftermath of a global surface erasure event \citep[e.g.][]{strom1994} or whether the crater population is in steady state \citep[e.g.][]{phil1992}. Many attributes of Venus craters can be found in Release~3 of the Venus Crater Database, which is an on-line update to the database described in \citet{herrick1997}. 

A complicating issue with interpreting Venusian craters is that its thick atmosphere prevents small projectiles from reaching its surface \citep[e.g.,][]{kory2000,kory2002,kz2003,kz2004,kz2005}. If an object breaks up while passing through the atmosphere but does not completely disintegrate, the separation produced by the surviving fragments may make an irregular crater, a multi-floored crater, or a crater field \citep{herrick1997}. There are 29 craters with $D_{\rm crat} \ge 20$~km that have one of these definitions, 9\% of the total observed (331 in all). The number of projectiles that leave no detectable surface expression is unknown. 

\begin{figure}[t!]
 \begin{center}
 \includegraphics[width=0.47\textwidth]{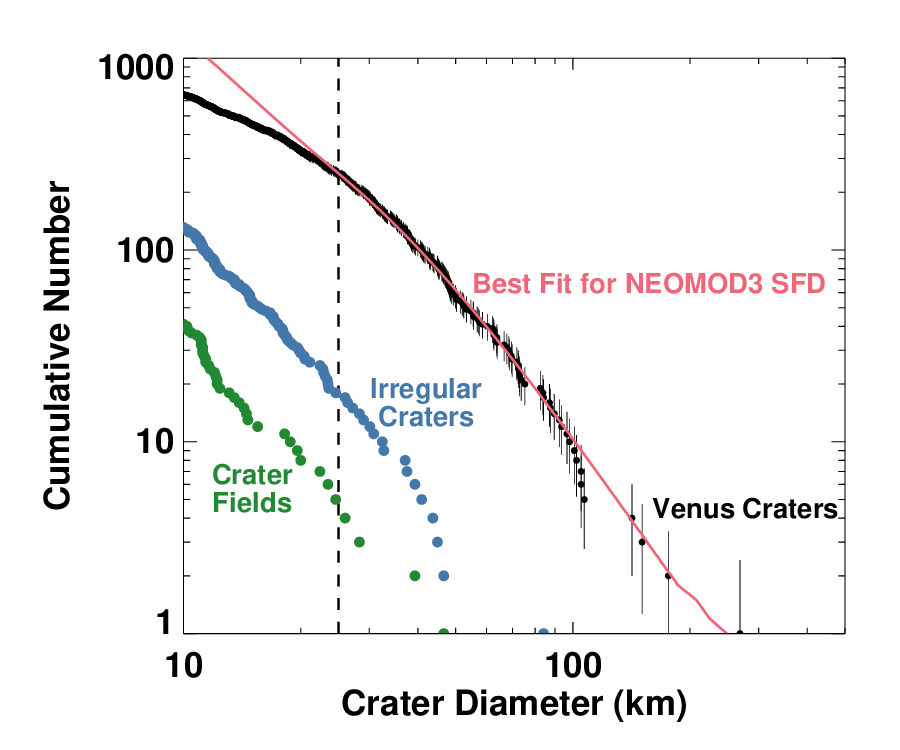} 
 \end{center} 
 \caption{Comparisons between the observed crater SFD on Venus and the shape of the NEOMOD3 SFD. The black dots are Venus craters (Release 3 of Venus Crater Database; see text). Error bars are the square root of the number of objects. The red line is the NEO SFD as defined by the NEOMOD3 model \citep {neomod3}. We have fit its shape to that of the crater SFD. The blue dots are irregular or multi-floored craters, while the green dots are crater fields \citep {herrick1997}. The dashed black line at $D_{\rm crat} \ge 25$~km indicates the crater size at which the observed crater SFD begins to diverge from the NEOMOD3 SFD, likely due to the effects of atmospheric breakup.} 
 \label{venus_crater}
\end{figure}

To account for atmospheric screening, we took the SFD derived for NEOs from \citet{neomod3} and fit its shape to that of the larger craters in the Venus crater SFD (Fig.~\ref {venus_crater}). We find a reasonably good match down to $D_{\rm crat} \ge 25$~km craters. Given that, we estimate that number of $D_{\rm crat} \ge 20$~km that would have formed on Venus without atmospheric screening over $650$~Myr was $\sim 380$. That falls close to our mean estimate, namely $355^{+145}_{-80}$ craters with $D_{\rm crat} \ge 20$~km. Accordingly, to reasonable approximation, we predict that the surface age of Venus is approximately the same as that derived using Earth's $D_{\rm crat} \ge 20$~km crater population, or $650$~Myr, give or take the order of $100$~Myr.  

One important consideration regarding this age estimate is that, as shown in Fig.\ref{fig_craters} \citep[see also][]{maz2019, ter2020}, the production rate of large craters was relatively low between $300$~Ma and the time of the Eulalia impact shower. Consequently, our age uncertainties should be adjusted to favor older surface ages. If Venus's surface were indeed modestly older than our current estimates suggest, it would shift a hypothesized global resurfacing event closer in time to the Eulalia impact shower. This in turn would raise the question of whether such timing was purely coincidental.

While it is intriguing to postulate a potential connection between impacts from the Eulalia family-forming event and the surface age of Venus, the existing physical evidence does not appear to support this hypothesis. Consider that over the last $\sim 650$~Ma, Venus has experienced numerous large impacts without any obvious surface-altering effects, including one that formed the enormous 280 km diameter Mead crater. We do not expect any impact from the Eulalia family to be comparable to the one that created Mead crater.  

We argue that the only way a connection could exist is if Venus was already predisposed to a catastrophic resurfacing event around $800$~Ma \citep[e.g.,][]{tur1999,romeo2010,tian2023}. For example, given that Venus currently lacks plate tectonics, which could release internal heat, Venus's mantle likely grew increasingly hot over time. Perhaps large impacts, occurring at a critical location and time, ruptured Venus's conductive lithospheric lid, resulting in a catastrophic, planet-wide volcanic eruption that coated the entire surface in lava. It is also possible that a significant impact triggered an upwelling plume originating from the core-mantle boundary \citep[e.g.,][]{oneill2020}, and that prompted some kind of runaway effect. For the time being, though, such ideas remain in the realm of speculation.

\section{Conclusions}\label{conclusions}

Here we summarize our findings regarding the Eulalia impact shower.

The earliest suggestion of an impact shower around $\sim 800$~Ma comes from \citet{zellner2009}. Through their analysis of $^{40}$Ar/$^{39}$Ar age profiles of lunar impact glasses, microscopic bead-like particles formed during cratering events and returned by the Apollo missions, they discovered that numerous glasses exhibit ages clustering around $800$~Ma. Additionally, the diverse compositions of these glasses indicated they originated from multiple distinct craters (Sec.~\ref{back3}).

Supporting evidence for an impact shower scenario was provided by \citet{ter2020}. They identified a significant increase in the production of large lunar craters ($D_{\rm crat} > 20$~km) approximately $800$~Ma. From their dataset, we identified seven craters whose mean ages range from $760$ to $820$~Ma: Al-Khwarizmi (22.5~km), Stefan (26~km), Saha (29.4~km), Godin (35.1~km), Das (36.6~km), Lowell (65.5~km), and Copernicus (93.1~km).  We consider them to be our baseline lunar impact shower (Sec.~\ref {back1}).  Using crater ages derived from \citet{ter2020} data, we also verified that during this putative impact shower, the production of large craters from the background impactor population was at least two times lower than present-day levels \citet{maz2019} (Sec.~\ref {back2}). 

We proposed that a plausible source for this putative shower was the disruption of the Eulalia asteroid parent body, which formed the Eulalia asteroid family near 800 Ma. Located near the J3:1, the Eulalia family consists of primitive, low-albedo C-complex asteroids and is thought to have originated from the breakup of a large parent body ($D > 100$~km). This family has also been identified as a plausible source for Bennu and Ryugu, the targets of the OSIRIS-REx and Hayabusa2 sample return missions, respectively. Both families have similar spectral signatures (and possibly similar compositions), so the Bennu and Ryugu could possibly come from either one.   

The potential link between the Eulalia family and the impact shower is further strengthened by the composition of the Copernicus impactor, which appears consistent with the characteristics of the samples returned by these missions (Sec.~\ref{methods1}).

The Eulalia family overlaps with other asteroid families, including the spectrally similar New Polana family and the S-complex Nysa/Hertha family. Using the best available data, including Sloan Digital Sky Survey color observations and albedo measurements from WISE, we worked to disentangle the Eulalia family from interlopers, thereby refining its characteristics (Sec.~\ref{ident}). Our work also addressed the impact of observational biases present in these datasets and developed methods to mitigate their influence on the apparent incompleteness of the Eulalia family (Sec.~\ref{biases}).

Using our refined definition of the Eulalia family, we used dynamical models to trace the evolutionary pathways of its members. By testing various estimates for the family's central location, we determined the fraction of asteroids that were directly injected into the J3:1 as well as those that migrated into the J3:1 through the combined effects of Yarkovsky thermal drift forces and YORP-induced rotational torques. Our model also allowed us to estimate the age of the family-forming event. 

Our preferred solution, called Solution~5, dates the event to approximately $T = 865$~Ma, with the Eulalia parent body residing at semimajor axis $a_{\rm c}=2.48$~au. Our analysis suggests that nearly three-fourths of the family was lost to the J3:1 over a timescale of $150$~Ma (Sec.~\ref{results}). The number of Eulalia family members reaching the J3:1 via Solution~5  are found in Table~\ref{bins} (Sec.~\ref{model1bis}). 

Our next goal was to analyze how the Eulalia family SFD was shaped by collisional evolution over its lifetime. Here we focused on the ``remnant Eulalia family" that remained behind in the main belt. Using collisional models, we simulated the changes in the remnant Eulalia family's SFD over the last $\sim 865$~Myr via impacts from the main belt population. Our results indicate that $3 < D < 8$~km asteroids in the remnant Eulalia family SFD lost approximately 1.5 times their population to comminution (Sec.~\ref{model2}).  

By combining these values with the collision probabilities for objects striking the terrestrial planets from the J3:1 (Sec.~\ref{model3}), we were able to estimate the total number of impacts occurring on Venus, Earth, the Moon, and Mars (Sec.~\ref{model4}). As shown in Fig.~\ref{fig_impacts}, our calculations indicate that the largest impactor to strike Earth and Venus during this event had an average diameter of approximately $D \approx 8$~km. This value is comparable to the estimated $D \sim 10$~km projectile that created Chicxulub crater 66 Ma on Earth. For Mars, the largest impactor was on average $4.5$~km, while for the Moon, it was $D > 3$~km.  While this size is insufficient to produce Copernicus crater, our analysis suggests the probability of generating Copernicus during the Eulalia impact shower remains reasonably plausible (Sec.~\ref{model4}).  

To evaluate the statistical likelihood of the Eulalia impact shower accounting for the large lunar craters formed around 800 Ma, we conducted a Monte Carlo simulation. As shown in Fig.~\ref{fig_monte}, our simulations fall largely within the $1\sigma$ envelope for reproducing the observed lunar craters associated with this period. Given existing uncertainties, we interpret this as encouraging evidence supporting the hypothesis that the Eulalia family disruption significantly contributed to the lunar impact record near $800$~Ma (Sec.~\ref{model4})

One of the most intriguing questions about the Eulalia asteroid breakup event is whether its impact shower left evidence on Earth (Sec.~\ref{imp1}). Unfortunately, the terrestrial crater record is challenging to analyze at these times due to extensive erosion from Snowball Earth glaciations ($650–720$~Ma). They likely obliterated most craters older than $650$~Ma (Sec.~\ref{imp1a}). 

Indirect evidence for large impacts may lie in the Bitter Springs Anomaly (BSA), a pronounced negative carbon isotope shift $\sim 800–810$~Ma, which suggests a global disruption in carbon cycling and primary productivity. Although the BSA has been associated with Rodinia's breakup, we instead propose that debris from the Eulalia breakup may have influenced atmospheric and oceanic chemistry, leading to widespread cooling and enhanced nutrient deposition.

In Sec.~\ref{imp1b}, we argue the BSA and its potential link to the Eulalia asteroid shower share intriguing parallels with both the Cretaceous–Paleogene (K–Pg) impact event on the Yucatán Peninsula 66 Ma and the breakup of the L chondrite parent body 466 Ma. Both analogies are discussed in some detail.  

The K–Pg event caused a significant $\sim 2$\textperthousand~negative excursion in $\delta\,^{13}$C in both terrestrial organic carbon and marine inorganic carbon. This anomaly was driven by the release of large amounts of $\text{CO}_2$, sulfate aerosols, and other gases, leading to global cooling, acid rain, and a disruption in the carbon cycle. The collapse of oceanic primary productivity, along with the release of $^{12}\text{C}$-enriched carbon from decaying organic matter and methane hydrate destabilization, contributed to the observed negative $\delta\,^{13}$C anomaly. We find it plausible that the largest projectile from the Eulalia asteroid shower caused similar behavior $\sim 800$~Ma. While Earth's environmental conditions were vastly different at that time, such an impact could have released large amounts of dust, debris, and aerosols into the atmosphere, leading to global cooling and widespread ecological disruptions. 

The breakup of the L chondrite parent body $\sim 466$~Ma, which like the Eulalia asteroid shower may have  occurred near a powerful resonance, produced an enhanced influx of dust and larger projectiles. Possible consequences include ice age conditions, stimulated biodiversity changes as part of the Great Ordovician Biodiversification Event (GOBE), and contributions to the subsequent Late Ordovician Mass Extinction (LOME). We propose that similar mechanisms from the Eulalia impact shower may explain the BSA while also contributing to putative eukaryotic diversification around $800$~Ma. Although our claims remain speculative, we consider it plausible that extraterrestrial breakups around $\sim 466$~Ma and $\sim 800$~Ma both played substantial roles in shaping key evolutionary and climatic milestones on Earth.

The Snowball Earth events of the Neoproterozoic era, including the Sturtian ($660–717$~Ma), Marinoan ($635–641$~Ma), and Gaskiers ($\sim 580$~Ma) glaciations represent some of the most extreme ice ages in Earth's history. These events were likely triggered by a combination of geological and climatic factors, though no single factor has been definitively identified as their primary cause. 

In Sec.~\ref{imp1c}, we speculate that large asteroid impacts from the Eulalia impact shower might have contributed to global cooling and initiated some of the glaciations by injecting vast quantities of aerosols and debris into the atmosphere. While no direct geological evidence has been found that can confirms this link, iridium anomalies found in cap carbonates from these glaciation periods may be a consequence of extraterrestrial impacts or natural cosmic dust accumulation. Additional research and data are needed to evaluate the role of asteroid impacts in triggering these extreme climate events.  

An examination of Mars reveals coincidental evidence of a pronounced peak around $\sim 800$~Ma in the formation ages of calderas associated with its largest volcanoes. This suggests a possible connection with the timing of the Eulalia asteroid shower. Other peaks in the Martian caldera timeline correspond to periods of known asteroid disruptions and associated asteroid showers, such as the L chondrite parent body breakup ($466$~Ma) and the Flora family disruption ($\sim 1300$~Ma). We postulate that asteroid impacts may have influence Martian volcanism in the Amazonian era via seismic shaking, which can destabilize magma reservoirs, enhance magma ascent, and trigger volcanic eruptions. This sets the stage to explore the interplay between asteroid impacts and volcanic activity on terrestrial worlds (Sec.~\ref{imp2}).

The potential effects of the Eulalia asteroid shower on Venus remain unclear, primarily due to the uncertainty surrounding the planet's surface age. Based on Earth's crater record and inner solar system impact flux models, we estimate Venus's surface age to be approximately $650$~Ma, but with age uncertainties of roughly 100 Myr. This timing is sufficiently close to that of the Eulalia impact shower ($800$~Ma) to warrant consideration of a potential link to a global resurfacing event on Venus. We propose that for any such link to exist, Venus must have been predisposed to resurfacing at that time, with large impacts acting as triggers for widespread tectonic and volcanic activity. While this scenario is thought-provoking, we emphasize that it remains highly speculative at this stage (Sec.~\ref{imp3}).


\acknowledgments
The work in this paper was supported by NASA’s Solar System Workings program through grant 80NSSC21K1829 and the Center for Lunar Origin and Evolution (CLOE), a team in NASA’s Solar System Exploration Research Virtual Institute (SSERVI) program (cooperative agreement 80NSSC23M0176). The work of DV was partially supported by the Czech Science Foundation (grant 25-16507S). We extend our gratitude to our two anonymous referees for their valuable and constructive feedback.  

\bibliographystyle{aasjournal}

\appendix

\section{Additional Background on the 800 Ma Impact Shower}

\subsection{Impact Signatures on Mars}\label{appa1}

Given that Mars and the Moon are both struck by components of the same terrestrial planet-crossing  population \citep {bot2002,neomod1}, features seen in Fig.~\ref{fig_craters} should presumably also be seen on Mars. At present, the best available results come from \citet{lag2022}. In their study, \citet{lag2022} analyzed superposed crater counts on 521 Martian impact craters with diameters of $D_{\rm crat} \ge 20$~km. All of their large craters were located within a $35^\circ$ latitudinal band and were selected based on the \citet{rh2012a,rh2012b} database, which was later updated by \citet{lag2022}. Craters were excluded from their study if they lacked an ejecta blanket, were poorly preserved, or had ejecta blankets that had been significantly modified by tectonic/geologic activity, volcanic processes, or large impacts.  

The methodology used by \citet{lag2022} was similar to that applied to the Moon by \citet{ter2020}, but with a key difference, namely that they employed automated crater identification techniques to measure the spatial densities of small craters on the crater floors and the ejecta blankets of large craters. This practice is potentially powerful, but it means that the algorithm must deal with several issues that are complicated for human geologists, such as selecting an appropriate surface for crater counting, how to deal with rugged terrains, the issue of sifting out secondary craters, and so on. 

Their Fig.~2 provides an example of the methodology applied to a $40$~km crater, and it highlights some of the challenges inherent to such automated methods. For example, approximately one-third of the predicted diameters of craters in their Fig.~2b appear to be slightly overestimated. Additionally, the small crater population seems to exhibit signs of secondary crater contamination; only primary craters should be used for accurately determining crater ages. Another challenge arises from the subtle distinctions between detected craters and those that are partially buried. For instance, one marginal crater-like feature, located between the 9 and 10 o'clock positions in Fig.~2, was not identified by the algorithm as a crater. Although crater counts performed by trained geologists are not free from biases \citep[e.g.,][]{rob2014}, automated algorithms have yet to demonstrate comparable or superior reliability and accuracy in their results.

With these caveats in mind, we turn our attention to Fig.~6 in \citet{lag2022}, which displays the estimated ages of $D_{\rm crat} \ge 20$~km Martian craters. The range of their ages covers a time span from the present epoch to $3.8$~Ga. The authors identified 49 craters younger than $600$~Ma, and about half that number in the $600–1000$~Ma range. Many of the craters have formal age uncertainties exceeding $100$~Myr. They observe modest increases in the impact flux near $600$~Ma, $800$~Ma, and $1000$~Ma, with the $800$~Ma feature notable for our hypothesis.  With that said, the substantial age uncertainties associated with the craters contributing to these peaks leave the significance and interpretation of these trends unclear.

On the positive side, this figure shows an approximately 1.9-fold increase in the production rate of large Martian craters around $300$~Ma. These findings agree with the results of \citet{maz2019} and \citet{ter2020}. Curiously, \citet{lag2022} dismiss their result because they argue it lacks statistical significance. They do not discuss the likelihood that the Earth, the Moon, and Mars all happened to show a comparable impact flux increase at approximately the same time. Although it is beyond the scope of this study, we suggest that combining and analyzing these three datasets as an ensemble would yield a more robust and statistically significant set of conclusions.

We suggest that a follow-up campaign to validate the ages of Martian craters reported in \citet{lag2022} may provide useful results. The objective would be to evaluate whether the potential biases discussed above had a significant impact on their results and, if so, to determine whether they could be mitigated through detailed geologic mapping and analysis.

\subsection{Understanding the nature of the lunar impact flux for sub-100 m impactors}\label{appa2}

There is a potential inconsistency in our hypothesis when all crater sizes are considered. Studies of the spatial densities of small craters ($\sim 0.1 < D_{\rm crat} < 1-2$~km diameter) on dated lunar terrains suggest that the impact rate of asteroids ranging in size from a few meters to a few tens of meters has remained relatively constant over the past $\sim 3$ to $3.5$~Gyr \citep[e.g.,][]{neu2001,sr2001,stoff2006}. At first glance, this finding seems to contradict the possibility of episodic asteroid showers impacting the Moon and other terrestrial planets. Crucially, however, these studies do not tell us what is going on with the flux of larger impactors, which we argue are more likely to be influenced by episodic surges in asteroid delivery.

To gain a better understanding of this issue, it is useful to consider the fundamentals of asteroid delivery from the main belt. Collisional and dynamical models indicate that most of the impactors for the terrestrial planets originate from the main belt \citep[see reviews by][]{bot2006,betal2015}. Asteroids with diameters $D < 30$~km are meaningfully affected by the Yarkovsky and YORP effects, thermal radiation forces and torques that alter a small body's semimajor axis and spin vector, respectively, depending on its spin state, orbital dynamics, and material properties. Over time, these mechanisms work together to gradually push asteroids into dynamical resonances with the planets, from which many can be driven onto planet-crossing orbits. This process takes time, though, and that means we have to consider how the escaping objects are affected by collisional processes.

As a thought experiment, consider what happens after the disruption of a hypothetical 100 km main belt asteroid, as shown in Figure~10 of \citet{bot2015}. For illustration purposes, the fragment SFD was chosen to span a wide range of sizes, from meter-scale fragments to bodies several tens of kilometers in diameter, and was given a steep power law slope. Using the collisional evolution model from \citet{coddem1,coddem2}, it was found that the smallest fragments were destroyed within a few Myr, while the multikilometer-sized and larger fragments survived for much longer timescales, ranging from hundreds of Myr to several Gyr. Although disruptions of larger fragments help replenish the smaller fragments, this process occurs too slowly to maintain the initial steep power-law slope of the SFD. As a consequence, the population of smaller fragments steadily decreases over time.

Eventually, the population of smaller bodies stabilizes into a quasi-equilibrium state, with an SFD whose shape is comparable to that of the background main belt. In contrast, the multikilometer-sized and larger fragments, which have the longest collisional lifetimes, experience only a modest decline. In practice, this means the SFD of the larger fragments often stay more or less consistent with the original power-law slope.

We can draw several conclusions from these results. First, small bodies generated by family-forming events are rapidly eroded by collisions, limiting their ability to substantially influence the background SFD of the main belt over extended timescales. Consequently, any temporary increase in the small body flux reaching resonances that supply the planet-crossing population will be relatively short-lived. This means the background main belt SFD should dominate the small body flux reaching terrestrial planet-crossing orbits over the past $\sim 3$ to $3.5$~Gyr. This helps explain why the spatial density of small lunar craters reliably functions as a ``clock" for dating surface ages on Mercury, the Moon, and Mars.    

Second, a steady flux is less applicable to kilometer-sized to multikilometer-sized asteroids, depending on the timescale involved. For certain asteroid families, the escape times for such fragments may be shorter than their collisional lifetimes \citep{coddem2,betal2015}. This implies that large and strategically positioned asteroid disruption events can generate a noticeable increase in the impact flux of large asteroids to the terrestrial planets over an extended period, provided the flux from the asteroid family exceeds that from the background main belt.

Third, the difference in the collisional evolution of small and large asteroids helps resolve an apparent paradox that could otherwise challenge the concept of asteroid showers. Imagine that the lunar impact flux was controlled by a metaphorical dial, with the populations of small and large projectiles always maintaining the same ratio. If our only means of measuring impact rates were the spatial densities of small craters, we would be unable to distinguish between a truly constant flux and one where periods of increased flux (``dial ups") were balanced by periods of decreased flux (``dial downs"). The only way to  distinguish between the two scenarios would be in obtaining ground truth. This would involve dating numerous lunar surfaces and/or large lunar craters through sample analysis and correlating these results with the spatial density of small craters superimposed on those larger craters.

Fortunately, we appear to have enough well-dated surfaces on the Moon that we can infer that the small body flux has remained relatively constant over the last $\sim 3$ to $3.5$~Gyr \citep[e.g.,][]{neu2001,sr2001,stoff2006}. As discussed above, we also have a theoretical framework that supports this view \citep[e.g.,][]{betal2015}. Specifically, collisional and dynamical modeling work demonstrates that family-forming events cannot sustain an elevated small body flux for long enough period to meanigfully affect lunar chronology. At best, a family-forming event may marginally increase the small body flux for just a few million years. 

In support of this, let us consider a second thought experiment, namely that the Eulalia breakup produced a ten-fold increase in the small body flux to the Moon. For the sake of argument, we will assume that Copernicus formed early enough in the shower that its surfaces received virtually all of this increased small body flux, and that the ten-fold increase lasted for 5 Myr. That would be the equivalent to $50$~Myr of extra impacts. That value can be compared to nearly $800$~Myr of small crater production on Copernicus. Our putative increase would only produce a 6\% change in the spatial density of small craters on Copernicus, not enough to yield a meaningful age change. Accordingly, for most large craters, the background production population of small craters will dominate any transient increases.

Given this, we consider it reasonable to expect the production rate of large craters on Earth and the Moon to exhibit periods of surges and lulls, as shown by \citet{maz2019} and \citet{ter2020} (Fig.~\ref{fig_craters}). Our theoretical framework for the delivery of asteroids onto planet-crossing orbits is consistent with this scenario.

\begin{figure}[t!]
 \begin{center}
 \includegraphics[width=0.47\textwidth]{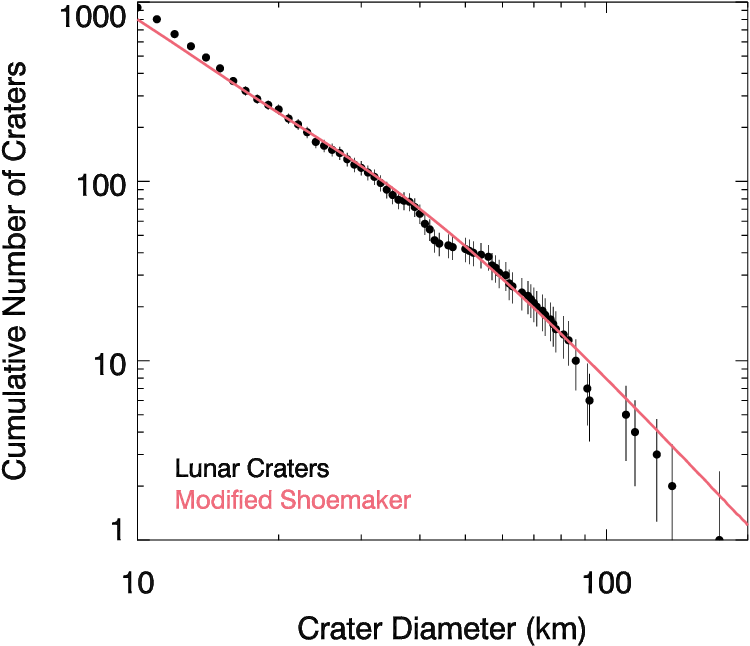} 
 \end{center} 
 \caption{Comparison between modeled and observed crater SFDs on the Moon. The red curve represents the modeled crater SFD, calculated by applying the Shoemaker crater scaling law \citep[see][]{shoemaker1990, stuart2004} to the NEO SFD from the NEOMOD3 model \citep{neomod3}. The black dots are Copernican- and Eratosthenian-era craters on the Moon as determined by \cite{wilhelms1978} \citep[see also][]{mcewen1997}. The red curve was fit to crater data at $D_{\rm crat} = 20$~km.} 
 \label{crater_scale}
\end{figure}

\subsection {Crater scaling for the Moon}\label{appa3}

There are many crater scaling laws in the literature used to convert projectiles into crater diameters, but it is not clear which ones provide the best results. In this study, we adopted the Shoemaker crater scaling law \citep {shoemaker1990}, as described in \citet{stuart2004} to convert projectile sizes into lunar crater diameters. We chose this one because it could be verified against real world test cases. Specifically, it enabled \citet{stuart2004} to replicate the lunar crater SFDs reported in \citet{hartmann1981} \citep[see also][]{ivanov2001} using the NEO SFD characterized by \citet{stuart2004}. A version of this scaling law has also been used to model meteoroid impacts on the Moon \citep {sheward2025}.

As a check on the results in \citet{stuart2004}, we conducted our own tests using the Shoemaker crater scaling law. Using the NEO SFD defined by the NEOMOD3 model \citep{neomod3}, we input it into Shoemaker scaling law and compared the resultant SFD with the crater SFD of combined Copernican and Eratosthenian-era craters as determined by \citet{wilhelms1978} \citep[see also][]{mcewen1997}. These craters are considered younger than $3.5$~Ga. Details on the counting surface and how the age definitions were made can be found in \citet{wilhelms1978}. They counted 251 craters with $D_{\rm crat} > 20$~km over $3.1 \times 10^7$~km$^2$ and 278 craters with $10 < D_{\rm crat} < 30$~km over $1.07 \times 10^{7}$~km$^2$. The impact velocity on the Moon was assumed to be $19.2$ km~s$^{-1}$, using dynamical results from \citet{neomod1, neomod2}. The projectile and target surface were assumed to have the same density.  

The model crater SFD derived from these data is presented in Fig.~\ref{crater_scale}. We find a reasonable match between the modeled and observed crater SFDs. This gives us increased confidence that the Shoemaker crater scaling law works reasonably well for lunar impacts over the last $\sim 3$~Gyr. We caution, however, that this scaling law has yet to be tested for craters on other worlds, so its use should be limited until further testing is completed.

\end{document}